\documentclass[fleqn,usenatbib]{mnras}

\usepackage{newtxtext,newtxmath}

\usepackage[T1]{fontenc}
\usepackage{longtable}
\usepackage{appendix}

\DeclareRobustCommand{\VAN}[3]{#2}
\let\VANthebibliography\thebibliography
\def\thebibliography{\DeclareRobustCommand{\VAN}[3]{##3}\VANthebibliography}

\usepackage{graphicx}	
\usepackage{amsmath}	
\usepackage{chemformula}
\usepackage{caption}
\usepackage{subcaption}

\title[GUAPOS V]{The GUAPOS project. V: The chemical ingredients of a massive stellar protocluster in the making}

\author[Á. López-Gallifa et al.]{
Á. López-Gallifa$^{1}$\thanks{E-mail: alvarolg@cab.inta-csic.es},
V. M. Rivilla$^{1}$,
M. T. Beltrán$^{2}$,
L. Colzi$^{1}$,
C. Mininni$^{3}$,
Á. Sánchez-Monge$^{4,5}$,
\newauthor 
F. Fontani$^{2}$,
S. Viti$^{6,7}$,
I. Jiménez-Serra$^{1}$,  
L. Testi$^{8}$,
R. Cesaroni$^{2}$,
and A. Lorenzani$^{2}$
\\
$^{1}$Centro de Astrobiología (CAB) CSIC-INTA, Ctra. de Ajalvir, km. 4, Torrejón de Ardoz, E-28850 Madrid, Spain\\
$^{2}$INAF-Osservatorio Astrofisico di Arcetri, Largo E. Fermi 5, I-50125, Florence, Italy\\
$^{3}$INAF-IAPS, via del Fosso del Cavaliere 100, I-00133 Roma, Italy\\
$^{4}$Institut de Ci\`encies de l'Espai (ICE, CSIC), Carrer de Can Magrans, s/n, E-08193 Bellaterra, Barcelona, Spain\\
$^{5}$Institut d'Estudis Espacials de Catalunya (IEEC), E-08034 Barcelona, Spain\\
$^{6}$Leiden Observatory, Leiden University, Huygens Laboratory, Niels Bohrweg 2, NL-2333 CA Leiden, The Netherlands\\
$^{7}$Department of Physics and Astronomy, University College London, Gower Street, WC1E 6BT, London, UK\\
$^{8}$Dipartimento di Fisica e Astronomia, Universitá di Bologna, Via Gobetti 93/2, 40122 Bologna, Italy
}

\date{Accepted XXX. Received YYY; in original form ZZZ}

\pubyear{2022}

\begin{document}
\label{firstpage}
\pagerange{\pageref{firstpage}--\pageref{lastpage}}
\maketitle

\begin{abstract}
Most stars, including the Sun, are born in rich stellar clusters containing massive stars. Therefore, the study of the chemical reservoir of massive star-forming regions is crucial to understand the basic chemical ingredients available at the dawn of planetary systems. We present a detailed study of the molecular inventory of the hot molecular core G31.41+0.31 from the project GUAPOS (G31.41+0.31 Unbiased ALMA sPectral Observational Survey). We analyze 34 species for the first time plus 20 species analyzed in previous GUAPOS works, including oxygen, nitrogen, sulfur, phosphorus, and chlorine species. We compare the abundances derived in G31.41+0.31 with those observed in other chemically-rich sources that represent the initial and last stages of the formation of stars and planets: the hot corino in the Solar-like protostar IRAS 16293–2422 B, and the comets 67P/Churyumov-Gerasimenko and 46P/Wirtanen. 
The comparative analysis reveals that the chemical feedstock of the two star-forming regions are similar. The abundances of oxygen- and nitrogen-bearing molecules exhibit a good correlation for all pair of sources, including the two comets, suggesting a chemical heritage of these species during the process of star formation, and hence an early phase formation of the molecules. However, sulfur- and phosphorus-bearing species present worse correlations, being more abundant in comets. This suggests that while sulfur- and phosphorus-bearing species are predominantly trapped on the surface of icy grains in the hot close surroundings of protostars, they could be more easily released into gas phase in comets, allowing their cosmic abundances to be almost recovered.

\end{abstract}




\section{Introduction}

One of the fundamental open questions in astrochemistry is to understand which is the chemical feedstock of star and planet birthsites, and how it is transferred during the different phases of evolution of star and planetary systems: from parental molecular clouds, to starless/prestellar cores, to protostars, to planet-forming disks, and to planetary objects (planets, comets and asteroids). Nowadays, it is debated to what extent the molecular content is conserved or significantly reprocessed between the different evolutionary stages. 

It has been shown that water ices from the Solar Nebula were widely available for the formation of the subsequent planetary system \citep{Cleeves2014,Altwegg2017a}, and that at least most of them could survive to the process of accretion in planet-forming disks \citep{Oberg2011, Visser2009}. The detection of molecules in protoplanetary disks also suggest that
the chemical reservoir found in planet-forming sites might be directly inherited from earlier phases (e.g., \citealt{Loomis2018, Bianchi2019, Booth2021, vanderMarel2021, Ceccarelli2023, Tobin2023}). Moreover, the molecular abundances of some species derived in planet-forming disks are similar to those found in comets \citep{Oberg2015}.
Supporting this chemical heritage, \cite{Bockelee-Morvan2000} found that the abundances of some selected molecules present in the comet Hale-Bopp are similar to those of several interstellar high-mass star-forming regions. More recently, 
\cite{Drozdovskaya2019} showed that the molecular abundances of the comet 67P/Churyumov-Gerasimenko and the Solar-like protostellar environment IRAS 16293-2422 B are reasonably well correlated. Furthermore, \cite{Coletta2020} found that the molecular ratios of two complex organic molecules (COMs, species with $>$5 atoms; \citealt{Herbst_van_dishoeck_2009}), methyl formate and dimethyl ether, are nearly constant in multiple sources at different evolutionary stages of star and planetary system formation (molecular clouds, prestellar cores, low-, intermediate- and high-mass star-forming regions, protostellar shocks, and comets).

Since there is a growing evidence that our Sun was born in a dense stellar cluster that included also massive stars \citep{Carpenter2000, Lada2003, Porras2003, Reach2009, Adams2010, Oberg2011, Dukes2012, PfalznerVincke2020, Korschinek2020, Brinkman2021}, the study of the chemical composition of massive star-forming regions can give us information about the molecular feedstock of an environment that might resemble the birthsite of our Solar System. 
This is the main purpose of the project GUAPOS (G31.41+0.31 Unbiased ALMA sPectral Observational Survey), which carried out a spectral survey of the whole ALMA Band 3 towards the massive star-forming region G31.41+0.31 (hereafter G31.41). This region, with a mass of 70 \(\textup{M}_\odot\) \citep{Cesaroni2019}, harbors a hot molecular core (HMC) where at least four massive protostars are forming \citep{Beltran2021}. 
G31.41 is located at 3.75 kpc distance \citep{Immer2019}, and has a luminosity of $\sim 4.4 \times 10^{4}$ \(\textup{L}_\odot\) \citep{Osorio2009}. The G31.41 hot core is one of the most chemically-rich sources in the Galaxy (\citealt{Beltran2009, Rivilla2017, Suzuki2023}).
The initial exploitation of GUAPOS data were focused on selected categories of COMs, like the isomers of glycolaldehyde (\ch{HCOCH2OH}; \citealt{Mininni2020}; hereafter GUAPOS I), peptide-like bond molecules (\citealt{Colzi2021}; hereafter GUAPOS II), several oxygen- and nitrogen-bearing COMs (\citealt{Mininni2023}; hereafter GUAPOS III); and phosphorus-bearing molecules (\citealt{Fontani2024}; hereafter GUAPOS IV).

The aim of this work, GUAPOS V, is to extend the chemical chemical census of the G31.41 massive star-forming region already presented in previous GUAPOS works, with the aim of comparing it with those of other well-studied astronomical objects at different evolutionary stages of the star- and planet-formation process. We have chosen sources that have been extensively studied in a homogeneous way, i.e. using the same dataset analyzed with the same method, and providing chemical censuses of tens of species, which will allow us to study the chemical budgets of different chemical families in a statistical way.
The chosen sources to perform the comparison are: the low-mass proto-Solar analogue IRAS 16293-2422 B (hereafter IRAS16B; \citealt{Jorgensen2016, Drozdovskaya2019}), and the comets 67P/Churyumov-Gerasimenko (hereafter 67P/C-G; \citealt{Rubin2019, Drozdovskaya2019}), and 46P/Wirtanen (hereafter 46P/W; \citealt{Biver2021}). In Sect. \ref{sec:observations} we present the ALMA observations of G31.41 used in our analysis. 
In Sect. \ref{sec:sample} we describe the census of molecules that we have used to compare the chemical reservoir of G31.41 with those of other sources.
In Sect. \ref{sec:data analysis} we present the molecular data analysis of the G31.41 spectra. In Sect. \ref{sec:comparison}, we quantitatively compare the molecular abundances derived in G31.41 with those of other three sources using three complementary statistical tests. In Sect. \ref{sec:discuss} we discuss the implications of our results, and in Sect. \ref{sec:conclusions} we summarize the main conclusions of this work.

\begin{table*}
\caption{\label{tab:Molecules_G31_GUAPOS_works} Molecules analyzed on G31.41 throughout all GUAPOS publications that are used in this work. The molecules that are not detected, for which we provide column density upper limits, are indicated in italics.}

\begin{tabular}{ccccccc}
\hline

\multicolumn{7}{c}{This work}
\tabularnewline \hline
  HCN & HNC & CO & \ch{H2CO}  & \textit{\ch{H2S}} & \ch{CH3CCH} & \ch{CH3NC} \\
  \ch{NH2CN} & \textit{\ch{C3H6}} & \textit{HOCN} & c-\ch{C2H4O} & \textit{syn-\ch{CH2CHOH}} & CS & \textit{\ch{PN}} $\dagger$ \\
 \ch{H2CS} & trans-HONO* & \textit{\ch{PO}} $\dagger$ & \ch{CH3SH} & SO & \textit{\ch{CH3Cl}} & \ch{HC3N}\\
\textit{\ch{HC2NC}} & \textit{trans-\ch{C2H3CHO}} & \textit{\ch{HOCH2CN}} &  \textit{\ch{CH3CHCH2O}} & \ch{C2H5CHO}* & OCS & \textit{gauche-\ch{C2H5SH}} \\
 \ch{CH3OCH2OH}* & \textit{\ch{S2}} & \ch{SO2} & \textit{\ch{HS2}} & \textit{\ch{H2S2}} & \textit{\ch{NH2CH2COOH} (conf. I)} \\
  \hline
\multicolumn{7}{c}{Previous publications}
\tabularnewline \hline
\ch{NH3} $^e$ & \ch{CH3OH} $^c$ & \ch{CH3CN} $^c$ & HNCO $^b$ & \ch{CH3CHO} $^c$ & \ch{NH2CHO} $^b$ & \ch{C2H5OH} $^c$ \\
\ch{CH3OCH3} $^c$ & trans-HCOOH $^d$ & \ch{CH3NCO} $^b$ & \ch{C2H3CN} $^c$ & \ch{CH3COCH3} $^c$ & \ch{CH3C(O)NH2} $^b$ & \ch{CH3NHCHO} $^b$ \\
\ch{C2H5CN} $^c$ & \ch{CH3OCHO} $^a$ & \ch{CH2OHCHO} $^a$ & \ch{CH3COOH} $^a$ & aGg’-(CH$_2$OH)$_2$ $^c$ &  gGg’-(CH$_2$OH)$_2$ $^c$ \\ \hline
\end{tabular}
References. $a$) GUAPOS I $-$ \protect\cite{Mininni2020}; $b$)  GUAPOS II $-$ \protect\cite{Colzi2021}; $c$) GUAPOS III $-$ \cite{Mininni2023}; $d$) \protect\cite{Garcia-de-la-concepcion-2022}; $e$) \protect\cite{Cesaroni1994}. Molecules marked with * are tentative detections. $\dagger$ We note that PN and PO are not detected towards the position of the hot molecular core analyzed in this work, but that they are detected towards shocked gas in the region that is analyzed in GUAPOS IV (\citealt{Fontani2024}).
\end{table*}

\section{ALMA Observations of G31.41+0.31}
\label{sec:observations}

The observations were carried out with ALMA (Atacama Large Millimeter Array) during the Cycle 5 as part of the project 2017.1.00501.S (PI: M. T. Beltrán). The phase center was $\alpha_{\mathrm{J}2000} = 18^{\mathrm{h}}45^{\mathrm{m}}34^{\mathrm{s}}$ and $\delta_{\mathrm{J}2000} = - 0.1^{\circ}12'45''$. We fully covered the Band 3 (86.05 GHz - 115.91 GHz) obtaining an unbiased spectral survey. The frequency  resolution of the correlator setup was 0.49 MHz equivalent to a velocity resolution of $\sim$ 1.6 km s$^{-1}$ at 90 GHz. The datacubes of the whole survey were created using a common restoring beam of 1.2'', about 4500 au. 
The uncertainties in the flux calibration are $\sim$5$\%$ (from Quality Assesment 2 reports), which is in good agreement with flux uncertainties at ALMA band 3 reported in \cite{Bonato2018}.
For more details regarding the observations we refer to  \cite{Mininni2020}.

\section{Sample of molecules}
\label{sec:sample}

The molecules considered in this study were selected because they have been detected or searched for (providing an upper limit for their column density) towards at least two of the four sources considered (the high- and low-mass star-forming regions G31.41 and IRAS16B, respectively, and the comets 67P/C-G and 46P/W). The total number of molecules considered for the comparative study are 57 and the information about the detection or non detection of the molecules towards the four astronomical sources are listed in Table \ref{tab:Molecules_used_each_source}.
For IRAS16B, we have used the abundances and upper limits reported in \citet{Drozdovskaya2019} (see also references therein), \cite{Martin-Domenech2017}, \cite{Coutens2018,Coutens2019}, \cite{Calcutt2019} and \cite{Manigand2021}.
Regarding 67P/C-G, we used the results presented in \cite{Calmonte2016}, \cite{Dhooghe2017}, \cite{Fayolle2017}, \cite{Altwegg2019}, \cite{Hadraoui2019}, \cite{Rubin2019}, \cite{Schuhmann2019} and \cite{Rivilla2020}; while for 46P/W we considered the information reported in \citet{Biver2021}.
For G31.41, we have used the molecules already reported in previous works (GUAPOS I, \cite{Mininni2020}; II \cite{Colzi2021}; III, \cite{Mininni2023}; \citealt{Garcia-de-la-concepcion-2022}; and \citealt{Cesaroni1994}), and the ones analyzed in this work (see Sect. \ref{sec:data analysis}). For \ch{NH3}, we used the data presented by \cite{Cesaroni1994}, based on VLA observations with a beam of ~1.3" x 1", which is very similar to that of the GUAPOS ALMA observations, and hence allows a fair comparison.

\section{Data analysis of the G31.41+0.31 spectra}
\label{sec:data analysis}

We present here the analysis of the molecular emission of 34 molecules in the GUAPOS spectral survey, along with other 20 molecules analyzed in previous works (see Table \ref{tab:Molecules_G31_GUAPOS_works}). 
The ALMA data were calibrated and imaged with CASA\footnote{https://casa.nrao.edu} (the Common Astronomy Software Applications package, \citealt{McMullin2007}). Since G31.41 presents an extremely rich spectrum, with no line-free channels, we did not perform the continuum substraction in the uv-space before imaging. For each observed basebands, the spectrum including the continuum level was extracted from an area equal to the synthesized beam (1.2") and centered on the continuum peak of our source. The root mean square (rms) noise of the spectra is 7-27 mK. More details are described in \cite{Mininni2020}. To subtract the continuum in the extracted spectra, as described in \cite{Colzi2021}, we applied to each baseband spectrum the sigma clipping method (c-SMC) of the python based tool STATCONT \footnote{https://hera.ph1.uni-koeln.de/~sanchez/statcont} \citep{Sanchez-Monge2018}. 

To perform the spectral analysis (line identification and fitting) we use the Madrid Data Cube Analysis (MADCUBA) software\footnote{MADCUBA is developed at the Centro de Astrobiología (CAB) in Madrid, free access: http://cab.intacsic.es/madcuba/Portada.html} \citep{Martin2019}. We searched for different molecules in the spectrum, and we fit them using the SLIM (Spectral Line Identification and Modeling) tool of MADCUBA, which incorporates the Cologne Database for Molecular Spectroscopy \footnote{http://cdms.astro.uni-koeln.de/classic/}(CDMS, \citealt{Muller2001, Muller2005, Endres2016}) and the Jet Propulsion Laboratory molecular catalogs\footnote{https://spec.jpl.nasa.gov/ftp/pub/catalog/catdir.html} (JPL, \citealt{Pinckett1998}). 

SLIM produces a synthetic spectrum assuming local thermodynamic equilibrium (LTE) conditions, and taking into account the line opacity (see formalism in \citealt{Martin2019}). 
Considering the high density of the G31.41 molecular core, which is $\sim$ $10^8$ cm$^{-3}$ \citep{Mininni2020}, LTE is a good approximation. SLIM compares the synthetic LTE spectrum with the observed spectrum, and finds the best fit applying the AUTOFIT tool based on a Levenberg-Marquardt algorithm. The free physical parameters of the fit are: the column density ($N$), the excitation temperature ($T_{\rm ex}$), the velocity ($v-v_{\rm 0}$), where $v_{\rm 0}$ is the systemic velocity of G31.41 (96.5 km s$^{-1}$, \citealt{Beltran2018}), and the full width at half maximum (FWHM). As discussed in \cite{Mininni2020} and \cite{Colzi2021}, the molecular emission fills the beam, and hence no beam dilution correction was applied. 

G31.41 is a chemically-rich source, so the blending among lines of different species is frequent. To take this into consideration,
we performed the fit of each molecule accounting also for the emission of all other species detected. To run AUTOFIT, we have selected the most unblended transitions of each molecule, and also those transitions that, together with the emission from other known species, well reproduce the observed spectrum. If possible, we have left free the four physical parameters of the fit. In case the AUTOFIT algorithm did not converge, we fixed some of the parameters. 
Unfortunately, in many cases AUTOFIT does not converge if $T_{\rm ex}$ is set as a free parameter due to the lack of enough transitions free of contamination and covering a broad range of energy levels. Thus, $T_{\rm ex}$ cannot be directly derived, and we assumed a value for it. For the most complex molecules with $\geq$6 atoms, which are expected to trace the hot molecular core, we have assumed 150 K (similar to the values derived by \citealt{Rivilla2017,Mininni2020,Colzi2021}); while for molecules with  $<$ 6 atoms, which likely trace colder gas, we have assumed 50 K. To evaluate the impact of these assumptions on the derived column densities, we have also repeated the fits using a range of temperatures: 100 and 200 K for molecules with $\geq$6 atoms, and 25 and 75 K for molecules with $<$ 6 atoms, respectively. As shown in Table \ref{tab:Poperties_molecules_on_G31} and in Table \ref{tab:Fits_isotopologues_molecules_on_G31}, the resulting column densities in these temperature ranges are, in most cases, within less than a factor of 2 with respect to the values derived using the assumed values of 50 K or 150 K.
In addition, to allow the convergence of AUTOFIT in some cases, $v-v_{\rm 0}$ and/or FWHM were fixed to 0 km s$^{-1}$ and 7 km s$^{-1}$, respectively, which  fit properly the most unblended transitions.

Some of the molecules analysed present high abundances and thus they are optically thick. As a consequence, the derived $N$ should be considered as a lower limit. In these cases, we have searched for and analysed their optically-thinner isotopologues to obtain a better estimation of the column density. If more than one isotopologue is well detected, we have adopted the optically thinnest to derive the $N$ of the main isotopologue. We used the following values of the isotopic ratios, which depends on the galactocentric distance ($D_\mathrm{GC}$), assuming $D_\mathrm{GC}$ for G31.41 of 5.02 kpc (calculated from the heliocentric distance of 3.75 kpc, \citealt{Immer2019}):
$\mathrm{^{12}C/^{13}C}$ = 39.5 $\pm$ 9.6 (\citealt{Milam2005, Yan2019}), $\mathrm{^{14}N/^{15}N}$ = 340 $\pm$ 90 (\citealt{Colzi2018}), $\mathrm{^{32}S/^{34}S}$ = 14.6 $\pm$ 2.1 (\citealt{Yu2020}) and $\mathrm{^{16}O/^{18}O}$ = 330 $\pm$ 140 (\citealt{Wilson1999}).
Moreover, for isotopic ratios that are not sensitive to $D_\mathrm{GC}$, we used:
$\mathrm{^{34}S/^{33}S}$ = 5.9 $\pm$ 1.5 (\citealt{Yu2020}), $\mathrm{^{34}S/^{36}S}$ = 115 $\pm$ 17 (\citealt{Mauersberger1996}) and $\mathrm{^{18}O/^{17}O}$ = 3.6 $\pm$ 0.2 (\citealt{Wilson1999}, taking the value of local ISM).

The molecular abundances were obtained by using the column density of $N$(H$_{2}$) = (1.0 $\pm$ 0.2)$\times$10$^{25}$ cm$^{-2}$ derived from the continuum emission within the same region used to extract the spectrum (more details in \citealt{Mininni2020}).

In the following sections we describe in detail the fitting procedure and the results obtained for each molecular species (including isotopologues) detected, the tentative detections and the upper limits derived from the non-detected molecules.
The spectroscopic information of all the molecules used in this work is summarized in Table \ref{tab:Entries_CDMS_JPL_molecules_G31}. The transitions used for fitting each species are listed in Table \ref{tab:Transitions_Molecules_on_G31}.

\subsection{Detected molecules}
\label{sec:detceted_mols}

We summarize in Table \ref{tab:Poperties_molecules_on_G31} the results of the line fitting of the different molecules. For optically thick molecules, we show in the table only the result of the optically thin isotopologue used to derive the column density of the main isotopologue. The results of the fits of the remaining isotopologues, including the main ones, are shown in Table \ref{tab:Fits_isotopologues_molecules_on_G31} instead. The detailed analysis of each molecule (and its isotopologues) are described in the following sections.


\subsubsection{Hydrogen cyanide (HCN)}
\label{mols:HCN}

The spectral profiles of the $J$=1$-$0 rotational transitions of HCN and H$^{13}$CN show clear absorptions (left and middle upper panels of Fig. \ref{fig:HCN_HNC}), which are likely due to filtering of extended emission by the interferometer and/or possible infall motions. To derive the molecular abundance of HCN, we have then used the $^{15}$N isotopologue, which shows a Gaussian emission profile (upper right panel of Fig. \ref{fig:HCN_HNC}). The results of the fit are shown in Table \ref{tab:Poperties_molecules_on_G31}. The derived line opacity ($\tau$) is $\sim$1.7, which indicates that even the $^{15}$N isotopologue is optically thick. The column density of HCN derived, assuming the $^{14}$N/$^{15}$N ratio, is (1.5 $\pm$ 0.4)$\times$10$^{18}$ cm$^{-2}$. We warn that, due to the opacity, this value is likely a lower limit of the true column density.

\subsubsection{Hydrogen isocyanide (HNC)}
\label{mols:HNC}

Similarly to HCN, the spectral profiles of the $J$=1$-$0 rotational transitions of HNC and HN$^{13}$C show absorptions (left and middle lower panels of Fig. \ref{fig:HCN_HNC}). Unlike HC$^{15}$N, the $J$=1$-$0 transition of H$^{15}$NC is heavily contaminated with a transition of \ch{CH3NH2} (right lower panel of Fig. \ref{fig:HCN_HNC}). 
Therefore, we derived an upper limit for its column density (procedure explained in Sect. \ref{sec:not_detceted_mols}) using the H$^{15}$NC isotopologue, and we retrieve the upper limit of the HNC column density applying the $^{14}$N/$^{15}$N ratio (Table \ref{tab:Poperties_molecules_on_G31}). We obtain a high HCN/HNC ratio >136, as expected in hot gas \citep{Colzi2018, Hacar2020}.

\subsubsection{Carbon monoxide (CO)}
\label{mols:CO}

Similarly to HCN and HNC, the spectral profiles of the $J$=1$-$0 rotational transitions of CO and $^{13}$CO show strong absorptions (Fig. \ref{fig:CO}), so we searched for optically thinner isotopologues. The isotopologue C$^{17}$O is heavily blended with two bright transitions of \ch{CH3COOH} (Fig. \ref{fig:CO}), and thus we report an upper limit (procedure explained in Sect. \ref{sec:not_detceted_mols}). The transition of the C$^{18}$O isotopologue is detected, although partially blended with emission of \ch{CH3OCHO} (see Fig. \ref{fig:CO}). The transition of the optically thinner double isotopologue $^{13}$C$^{18}$O is also detected (Fig. \ref{fig:CO}), only slightly blended with a transition of \ch{CH3NCO}. 
We used the $^{13}$C$^{18}$O isotopologue to derive the column density of CO taking into account the slight contribution of \ch{CH3NCO}, and using the corresponding carbon and oxygen isotopic ratios. The results of the fits are shown in Table \ref{tab:Poperties_molecules_on_G31} and Table \ref{tab:Fits_isotopologues_molecules_on_G31}.
The derived CO/\ch{H2} ratio is 2.6$\times$10$^{-4}$, which is consistent with the typical value found in molecular clouds (e.g. \citealt{Dickman1978, Lis1988, Duan2023}).

\subsubsection{Formaldehyde (\ch{H2CO})}
\label{mols:H2CO}

The brightest transition of \ch{H2CO}, $6_{1,5} \rightarrow 6_{1,6}$ at 101.3330 GHz (Table \ref{tab:Transitions_Molecules_on_G31}), is detected and appears unblended, but it is heavily optically thick ($\tau$ = 4), while other weaker transitions appear heavily blended (Fig. \ref{fig:H2CO}). Thus, we searched for rarer isotopologues. The $^{13}$C and the $^{17}$O isotopologues are detected, but their transitions are partially optically thick ($\tau$ = 0.24 and 0.4, respectively, see Table \ref{tab:Fits_isotopologues_molecules_on_G31}). Moreover, the transition of HC$_2^{17}$O is partially blended with \ch{CH3OH}. The 6$_{1,5} \rightarrow $ 6$_{1,6}$ transition of the $^{18}$O isotopologue is unblended, and it is optically thin ($\tau$ = 0.053). Therefore, we used it to derive the column density of H$_2$CO, obtaining (2.6 $\pm$ 1.2)$\times$10$^{18}$ cm$^{-2}$ (Table \ref{tab:Poperties_molecules_on_G31}).

\subsubsection{Propyne (\ch{CH3CCH})}
\label{mols:CH3CCH}

We fitted the two $K$-ladders of the $J$=5$-$4 and $J$=6$-$5 rotational transitions showed in Fig. \ref{fig:CH3CCH}. Some of the transitions are blended with \ch{CH3COCH3}, aGg'-\ch{(CH2OH)2} and \ch{CH3OCHO}. The observed spectrum is well reproduced with the combined emission of \ch{CH3CCH} and the blending species, as shown in Fig. \ref{fig:CH3CCH}. If we leave $T_{\mathrm{ex}}$ as a free parameter the AUTOFIT algorithm does not converge, hence we fixed it to 150 K.
We derived a column density of (1.20 $\pm$ 0.09)$\times$10$^{17}$ cm$^{-2}$ (see Table \ref{tab:Poperties_molecules_on_G31}).

\subsubsection{Methyl isocyanide (\ch{CH3NC})}
\label{mols:CH3NC}

We fitted the $K$-ladder of the $J$=5$-$4 rotational transition. As shown in Fig. \ref{fig:CH3NC}, the $K$=0 and $K$=1 transitions (at 100.5265 GHz and 100.5242 GHz respectively, see Table \ref{tab:Transitions_Molecules_on_G31}) are slightly blended with aGg'-($\mathrm{CH_2OH)_2}$. In addition, the $K$=2 transition (at 100.5174 GHz) is blended with \ch{CH3OCHO} and $K$=3 (at 100.5061 GHz) is blended with \ch{HCOCH2OH} and \ch{CH3COCH3}. The blending of the higher energy transitions of the $K$-ladder (see Table \ref{tab:Transitions_Molecules_on_G31}) does not allow convergence of the AUTOFIT algorithm leaving $T_{\mathrm{ex}}$ as a free parameter, so we fixed it to 150 K. We retrieved a column density of (1.1 $\pm$ 0.4)$\times$10$^{15}$ cm$^{-2}$ (see Table \ref{tab:Poperties_molecules_on_G31}).

\subsubsection{Cyanamide (\ch{NH2CN})}
\label{mols:NH2CN}

\begin{figure*}
\includegraphics[scale=0.5]{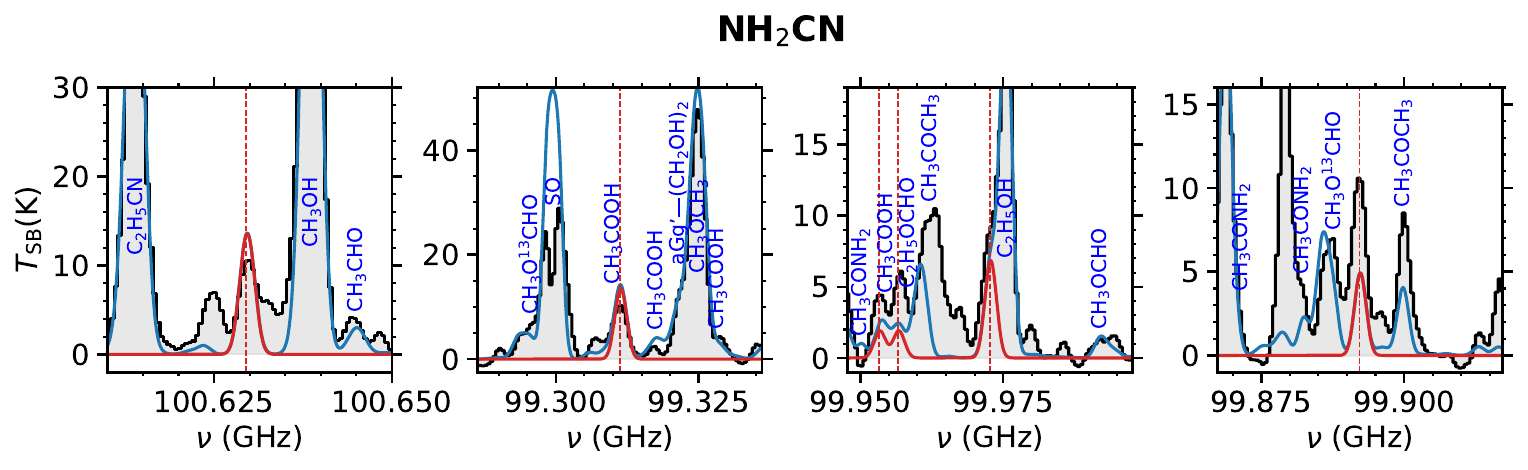}
\caption{\ch{NH2CN} described in Sect. \ref{mols:NH2CN}. The black histogram and its grey shadow are the observational spectrum. The red curve is the fit of the individual transition and the blue curve is the cumulative fit considering all detected species. The red dashed lines are the frequency of the transitions that we are fitting. The plots are sorted by decreasing intensity of the corresponding species (red line).} 
\label{fig:NH2CN}
\end{figure*}

We show in Fig. \ref{fig:NH2CN} the most intense and less blended transitions of this species (Table \ref{tab:Transitions_Molecules_on_G31}). The $5_{1,5} \rightarrow 4_{1,4}$ transition at 99.3112 GHz is only  marginally contaminated by \ch{CH3COOH}. The other transitions of the right panels of Fig. \ref{fig:NH2CN}, are partially blended with \ch{CH3COOH}, \ch{C2H5OCHO}, \ch{C2H5OH} and with a transition not identified, respectively.
We note that all other transitions predicted by the LTE model are consistent with the observed spectra, but they are heavily blended with emission from other species.
The transitions used for the fit are partially optically thick ($\tau$ = 0.04 - 0.34; Table \ref{tab:Poperties_molecules_on_G31}).
We searched for the $^{13}$C isotopologue, but it is not detected. The \ch{NH2CN}/\ch{NH2CHO} ratio in G31.41 is 0.007, using the value derived here and that reported by \citet{Colzi2021} for formamide. For comparison, in IRAS16B is 0.02 \citep{Coutens2018}, which is only a factor three of difference suggesting that the \ch{NH2CN} column density that we have derived for G31.41 is more or less reliable.

\subsubsection{Carbon monosulphide (\ch{CS})}
\label{mols:CS}

\begin{figure*}
     \centering
        \begin{subfigure}{0.24\textwidth}
              \centering
         \includegraphics[scale=0.5]{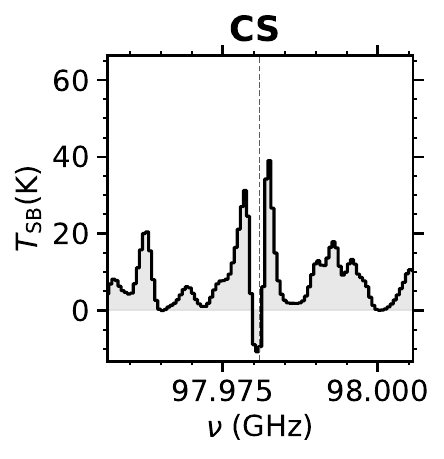}
     \end{subfigure}
     \begin{subfigure}{0.24\textwidth}
              \centering
         \includegraphics[scale=0.5]{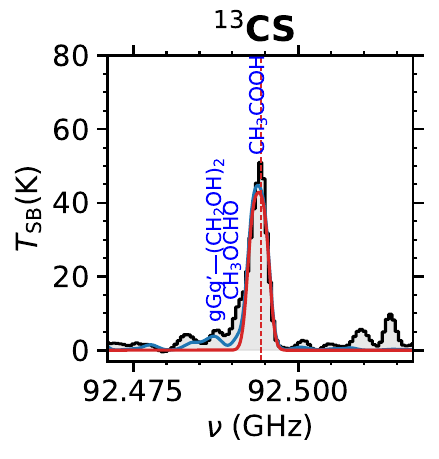}
     \end{subfigure}
     \begin{subfigure}{0.24\textwidth}
         \centering
        \includegraphics[scale=0.5]{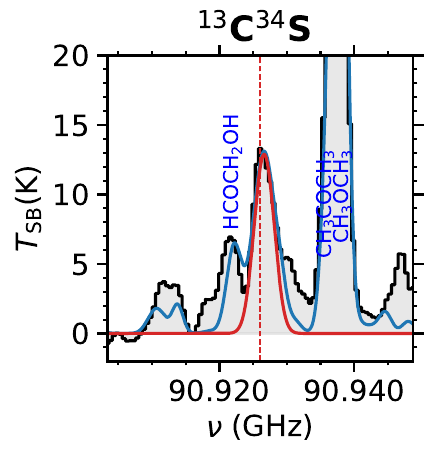}
     \end{subfigure}
     
     \begin{subfigure}{0.24\textwidth}
         \centering
         \includegraphics[scale=0.5]{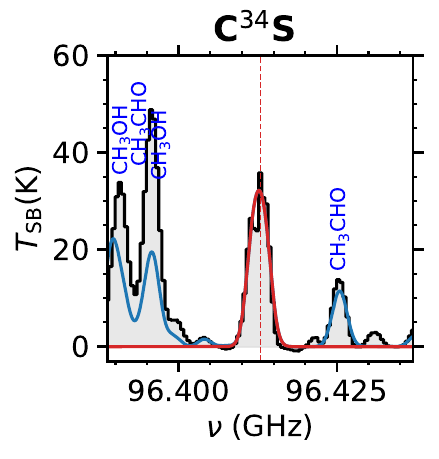}
     \end{subfigure}
               \begin{subfigure}{0.24\textwidth}
         \centering
         \includegraphics[scale=0.5]{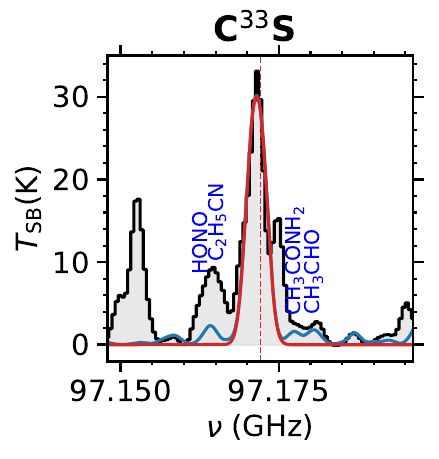}
     \end{subfigure}
          \begin{subfigure}{0.24\textwidth}
         \centering
         \includegraphics[scale=0.5]{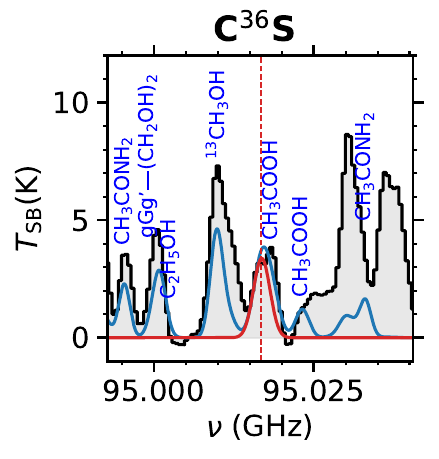}
     \end{subfigure}
        \caption{CS isotopologues described in Sect. \ref{mols:CS}. The black histogram and its grey shadow are the observational spectrum. The red curve is the fit of the individual transition and the blue curve is the cumulative fit considering all detected species. The red dashed lines indicate the frequency of the molecular transitions. The plots are sorted by decreasing intensity of the corresponding species (red line).}
        \label{fig:CS}
\end{figure*}

The only rotational transition of CS in the GUAPOS spectral range is the $2 \rightarrow 1$ at 97.9810 GHz (see Table \ref{tab:Transitions_Molecules_on_G31}), which is optically thick and it shows a clear absorption probably due to filtering of extended emission and/or infall (see Fig. \ref{fig:CS}). The transitions of the isotopologues $^{13}$CS, C$^{34}$S and C$^{33}$S are unblended, but they also are optically thick, with $\tau$ of 2.5, 1.17 and 1.0 respectively (see Table \ref{tab:Transitions_Molecules_on_G31}). The isotopologues C$^{36}$S and $^{13}$C$^{34}$S are also detected with no or very little contamination (Fig. \ref{fig:CS}). For the fit of C$^{33}$S and C$^{36}$S isotopologues, we fixed the FWHM to 9.5 km s$^{-1}$ (see Table \ref{tab:Poperties_molecules_on_G31} and Table \ref{tab:Fits_isotopologues_molecules_on_G31}), derived from the fit of C$^{34}$S, instead of the 7 km s$^{-1}$ generally used when no further information are available and the AUTOFIT do not converge. We used the optically thinner isotopologue, C$^{36}$S ($\tau$ = 0.08), to derive the column density of CS, which is shown in Table \ref{tab:Poperties_molecules_on_G31}.

\subsubsection{Ethylene oxide (\ch{c-C2H4O})}
\label{mols:c-C2H4O}

\begin{figure*}
\includegraphics[scale=0.55]{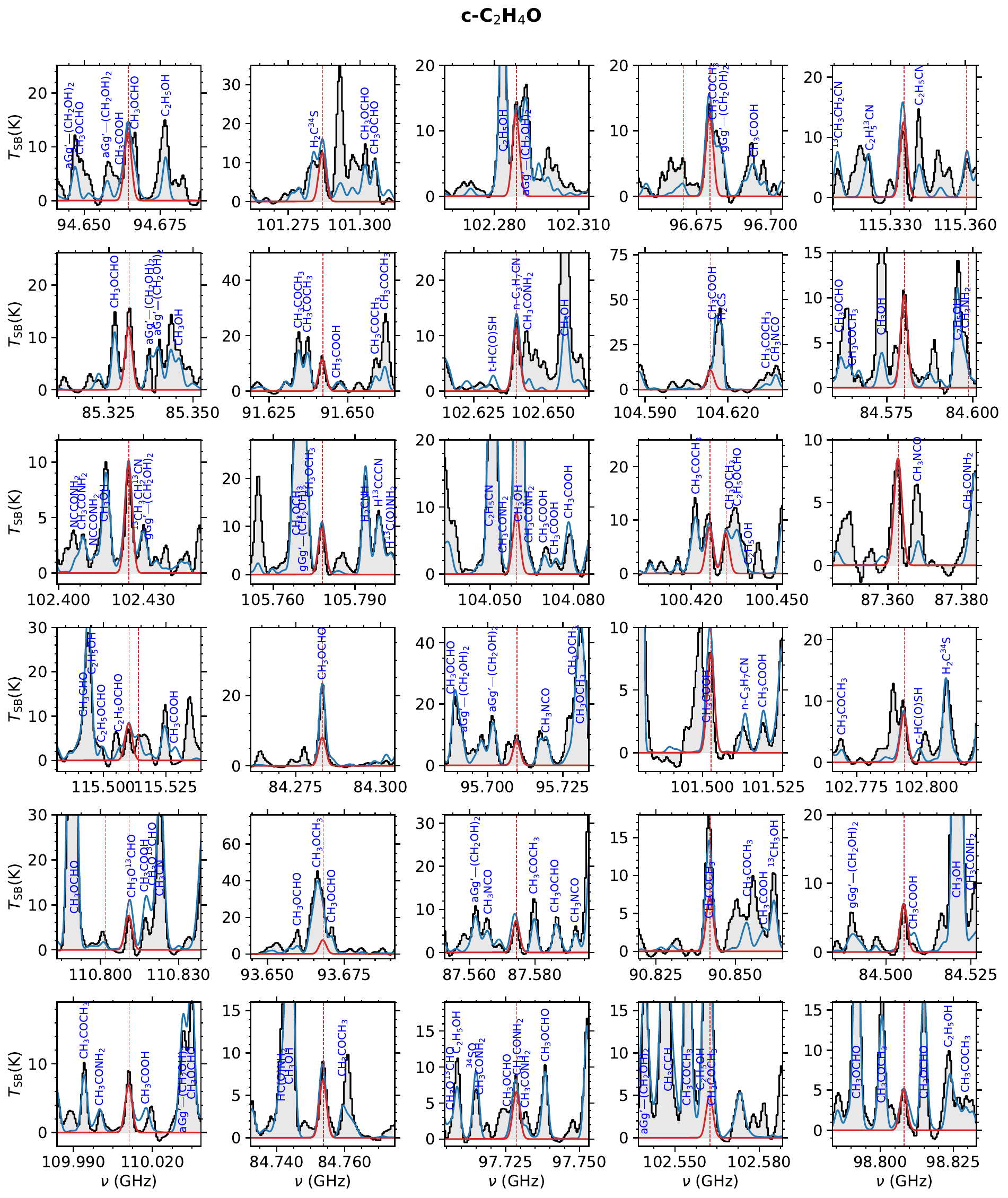}
\caption{\ch{c-C2H4O} described in Sect. \ref{mols:c-C2H4O}. Selected transitions of \ch{c-C2H4O} are shown, more transitions are detected but are not part of this figure. The black histogram and its grey shadow are the observational spectrum. The red curve is the fit of the individual transition and the blue curve is the cumulative fit considering all detected species. The red dashed lines indicate the frequency of the molecular transitions. The plots are sorted by decreasing intensity of the corresponding species (red line).}
\label{fig:c-C2H4O}
\end{figure*}

Figure \ref{fig:c-C2H4O} shows selected transitions of \ch{c-C2H4O} detected towards G31.41, which are unblended or partially blended and intense, with emission from other species whose joint contribution explains well the observed spectrum.
We note that the remaining transitions predicted by the LTE model are consistent with the observed spectrum, but appear more blended with transitions of other species. The transitions are optically thin ($\tau\leq$ 0.10; see Table \ref{tab:Transitions_Molecules_on_G31}). They cover a wide range of transitions energies from 9.5 K to 290 K, allowing us to leave $T_{\rm ex}$ as a free parameter. 
We found $T_{\rm ex}$ = 132 $\pm$ 7 K, similar to the assumed value of 150 K assumed for the molecules with > 5 atoms. The other derived parameters are shown in Table \ref{tab:Poperties_molecules_on_G31}.

\subsubsection{Thioformaldehyde (\ch{H2CS})}
\label{mols:H2CS}

The fit of the detected transitions of \ch{H2CS} (see Fig. \ref{fig:H2CS} and Table \ref{tab:Transitions_Molecules_on_G31}) gives a $T_{\rm ex}$ = 38 K (see Table \ref{tab:Fits_isotopologues_molecules_on_G31}), and indicates that the emission of the molecule is optically thick. We have thus searched for optically thinner isotopologues with $^{13}$C, $^{33}$S and $^{34}$S, which are also detected and shown in Fig. \ref{fig:H2CS}. Since AUTOFIT does not converge leaving $T_{\rm ex}$ as a free parameter, we fixed it. We do not use the value derived for the main isotopologue because of its optical thickness, but we used instead 50 K, following our general criteria. The transitions of the $^{13}$C isotopologue are partially blended, while the transitions of H$_2$C$^{33}$S are less blended with other species (Fig. \ref{fig:H2CS}). The isotopologue H$_2$C$^{34}$S is slightly optically thick ($\tau$ = 0.03-0.33), whereas H$_2$C$^{33}$S has a lower $\tau$ of 0.01-0.07. Therefore, we used the H$_2$C$^{33}$S isotopologue to retrieve the column density of \ch{H2CS} (Table \ref{tab:Poperties_molecules_on_G31}). 
We note that the \ch{H2CS} column density derived by using  H$_2$C$^{34}$S would be a factor 3.6 lower, confirming that it is partially optically thick.

\subsubsection{Nitrous acid  (\ch{HONO})}
\label{mols:HONO}

Some few transitions of this species are detected with no or slight blending (see Table \ref{tab:Transitions_Molecules_on_G31}), such as the 
the $6_{1,5} \rightarrow 6_{0,6}$ transition at 98.0877 GHz shown (fourth panel of Fig. \ref{fig:HONO}), 
the $4_{0,4} \rightarrow 3_{0,3}$ at 93.9528 GHz (slightly blended with \ch{CH3CONH2}; upper right panel of Fig. \ref{fig:HONO}), and the $8_{1,7} \rightarrow 8_{0,8}$ transition at 111.5519 GHz (second leftmost upper panel of Fig. \ref{fig:HONO}).
We note that all other transitions predicted by the LTE model are consistent with the observed spectra, but they are heavily blended, with only one possible discrepancy of the $3_{1,2} \rightarrow 3_{0,3}$ at 85.7329 GHz, in which the LTE model overestimate the observed spectrum by $\sim$1.5 K (middle right panel of Fig. \ref{fig:HONO}). This difference might be due to a non optimal baseline substraction, considering that the uncertainty of the continuum level of the ALMA baselines in the GUAPOS survey is of the same order (1.2 K, as described in \citealt{Colzi2021}).
Since the number of unblended transition detected of this species is not enough to claim a robust detection, we consider it as a tentative detection.
We note that this species has been only detected previously towards IRAS16B by \cite{Coutens2019}. In this source the  \ch{HONO}/\ch{CH3OH} ratio is 9$\times10^{-5}$, whereas the ratio in G31.41 we derived a similar value of 6$\times10^{-5}$, which further supports that the molecule is tentatively detected.

\subsubsection{Sulphur monoxide (\ch{SO})}
\label{mols:SO}

Four intense and unblended transitions of SO are detected
(see Fig. \ref{fig:SO} and Table \ref{tab:Transitions_Molecules_on_G31}). 
The AUTOFIT converged leaving all parameters free (see Table \ref{tab:Fits_isotopologues_molecules_on_G31}). The results indicate that three of the transitions are optically thick ($\tau$ > 0.8; Table \ref{tab:Transitions_Molecules_on_G31}), and that only the $5_{4} \rightarrow 4_{4}$ transition is optically thin ($\tau$=0.15). Therefore, we used the isotopologue $^{34}$SO, whose detected transitions are shown in the lower panels of Fig. \ref{fig:SO}. Since the main isotopologue is optically thick, we do not use the $T_{\rm ex}$ derived for it, but assume a value of 50 K, following our general criteria.
The 2$_3\rightarrow$1$_2$ transitions at 97.7153 GHz (lower left panel) is completely unblended, while the other three are partially blended. These transitions are optically thin ($\tau$ $<$ 0.2). We applied the $^{32}$S/$^{34}$S ratio to derive the column density of SO (see Table \ref{tab:Poperties_molecules_on_G31}).

\subsubsection{Methyl mercaptan (\ch{CH3SH})}
\label{mols:CH3SH}

\begin{figure*}
\includegraphics[scale=0.5]{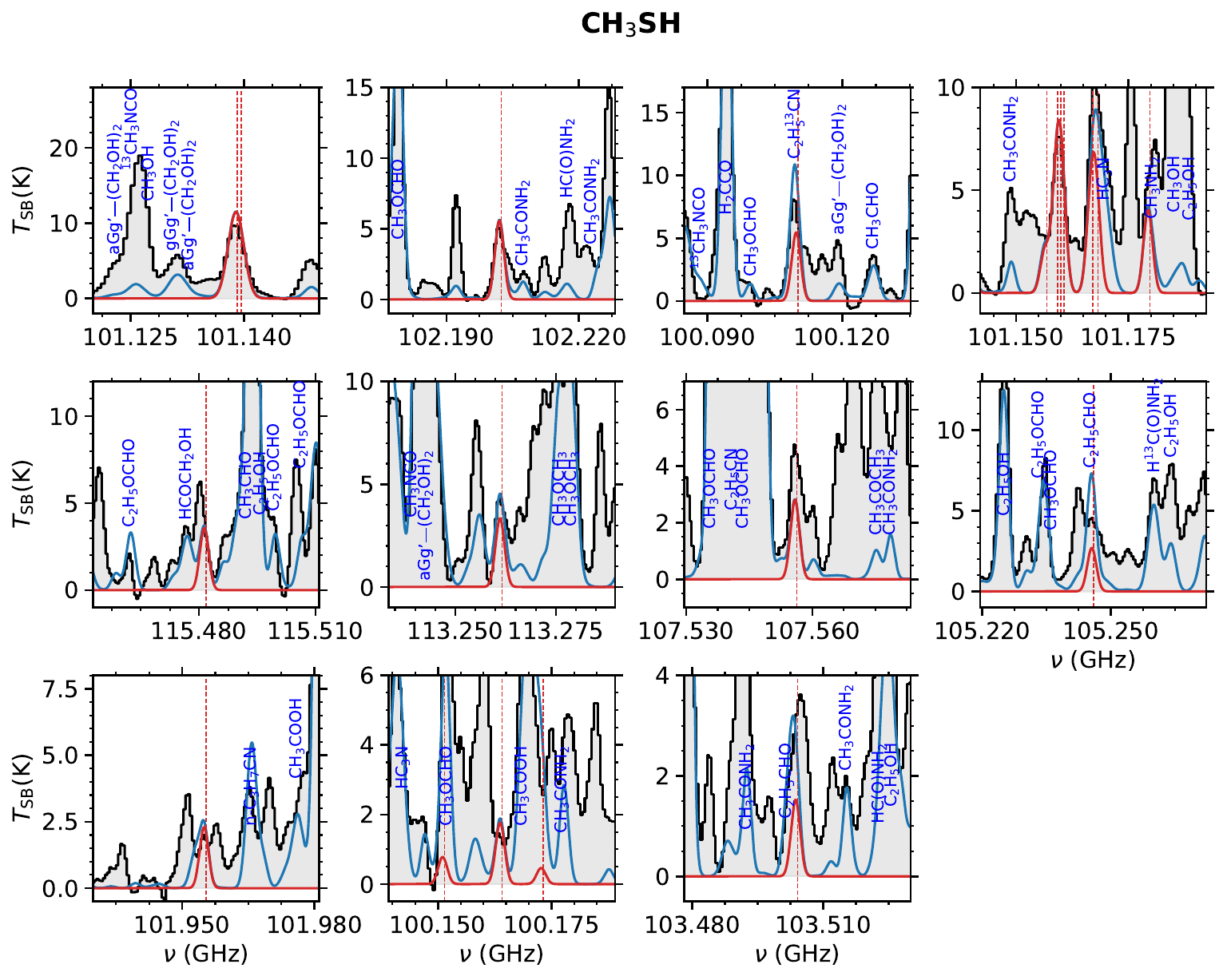}
\caption{\ch{CH3SH} described in Sect. \ref{mols:CH3SH}. The black histogram and its grey shadow are the observational spectrum. The red curve is the fit of the individual transition and the blue curve is the cumulative fit considering all detected species. The red dashed lines indicate the frequency of the molecular transitions. The plots are sorted by decreasing intensity of the corresponding species (red line).}
        \label{fig:CH3SH}
\end{figure*}

We have detected several unblended transitions of \ch{CH3SH}
(see Fig. \ref{fig:CH3SH} and Table \ref{tab:Transitions_Molecules_on_G31}). These transitions are optically thin ($\tau$ $<$ 0.042). 
The remaining \ch{CH3SH} transitions, along with the contribution from other species, such as C$_2$H$_5^{13}$CN or \ch{C2H5CHO}, are consistent with the observed spectra. The derived parameters of the fit are shown in Table \ref{tab:Poperties_molecules_on_G31}.

\subsubsection{Cyanoacetylene (\ch{HC3N})}
\label{mols:HC3N}

The spectral coverage of the GUAPOS survey contains the $J$=10$-$9, $J$=11$-$10 and $J$=12$-$11 transitions of \ch{HC3N}.
The transitions $J$=10$-$9 and $J$=12$-$11 are unblended, while the $J$=11$-$10 transitions is partially blended with \ch{CH3OCHO} (see Fig. \ref{fig:HC3N} and Table \ref{tab:Transitions_Molecules_on_G31}). By fitting these transitions we obtained a value for $T_{\rm ex}$ of 58 $\pm$ 8 K (see Table \ref{tab:Fits_isotopologues_molecules_on_G31}), and line opacities of $\tau$=1.6-2.0, which indicate that the main isotopologue is optically thick.
We have also searched for the three $^{13}$C isotopologues: H$^{13}$CCCN, HC$^{13}$CCN and HCC$^{13}$CN. We fixed the $T_{\mathrm{ex}}$ to 50 K because the values obtained from the \ch{HC3N} fit are not reliable due to the high optical depth.
The transitions of the three isotopologues are unblended, or they are blended with emission from other identified species that explain properly the observed spectrum.
The transtition $J$=12$-$11 of HCC$^{13}$CN is blended with \ch{HCOCH2OH}; the transtition $J$=11$-$10 of H$^{13}$CCCN is blended with \ch{C2H3CN} and \ch{CH3COOH}; and the transtition $J$=12$-$11 of HC$^{13}$CCN is blended with aGg'-(CH$_{2}$OH)$_{2}$.
The emission of the $^{13}$C isotopologues is optically thinner than that of the main isotopologue ($\tau$ = 0.26$-$0.40). The column densities obtained for the three isotopologues are similar (Table \ref{tab:Poperties_molecules_on_G31}). To derive the column density of \ch{HC3N} we have thus done the average of the column densities of the  $^{13}$C isotopologues, and applied the isotopic ratio.
Since the transitions have intermediate opacities, the \ch{HC3N} column density might be slightly underestimated. We also searched for optically thinner isotopologues ($^{15}$N or the double $^{13}$C isotopologues), but they are not detected. We note that the upper limits obtained for those isotopologues are consistent with the column density we report for \ch{HC3N} (Table \ref{tab:Poperties_molecules_on_G31}).

\subsubsection{Propanal (\ch{C2H5CHO})}
\label{mols:C2H5CHO}

The transitions $12_{0,12} \rightarrow 11_{1,11}$ E and $12_{0,12} \rightarrow 11_{1,11}$ A, both at 114.1011 GHz (Table \ref{tab:Transitions_Molecules_on_G31}), shown in upper middle panel of Fig. \ref{fig:C2H5CHO}, are unblended. Figure \ref{fig:C2H5CHO} also shows other several transitions of \ch{C2H5CHO} that reproduce well the observed spectra when the contribution of the emission from other species is considered. The other transitions predicted by the LTE model are consistent with the observed spectra, but they are heavily blended, and thus they are not used for the LTE model fit. 
Therefore, we consider that this species is tentatively detected. 
The results of the fit using the transitions shown in Fig. \ref{fig:C2H5CHO} are presented in Table \ref{tab:Poperties_molecules_on_G31}.
This molecule has been detected previously in other interstellar sources, including IRAS16B, in which \cite{Lykke2017} found a \ch{C2H5CHO}/\ch{CH3OH} ratio of 2.2$\times10^{-4}$, very similar to the value derived here for G31.41  (4.0$\times10^{-4}$), which supports the tentative detection.

\subsubsection{Carbonyl sulphide (\ch{OCS})}
\label{mols:OCS}

Several transitions of OCS, O$^{13}$CS, OC$^{34}$S are clearly detected with no contamination (see Fig. \ref{fig:OCS} and Table \ref{tab:Transitions_Molecules_on_G31}).
For OCS and O$^{13}$CS AUTOFIT converges leaving $T_{\rm ex}$ as a free parameter, giving 74$\pm$5 K and 35$\pm$16 K, respectively (Table \ref{tab:Fits_isotopologues_molecules_on_G31}). However, both
are optically thick ($\tau$ > 1.1), and hence they do not provide a reliable value for the temperature. Thus, we have fixed $T_{\rm ex}$ = 50 K for the other isotopologues.
The transitions of $^{18}$OCS and OC$^{33}$S are optically thinner ($\tau$ = 0.16-0.22), and also unblended, except for the $8 \rightarrow 7$ transition of OC$^{33}$S, which is blended with \ch{CH3OCHO}, and for the $10 \rightarrow 9$ transition of $^{18}$OCS which is blended with an unidentified species. 
The results of the fits for the isotpologues and the main molecule are shown in Table \ref{tab:Poperties_molecules_on_G31} and Table \ref{tab:Fits_isotopologues_molecules_on_G31}.
The column density of OCS using the isotopologue $^{18}$OCS or OC$^{33}$S are consistent within a factor of 1.3. We used the isotopologue $^{18}$OCS to estimate the column density of OCS because it is slightly optically thinner.

\subsubsection{Methoxymethanol (\ch{CH3OCH2OH})}
\label{mols:CH3OCH2OH}

The LTE model using the parameters shown in  Table \ref{tab:Poperties_molecules_on_G31} is depicted in Fig. \ref{fig:CH3OCH2OH}, which shows several transitions of this species that might be detected towards G31.41.
The inclusion of the contribution of the \ch{CH3OCH2OH} emission, along with that of other molecules previously identified, explain better the observed spectrum than without considering it. We note that the remaining transitions predicted by the LTE model are consistent with the observed data.
However, there is not enough unblended transitions of this species to provide a robust detection, and hence we consider it as tentative.
This molecule has been only previosly detected in the MM1 core of NGC6334-I by \cite{McGuire2017}, in Orion KL by \cite{Tercero2018}, and in IRAS16B by \cite{Manigand2020}. These works found a \ch{CH3OCH2OH}/\ch{CH3OH} ratios of 0.014, 0.014 and 0.029 respectively. 
In this work we have derived a very similar ratio of 0.034 towards G31.41, supporting the tentative detection.

\subsubsection{Sulphur dioxide (\ch{SO2})}
\label{mols:SO2}

Figure \ref{fig:SO2} shows the transitions of \ch{SO2}, $^{34}$SO$_2$ and $^{33}$SO$_2$ detected towards G31.41 (see also Table \ref{tab:Transitions_Molecules_on_G31}).
Several transitions of SO$_2$ and $^{33}$SO$_2$ are blended with other species, while the two transitions of $^{34}$SO$_2$ are unblended. 
Moreover, this isotopologue is the only for which AUTOFIT converges leaving $T_{\rm ex}$ as a free parameter, giving a value of 56$\pm$14 K, and a low opacity of $\tau$=0.15$\pm$0.05. 
We have then fixed this derived value of $T_{\rm ex}$ also for \ch{SO2} and $^{33}$SO$_2$. The main isotopologue \ch{SO2} is optically thick ($\tau$=0.64), while $^{33}$SO$_2$ is optically thin ($\tau$=0.08). In this case, although the opacity of the $^{34}$SO$_2$ is slightly higher than that of $^{33}$SO$_2$, we have used the former to derive the column density of \ch{SO2} because their detected lines are unblended (Fig. \ref{fig:SO2}).
The results are shown in Table \ref{tab:Poperties_molecules_on_G31} and Table \ref{tab:Fits_isotopologues_molecules_on_G31}.

\subsection{Non detected molecules}
\label{sec:not_detceted_mols}

Besides the 18 molecules with positive detections (including 3 tentative detections) towards G31.41 presented in the previous section, we have also searched for other 16 species (of the molecule sample described in Sect. \ref{sec:sample}) that were not detected in this or any other GUAPOS works (see Table \ref{tab:Molecules_G31_GUAPOS_works}). We have thus derived upper limits for their column densities, which are listed in Table \ref{tab:Poperties_molecules_on_G31}.
The transitions of these species from which we have obtained the upper limits are mostly blended. Therefore, to compute the upper limits, we have selected the brightest transition of each species according to the LTE model that do not appear heavily blended.
Since there are no line-free channels in the G31.41 spectra, to derive the upper limits we have fixed the values of $\textit{T}_{\textrm{ex}}$, $v-v_{\rm 0}$ and FWHM to the values previously explained in Sect. \ref{sec:data analysis}, and we have increased the value of $N$ that makes the LTE model (including also the contribution from other species) visually compatible with the observed spectra. We list in Table \ref{tab:Transitions_Molecules_on_G31} the transition used for each molecule, and in Fig. \ref{fig:upper-limits} we show the observed spectra at the corresponding frequency. To assess how constraining are the derived upper limits, in Fig. \ref{fig:uplims_g31_iras} (Appendix \ref{sec:Molecules_not_detected_G31}) we have compared them, normalized to \ch{CH3OH}, with the molecular ratios (or upper limits) obtained in IRAS16B. The details of this comparison are presented in Appendix \ref{sec:Molecules_not_detected_G31}.

\onecolumn
\setlength{\tabcolsep}{1pt}

\begin{longtable}{ccccccccccc}
\caption{\label{tab:Poperties_molecules_on_G31} Results of the SLIM fits of the molecules analysed towards G31.41, ordered by increasing molecular mass, from which we have retrieved their abundances. For optically thick molecules, we include the optically thin isotopologue that we have used to derive the column density of the main isotopologue (the results of the other isotopologues are presented in Table \ref{tab:Fits_isotopologues_molecules_on_G31} instead). The resulting physical parameters, along with their associated uncertainties, are presented. The values of the parameters fixed to perform some fits are shown without uncertainties. In the case of non detections, we indicated the upper limit in the column density with <. In the case of tentative detections, they are denoted with $\sim$. When $T_{\mathrm{ex}}$ is fixed, we calculate in what factor the value of $N$ changes when using the different temperatures  $T_{\mathrm{min}}$ and $T_{\mathrm{max}}$: 25-75 K and 100-200 K for fixed temperatures of 50 K and 150 K, respectively.
We also present the maximum line opacity of the transitions of each molecule ($\tau_{max}$), and the molecular abundances compared to H$_2$. The transitions used to obtain these values are in Table \ref{tab:Transitions_Molecules_on_G31}. 
}
\\

    \hline
     Formula &  $\textit{T}_{\textrm{ex}}$ & \textit{N} & \textit{N(T}$_{\textrm{min}}$)/\textit{N-} & $v-v_{\rm 0}$ & FWHM & $\tau_{max}^a$ & $N/N_{\textrm{H}_{2}}^b$ & Fig. \\ 
     & (K) & ($\times$10$^{16}$ cm$^{-2}$) & \textit{N(T}$_{\textrm{max}}$)/\textit{N} & (km s$^{-1})$ & (km s$^{-1})$ & & ($\times$10$^{-8}$) & \\ \hline \endfirsthead

     \caption{Continued.}\\
     \hline
     Formula & $\textit{T}_{\textrm{ex}}$ & \textit{N} & \textit{N(T}$_{\textrm{min}}$)/\textit{N-} & $v-v_{\rm 0}$ & FWHM & $\tau_{max}^a$ & $N/N_{\textrm{H}_{2}}^b$ & Fig. \\ 
     & (K) & ($\times$10$^{16}$ cm$^{-2}$) & \textit{N(T}$_{\textrm{max}}$)/\textit{N} & (km s$^{-1})$ & (km s$^{-1})$ & & ($\times$10$^{-8}$) & \\ \hline \endhead   
     
     \hline
     \endfoot

     \hline \noalign{\smallskip}
    \multicolumn{11}{p{15.5cm}}{Notes. Isotopic ratios used: $\mathrm{^{12}C/^{13}C}$ = 39.5 $\pm$ 9.6 (mean value of isotopic ratios from \citealt{Milam2005} and \citealt{Yan2019}), $\mathrm{^{14}N/^{15}N}$ = 340 $\pm$ 90 \citep{Colzi2018}, $\mathrm{^{32}S/^{34}S}$ = 14.6 $\pm$ 2.1 \citep{Yu2020}, $\mathrm{^{34}S/^{33}S}$ = 5.9 $\pm$ 1.5 \citep{Yu2020}, $\mathrm{^{34}S/^{36}S}$ = 115 $\pm$ 17 \citep{Mauersberger1996}, $\mathrm{^{16}O/^{18}O}$ = 330 $\pm$ 140 \citep{Wilson1999}, $\mathrm{^{18}O/^{17}O}$ = 3.6 $\pm$ 0.2 (\citealt{Wilson1999}, taking the value of local ISM). $^{a}$ Maximum $\tau$ of all the transitions detected in each species. $^{b}$ Value of $N$(H$_{2}$) = (1.0 $\pm$ 0.2)$\times$10$^{25}$ cm$^{-2}$ \citep{Mininni2020}. $^{c}$ The transition of H$_2$S used has a high value of $E_{\rm up}$ of 520 K (see Table \ref{tab:Transitions_Molecules_on_G31}), which is completely out of the range of the $T_{\rm ex}$ adopted in the fits (25, 50 and 75 K); and thus the derived $N$ is extremely sensitive to the value assumed. * This molecule, HNC, is detected, but the profiles of the main isotopologue and the $^{13}$C isotopologue show absorptions (see Fig. \ref{fig:HCN_HNC}). We have derived an upper limit using the $^{15}$N isotopologue (see details in Section \ref{mols:HNC}).
    }
     \endlastfoot

    \multicolumn{11}{c}{Detected molecules}  \tabularnewline \hline
    HCN & & 150 $\pm$ 40 &  1.6-1.2  & ~ & ~ & & 10 $\pm$ 6 & \ref{fig:HCN_HNC}\\
    HC$^{15}$N & 50 &  0.45 $\pm$ 0.02 & 1.6-1.2 & 0.3 $\pm$ 0.2 & 9.1 $\pm$ 0.3 & 1.7 $\pm$ 0.2 & (4.5 $\pm$ 0.9)$\times$$10^{-2}$ & \ref{fig:HCN_HNC}\\ \hline
    HNC* &  &  <1.1 & 0.7-1.6 & & &   & 0.073 & \ref{fig:HCN_HNC}\\ 
    H$^{15}$NC* &  50 & <0.003 & 0.7-1.6  & 0 & 7 & 0.017 & 3.2$\times$10$^{-4}$ & \ref{fig:HCN_HNC}\\ \hline
    CO & & (2.6 $\pm$ 1.5)$\times$10$^5$ & 1.0-2.1  & ~ & ~ &  & (2.6 $\pm$ 1.6)$\times$$10^4$ & \ref{fig:CO}\\
    $^{13}$C$^{18}$O  & 50 & 20 $\pm$ 6 & 1.0-2.1 & 0 & 7 & 0.13 $\pm$ 0.11 & 2.0 $\pm$ 0.7 & \ref{fig:CO}\\ \hline
    H$_{2}$CO &  & (2.6 $\pm$ 1.2)$\times$10$^2$ & 2.0-1.0 & ~ & ~ &  & 26 $\pm$ 13 & \ref{fig:H2CO}\\
    H$_{2}$C$^{18}$O & 50 & 0.78 $\pm$ 0.07 & 2.0-1.0 & -0.9 $\pm$ 0.3 & 7.4 $\pm$ 0.7 & 0.053 $\pm$ 0.007 & 0.078 $\pm$ 0.017 & \ref{fig:H2CO}\\ \hline
    CH$_{3}$CCH & 150 & 12.0 $\pm$ 0.9 & 0.6-1.7 & 1.3 $\pm$ 0.6 & 11.8 $\pm$ 1.2 & 0.06 $\pm$ 0.03 & 1.2 $\pm$ 0.3 & \ref{fig:CH3CCH}\\ \hline
    CH$_{3}$NC & 150 & 0.11 $\pm$ 0.04 & 0.8-1.8 & 0 & 7 & 0.03 $\pm$ 0.03 & (1.1 $\pm$ 0.5)$\times$10$^{-2}$ & \ref{fig:CH3NC}\\ \hline
    NH$_{2}$CN & 50 & 0.111 $\pm$ 0.008 & 0.6-1.6 & -0.5 $\pm$ 0.3 & 7 & 0.34 $\pm$ 0.08 & (1.1 $\pm$ 0.3)$\times$10$^{-2}$ & \ref{fig:NH2CN}\\ \hline
    CS & & 71 $\pm$ 22 & 0.6-1.4 & ~ & ~ & & 7 $\pm$ 3 & \ref{fig:CS}\\
    C$^{36}$S & 50 & 0.04 $\pm$ 0.01 & 0.6-1.4 & -0.5 $\pm$ 1.2 & 9.5 & 0.08 $\pm$ 0.04 & (4.2 $\pm$ 1.3)$\times$10$^{-3}$ & \ref{fig:CS}\\ \hline
    c-C$_{2}$H$_{4}$O & 132 $\pm$ 7 & 6.3 $\pm$ 0.3 & & 0.47 $\pm$ 0.15 & 8.6 $\pm$ 0.4 & 0.10 $\pm$ 0.03 & 0.63 $\pm$ 0.13 & \ref{fig:c-C2H4O}\\ \hline
    H$_{2}$CS & & 30.6 $\pm$ 11 & 0.6-1.6 & ~ & ~ &  & 3.1 $\pm$ 1.2 & \ref{fig:H2CS} \\
    H$_{2}$C$^{33}$S & 50 &  0.36 $\pm$ 0.06 & 0.6-1.6 & 0 & 7 & 0.07 $\pm$ 0.06 & (3.6 $\pm$ 1.0)$\times$10$^{-2}$ & \ref{fig:H2CS}\\ \hline
    trans-HONO & 50 &  $\sim$ 0.45 $\pm$ 0.06 & 0.6-1.7 & -1.1 $\pm$ 0.5 & 7 & 0.10 $\pm$ 0.06 & (4.5 $\pm$ 1.1)$\times$10$^{-2}$ & \ref{fig:HONO}\\ \hline
    SO & & 3.9 $\pm$ 0.8 & 0.6-1.4  & ~ & ~ &  & 0.39 $\pm$ 0.11 & \ref{fig:SO}\\
    $^{34}$SO & 50 & 0.27 $\pm$ 0.04 & 0.6-1.4 & 1.3 $\pm$ 0.6 & 7 & 0.20 $\pm$ 0.06 & (2.7 $\pm$ 0.7)$\times$10$^{-2}$ & \ref{fig:SO}\\ \hline
    CH$_{3}$SH & 150 & 7.2 $\pm$ 0.4 & 0.5-1.6 & 1.3 $\pm$ 0.3 & 7 & 0.042 $\pm$ 0.014 & 0.70 $\pm$ 0.15 & \ref{fig:CH3SH}\\ \hline
    HC$_{3}$N & & 2.8 $\pm$ 0.4 & 1.1-1.2 & ~ & ~ &  & 0.28 $\pm$ 0.07 & \ref{fig:HC3N}\\
    H$^{13}$CCCN & 50 & 0.062 $\pm$ 0.004 & 1.1-1.2  & 1.0 $\pm$ 0.3 & 9.2 $\pm$ 0.7 & 0.32 $\pm$ 0.08 & (6.2 $\pm$ 1.3)$\times$10$^{-3}$ & \ref{fig:HC3N}\\
    HC$^{13}$CCN & 50 & 0.070 $\pm$ 0.003 & 1.1-1.2 & 0.7 $\pm$ 0.3 & 10.4 $\pm$ 0.6 & 0.33 $\pm$ 0.05 & (7.0 $\pm$ 1.5)$\times$10$^{-3}$ & \ref{fig:HC3N}\\
    HCC$^{13}$CN  & 50 & 0.080 $\pm$ 0.004 & 1.2-1.2 & 2.2 $\pm$ 0.3 & 9.5 & 0.40 $\pm$ 0.07 & (8.0 $\pm$ 1.7)$\times$10$^{-3}$ & \ref{fig:HC3N}\\ \hline
    C$_{2}$H$_{5}$CHO & 150 & $\sim$ 3.00 $\pm$ 0.15 & 0.6-1.3 & 0.1 $\pm$ 0.2 & 7 & 0.02 $\pm$ 0.02 & 0.30 $\pm$ 0.07 & \ref{fig:C2H5CHO}\\ \hline
    OCS &  & (1.8 $\pm$ 0.8)$\times$10$^2$ & 0.9-1.2 & ~ & ~ &  & 18 $\pm$ 9 & \ref{fig:OCS}\\ 
    $^{18}$OCS  & 50 & 0.55 $\pm$ 0.07 & 0.9-1.2  & 2.9 $\pm$ 0.5 & 7 & 0.16 $\pm$ 0.06 & (5.5 $\pm$ 1.3)$\times$10$^{-2}$ & \ref{fig:OCS}\\ \hline
    CH$_{3}$OCH$_{2}$OH & 150 & $\sim$(2.55 $\pm$ 0.15)$\times$10$^2$ & 0.6-1.4 & 2.7 $\pm$ 0.3 & 7 & 0.02 $\pm$ 0.03 & 26 $\pm$ 6 & \ref{fig:CH3OCH2OH}\\ \hline
    SO$_{2}$ & & 10 $\pm$ 3 & ~ & ~ & ~ &  & 1.1 $\pm$ 0.4 & \ref{fig:SO2}\\
    $^{34}$SO$_{2}$  & 56 $\pm$ 14 & 0.68 $\pm$ 0.14 &  & 1.9 $\pm$ 0.4 & 7 & 0.15 $\pm$ 0.05 & (7 $\pm$ 2)$\times$10$^{-2}$ & \ref{fig:SO2}\\ \hline
    
    \multicolumn{11}{c}{Non detected molecules (upper limits)}  \tabularnewline \hline

    H$_2$S & 50 &  <7.6$\times$10$^{4}$  & 9120-0.07$^c$ & 0 & 7 & <9$\times$10$^{-3}$ & <8$\times$10$^{3}$ & \ref{fig:upper-limits}\\ \hline
    C$_{3}$H$_{6}$  & 150 & <3.6 & 0.6-1.6 & 0 & 7 & <1.4$\times$10$^{-3}$ & <0.4 & \ref{fig:upper-limits}\\ \hline
    HOCN & 50 &  <0.006 & 0.6-1.8 & 0 & 7 & <0.03 & <6$\times$10$^{-4}$ & \ref{fig:upper-limits}\\ \hline
    syn-CH$_{2}$CHOH & 150 & <4.0 & 0.6-1.5 & 0 & 7 & <9$\times$10$^{-3}$ & <0.4 & \ref{fig:upper-limits}\\ \hline
    PN &  50 & <0.005 & 0.6-1.4 & 0 & 7 & <0.02 & <5.0$\times10^{-4}$ & \ref{fig:upper-limits}\\ \hline
    PO &  50 & <0.019 & 0.6-1.4 & 0 & 7 & <0.015 & <1.9$\times10^{-3}$ & \ref{fig:upper-limits}\\ \hline
    CH$_{3}$Cl &  50 & <0.04 & 0.6-1.6 & 0 & 7 & <0.012 & <4$\times$10$^{-3}$ & \ref{fig:upper-limits}\\ \hline
    HC$_{2}$NC &  50 & <0.003 & 0.9-1.2  & 0 & 7 & <5$\times$10$^{-3}$ & <2.6$\times$10$^{-4}$ & \ref{fig:upper-limits}\\ \hline
    trans-C$_{2}$H$_{3}$CHO &  150 & <0.16 & 0.7-1.5 & 0 & 7 & <0.010 & <1.6$\times$10$^{-2}$ & \ref{fig:upper-limits}\\ \hline
    \ch{HOCH2CN} &  150 & <0.8 & 0.6-1.4 & 0 & 7 & <0.011 & <8$\times$10$^{-2}$ & \ref{fig:upper-limits}\\ \hline
    CH$_{3}$CHCH$_{2}$O &  150 & <1.9 & 0.8-1.2 & 0 & 7 & <0.011 & <0.19 & \ref{fig:upper-limits}\\ \hline
    gauche-C$_{2}$H$_{5}$SH &  150 & <1.7 & 0.6-1.8 & 0 & 7 & <9$\times$10$^{-3}$ & <0.17 & \ref{fig:upper-limits}\\ \hline
    S$_{2}$ &  50 & <290 & 1.2-1.1 & 0 & 7 & <0.04 & <29 & \ref{fig:upper-limits}\\ \hline
    HS$_{2}$ &  50 & <0.4 & 0.6-1.6 & 0 & 7 & <0.02 & <3$\times$10$^{-2}$ & \ref{fig:upper-limits}\\ \hline
    H$_{2}$S$_{2}$ &  50 & <1.7 & 0.5-1.7 & 0 & 7 & <0.06 & <0.17 & \ref{fig:upper-limits}\\ \hline
     NH$_{2}$CH$_{2}$COOH & 150 & <2 & 0.7-1.6 & 0 & 7 & <5$\times$10$^{-3}$ & <0.2 & \ref{fig:upper-limits}\\ 
        
\end{longtable}
\twocolumn

\section{Comparison with other interstellar sources}
\label{sec:comparison}

We have compared the chemical reservoir of G31.41, using the molecular abundances derived here and in previous works (\citealt{Cesaroni1994,Mininni2020,Colzi2021,Mininni2023}) with that of the hot corino surrounding the low-mass Solar-like protostar IRAS16B, and the two comets 67P/C-G and 46P/W. While G31.41 and IRAS16B are two archetypical examples of a hot core and a hot corino, respectively, namely, the natal environments where massive and low-mass (Solar-like) protostars are born, the two comets (67P/C-G and 46P/W) are representatives of the outcoming products of the process of star and planet formation.
We select these three sources to be compared with G31.41 because its chemical reservoir have been previously extensively studied, providing molecular catalogs of tens of molecules.
Their molecular abundances were derived from deep surveys based on homogeneous data, which were analysed using the same procedure, like we have performed in the GUAPOS project for G31.41.
This is a fundamental requisite to obtain reliable values of molecular abundance ratios, and fair comparisons of the chemical content of the four sources.

IRAS16B is the prototypical example of a low-mass star-forming region, and its rich molecular content has been obtained  using 
ALMA data in multiple works, especially from the Protostellar Interferometric Line Survey (PILS) (e.g. \citealt{Jorgensen2016}). 
IRAS16B is a member of a multiple protostellar system \citep{Maureira2020}, and it is commonly considered a proto-Solar analog given its low mass, which is 0.1 $\it M$$_{\odot}$ \citep{Jacobsen2018}. IRAS16B is the most chemically rich hot corino known, and several molecules have been detected towards it for the first time in the ISM (e.g. \citealt{Coutens2019, Zeng2019}), or in low-mass star-forming regions (e.g. \citealt{Martin-Domenech2017, Ligterink2017, Calcutt2018, Coutens2018}). 
For our comparison, we have used the molecular abundances of 55 molecular species towards IRAS16B (41 detected molecules; and 14 non detected molecules for which we used column density upper limits, see Table \ref{tab:Molecules_used_each_source}), obtained from \citet{Drozdovskaya2019} (see also references therein), \cite{Martin-Domenech2017}, \cite{Coutens2018,Coutens2019}, \cite{Calcutt2019} and \cite{Manigand2021}. 
For those molecules for which the uncertainties of the derived molecular column densities are not available, we have assumed an uncertainty of 20$\%$.

The comet 67P/C-G was visited by the European Space Agency (ESA) spacecraft Rosetta, which provided us with unique data thanks to the instrument on board ROSINA (Rosetta Orbiter Spectrometer for Ion and Neutral Analysis; \citealt{LeRoy2015}). This instrument has provided an unprecedented view of the volatile chemical composition of a comet \citep{Rubin2019} by obtaining {\it in-situ} data from the volatile emanated from the nucleus of the comet. The abundances were obtained by mass spectroscopy, thus isomer groups cannot be distinguished. We used here the data from \cite{Calmonte2016}, \cite{Dhooghe2017}, \cite{Fayolle2017}, \cite{Altwegg2019}, \cite{Hadraoui2019}, \cite{Rubin2019}, \cite{Schuhmann2019} and \cite{Rivilla2020}.
For the comparison presented in this work we have considered the molecular abundances of 32 isomeric groups (29 detections and 3 non detected species for which abundance upper limits were derived, see Table \ref{tab:Molecules_used_each_source}).

The chemical composition of the comet 46P/W was studied in detail with the IRAM 30-m telescope and the NOEMA interferometer when its distance was less than 0.1 au from Earth by \cite{Biver2021}. We have used the molecular abundances reported in \cite{Biver2021}, which provided information of 31 species (11 detections and 20 non detections for which upper limits are available, see Table \ref{tab:Molecules_used_each_source}).

\subsection{Comparison of molecular abundances}
\label{sec:Comparison of molecular abundances}

\begin{figure*}
\includegraphics[width=\textwidth]{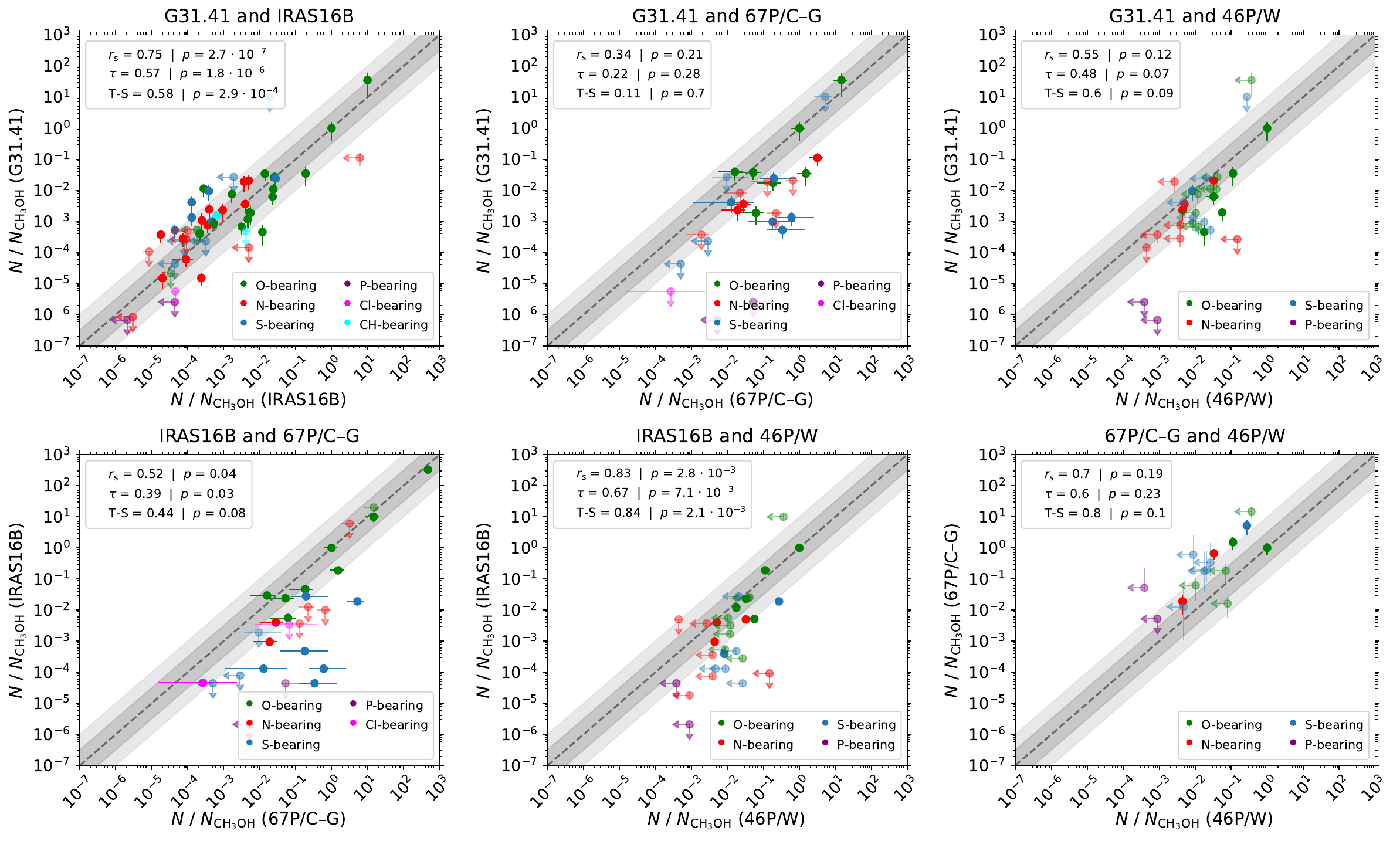}
\caption{Molecular abundances with respect to \ch{CH3OH} (of detected molecules and upper limits) classified by chemical families on different pair of sources. The errorbars marked with arrows are the upper limits. Grey dotted line is the 1:1 relation on each plot. Dark and light grey are half and one order of magnitude difference with respect to the grey dotted line. In the legend: $r_{\mathrm{s}}$ is the Spearman correlation coefficient, $\tau$ is the Kendall one, T-S is the Theil-Sen one, $p$ is the correspondent $p$-value.}
\label{fig:Correlaciones_fuentes_colores}
\end{figure*}

We have directly compared the molecular abundances derived towards the four sources in Fig. \ref{fig:Correlaciones_fuentes_colores}, in which we have normalized the column densities of all the species with respect to methanol (CH$_3$OH). 
To compute the molecular ratio uncertainties we have used error propagation for those column densities with symmetric uncertainties (e.g. those of G31.41 and IRAS16B). For the molecules with column densities reported with asymmetric uncertainties in the comet 67P/C-G (e.g. S-bearing from \citealt{Calmonte2016}), the error bars have been computed by considering the minimum and maximum ratios calculated considering the column density uncertainties, as also done by \citet{Drozdovskaya2019}.

Figure \ref{fig:Correlaciones_fuentes_colores} is composed of six panels that show the molecular abundance ratios for each possible pair of sources.
In the cases in which the molecules are not detected, but upper limits of their column densities have been reported (see Section \ref{sec:not_detceted_mols} for G31.41), we denote them with triangles.
Due to the use of the mass spectroscopy technique for the comet 67P/C-G, we summed the column densities of the molecules with the same molecular mass 
of the other three sources for the comparison.  
In the case that the column density of one of the species is an upper limit, we have considered the sum as an upper limit as well.

\begin{figure*}
\includegraphics[width=\textwidth]{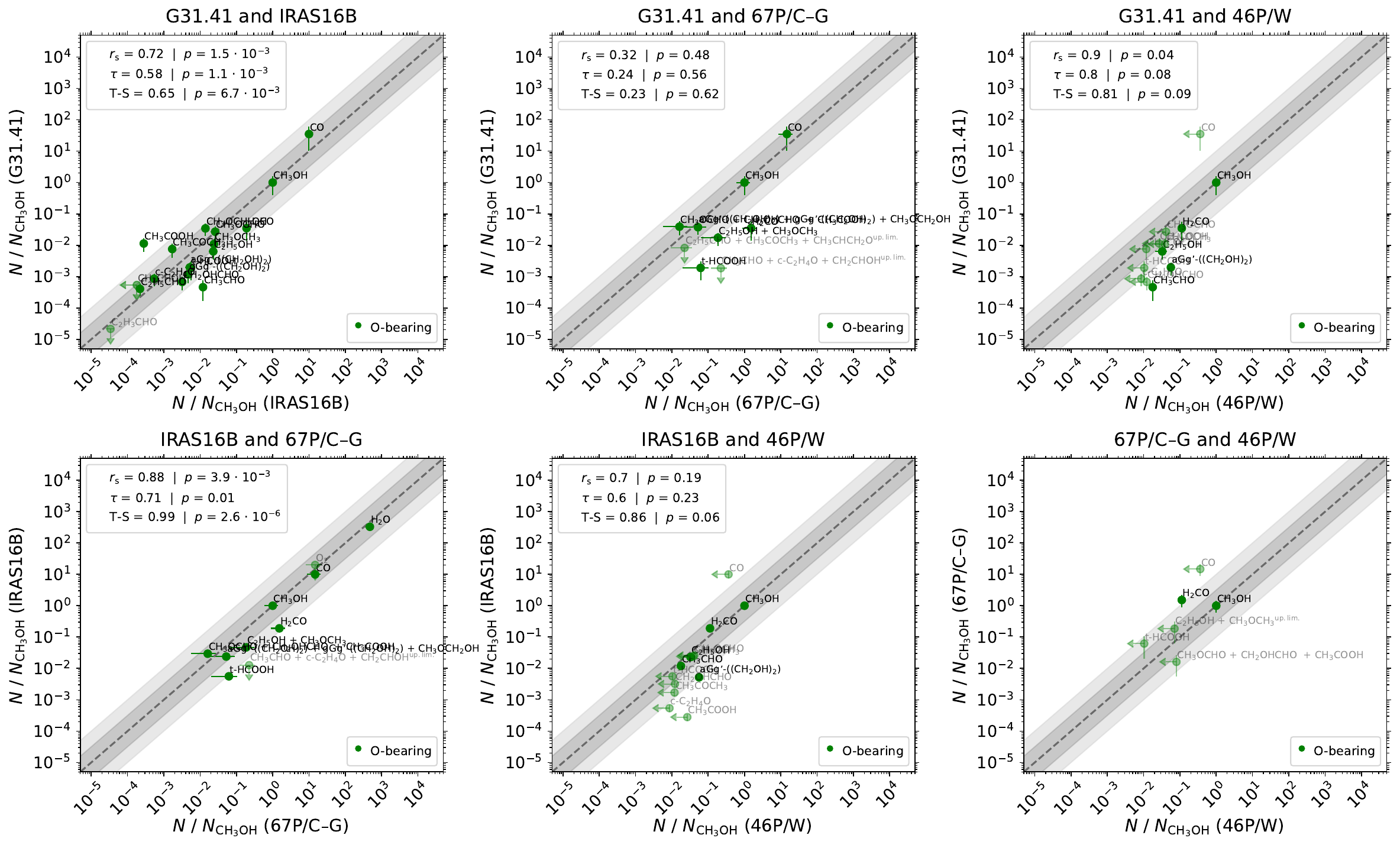}
\caption{Same as in Fig. \ref{fig:Correlaciones_fuentes_colores} but only plotting O-bearing molecules with respect to \ch{CH3OH}. The errorbars marked with arrows are the upper limits. Grey dotted line is the 1:1 relation on each plot. Dark and light grey are half and one order of magnitude difference with respect to the grey dotted line. In the legend: $r_{\mathrm{s}}$ is the Spearman correlation coefficient, $\tau$ is the Kendall one, T-S is the Theil-Sen one, $p$ is the correspondent $p$-value. We note that the limits of the axis are different to those shown in Fig. \ref{fig:Correlaciones_fuentes_colores} for display purposes.}
\label{fig:Correlaciones_fuentes_O}
\end{figure*}

\begin{figure*}
\includegraphics[width=\textwidth]{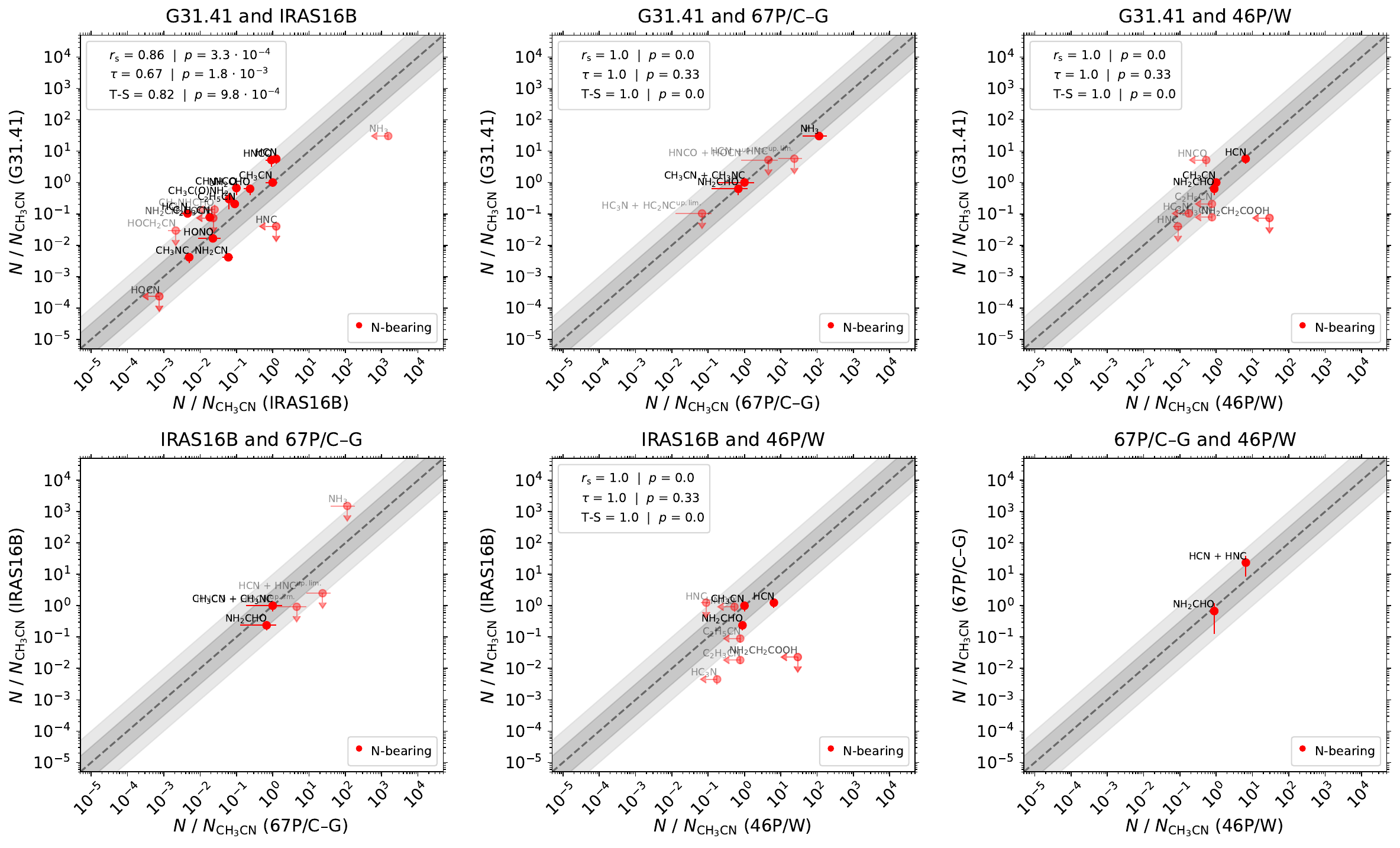}
\caption{Same as in Fig. \ref{fig:Correlaciones_fuentes_colores} but only for N-bearing molecules (including molecules containing O) with respect to \ch{CH3CN}. The errorbars marked with arrows are the upper limits. Grey dotted line is the 1:1 relation on each plot. Dark and light grey are half and one order of magnitude difference with respect to the grey dotted line. In the legend: $r_{\mathrm{s}}$ is the Spearman correlation coefficient, $\tau$ is the Kendall one, T-S is the Theil-Sen one, $p$ is the correspondent $p$-value. We note that the limits of the axis are different to those shown in Fig. \ref{fig:Correlaciones_fuentes_colores} for display purposes.}
\label{fig:Correlaciones_fuentes_N_con_O}
\end{figure*}

\begin{figure*}
\includegraphics[width=\textwidth]{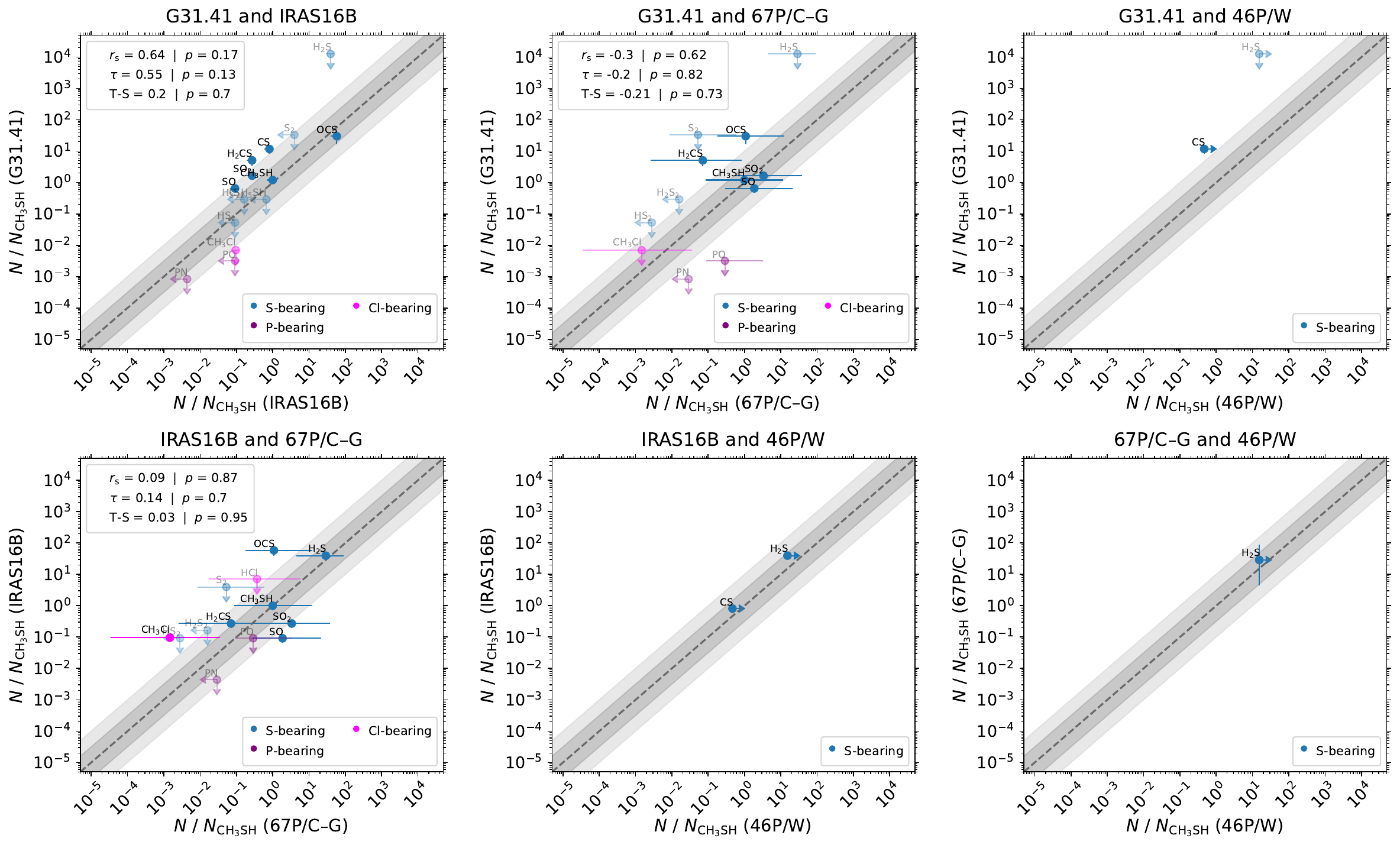}
\caption{Same as in Fig. \ref{fig:Correlaciones_fuentes_colores} but only for the P-, S-, and Cl- bearing molecules with respect to \ch{CH3SH}. The errorbars marked with arrows are the upper limits. Grey dotted line is the 1:1 relation on each plot. Dark and light grey are half and one order of magnitude difference with respect to the grey dotted line. In the legend: $r_{\mathrm{s}}$ is the Spearman correlation coefficient, $\tau$ is the Kendall one, T-S is the Theil-Sen one, $p$ is the correspondent $p$-value. We note that the limits of the axis are different to those shown in Fig. \ref{fig:Correlaciones_fuentes_colores} for display purposes.}
\label{fig:Correlaciones_fuentes_S_P_Cl}
\end{figure*}

In Figs. \ref{fig:Correlaciones_fuentes_O}, \ref{fig:Correlaciones_fuentes_N_con_O}, and \ref{fig:Correlaciones_fuentes_S_P_Cl} we have separated different families of molecules: i) oxygen(O)-bearing species; ii) nitrogen(N)-bearing species (including molecules with O); and iii) sulfur(S)-, phosphorus(P)- and chlorine(Cl)-bearing species. We have normalized the column density of each molecule by the reference molecule indicated in the axis label: \ch{CH3OH} for O-bearing, \ch{CH3CN} for N-bearing, and \ch{CH3SH} for S-, P- and Cl-bearing species. The column densities of \ch{CH3OH} and \ch{CH3CN} in G31.41 are calculated by \cite{Mininni2023} using CH$_3^{18}$OH and $^{13}$CH$_3$CN istopologues respectively, while \ch{CH3SH} has been analized in this work (Sect. \ref{mols:CH3SH}).

Figures \ref{fig:Correlaciones_fuentes_colores}, \ref{fig:Correlaciones_fuentes_O}, and \ref{fig:Correlaciones_fuentes_N_con_O} shows in general a good correlation of O- and N-bearing species for all pair of sources, spanning a remarkably broad range of molecular abundance ratios of up to six orders of magnitude. Most of the points fall in the $\pm$1 order of magnitude region from the 1:1 correlation (gray shaded area in the figures).

The case of S- and P-bearing species is different. Figure \ref{fig:Correlaciones_fuentes_colores} shows that these species, when normalized to \ch{CH3OH}, lie well below the 1:1 correlation, namely they are underabundant, in the star-forming regions (G31.41 and IRAS16B) compared to 67P/C-G (upper middle and lower left panels in Fig. \ref{fig:Correlaciones_fuentes_colores}). However, when the abundance ratios are computed by using \ch{CH3SH} instead of \ch{CH3OH}, they lie closer to the 1:1 line as shown in Fig. \ref{fig:Correlaciones_fuentes_S_P_Cl}.
This suggests that the chemical content of S- and P-bearing species, unlike O- and N-bearing, is different in the star-forming regions and the comet 67P/C-G. We will discuss in detail the implications of this in Sect \ref{sec:discuss}.

\subsection{Correlation between sources using statistical tests}
\label{sec:correlations_tests}

To quantify the goodness of the molecular abundance correlations between the different sources, and hence to investigate how their chemical contents behave, we used three statistical tests: Spearman ($r_{\mathrm{s}}$), Kendall ($\tau$) and Theil-Sen (T-S). Each test checks different properties of the correlations: while Spearman and Kendall report whether the data points are monotonically increasing or not (by comparing the ranges and by analyzing if the pairs of data are concordant or discordant, respectively), Theil-Sen indicates if the data can be fitted by a linear fit, similarly to the Pearson correlation coefficient, but with the advantage of being less sensitive to the outliers. 
We present a detailed description of the three tests in Appendix \ref{sec:statistical_tests}.

We applied the correlation tests using the molecular abundance ratios of the detected molecules.
For each statistical test we calculated its $p$-value, which quantifies the reliability of the derived correlation coefficient by taking into account the number of data points and the goodness of the correlation. If the correlation coefficient calculated is perfectly reliable, then the $p$-value is 0, whereas a value of 1 indicates that the correlation coefficient is not reliable at all. 
We considered that a value below 0.1 is good enough to trust the correlation coefficient.

\begin{figure}
	\includegraphics[width=\columnwidth]{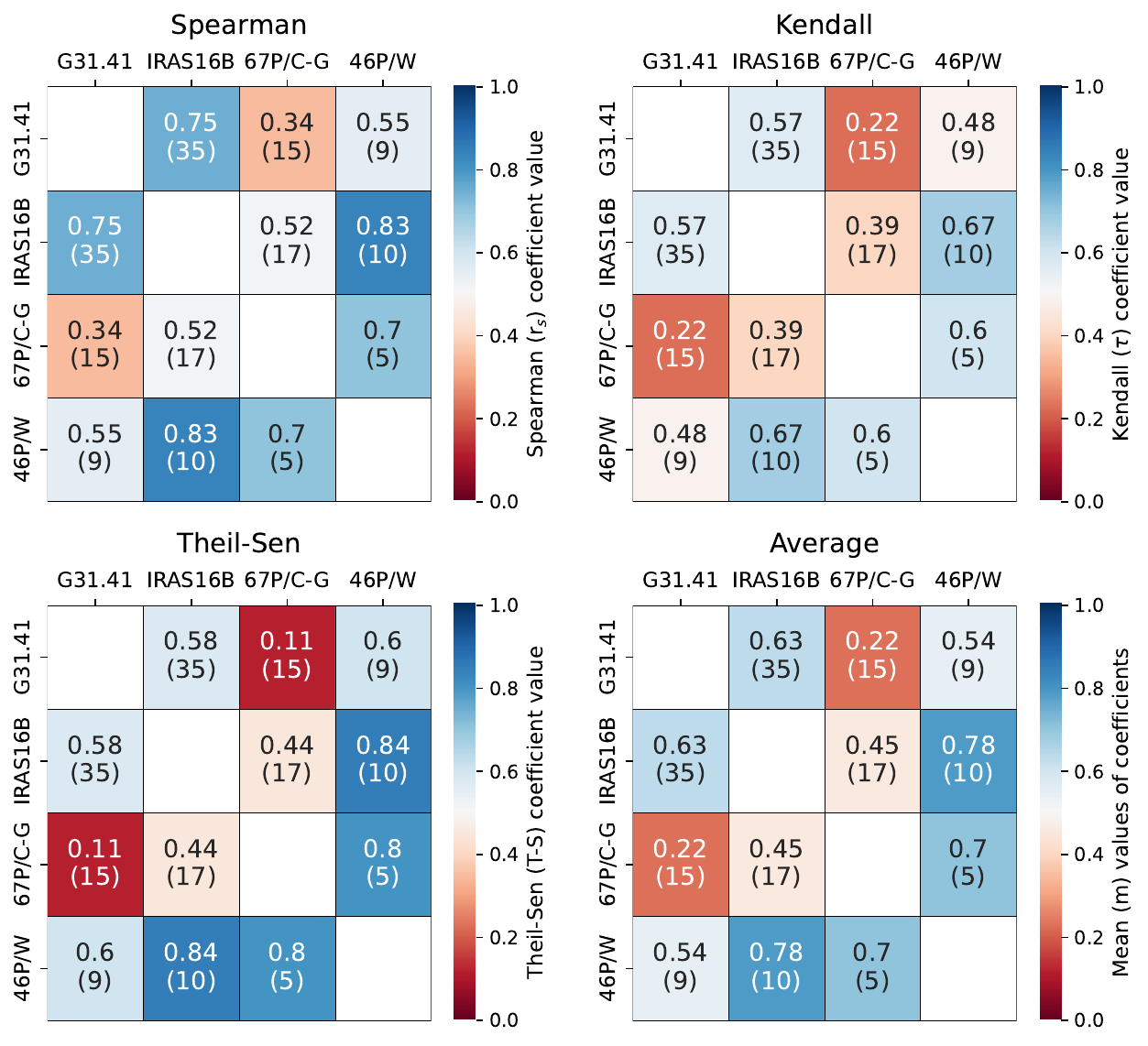}
    \caption{Correlation matrices considering all the molecules analysed in this work. The goodness of the correlation is color-coded, the bluer the color, the better the correlation. The value between brackets on each cell is the number of molecules considered.}
\label{fig:matriz_correlaciones_todas}
\end{figure}

\begin{figure}
	\includegraphics[width=\columnwidth]{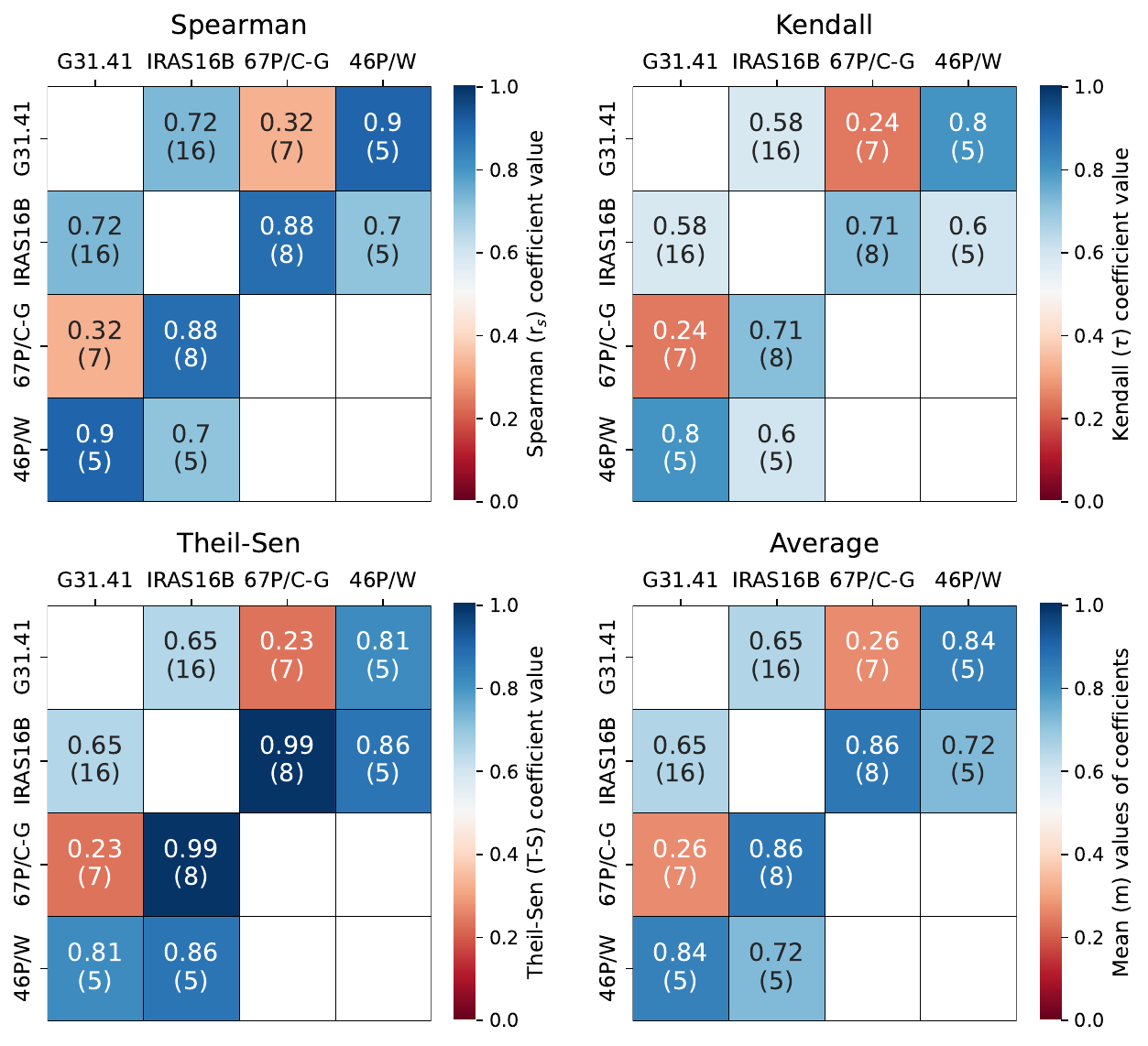}
    \caption{Correlation matrices for O-bearing molecules. The goodness of the correlation is color-coded, the bluer the color, the better the correlation. The value between brackets on each cell is the number of molecules considered.}
\label{fig:matriz_correlaciones_O}
\end{figure}

\begin{figure}
	\includegraphics[width=\columnwidth]{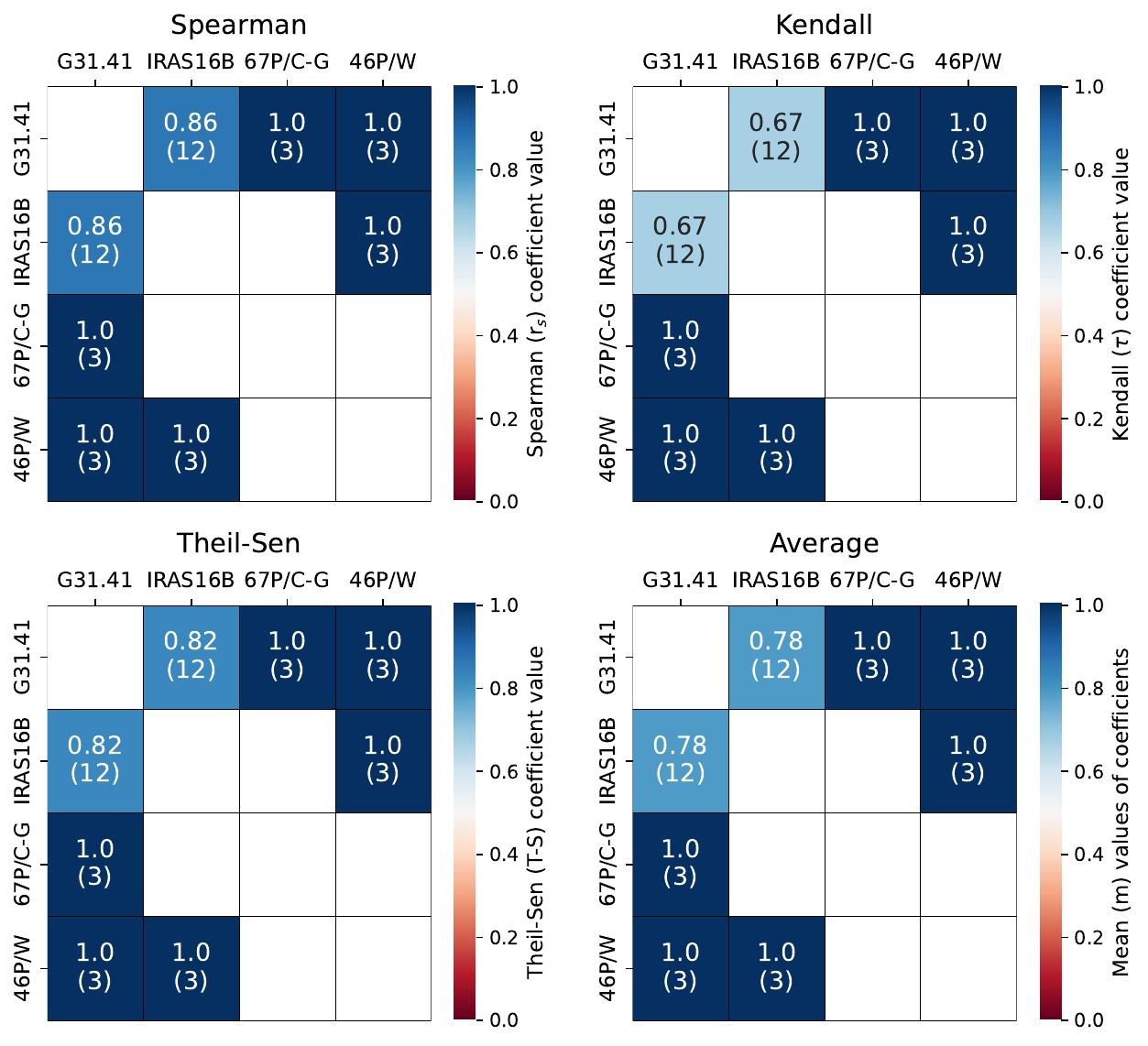}
    \caption{Correlation matrices for N-bearing molecules. The goodness of the correlation is color-coded, the bluer the color, the better the correlation. The value between brackets on each cell is the number of molecules considered.}
\label{fig:matriz_correlaciones_N_con_O}
\end{figure}

\begin{figure}
	\includegraphics[width=\columnwidth]{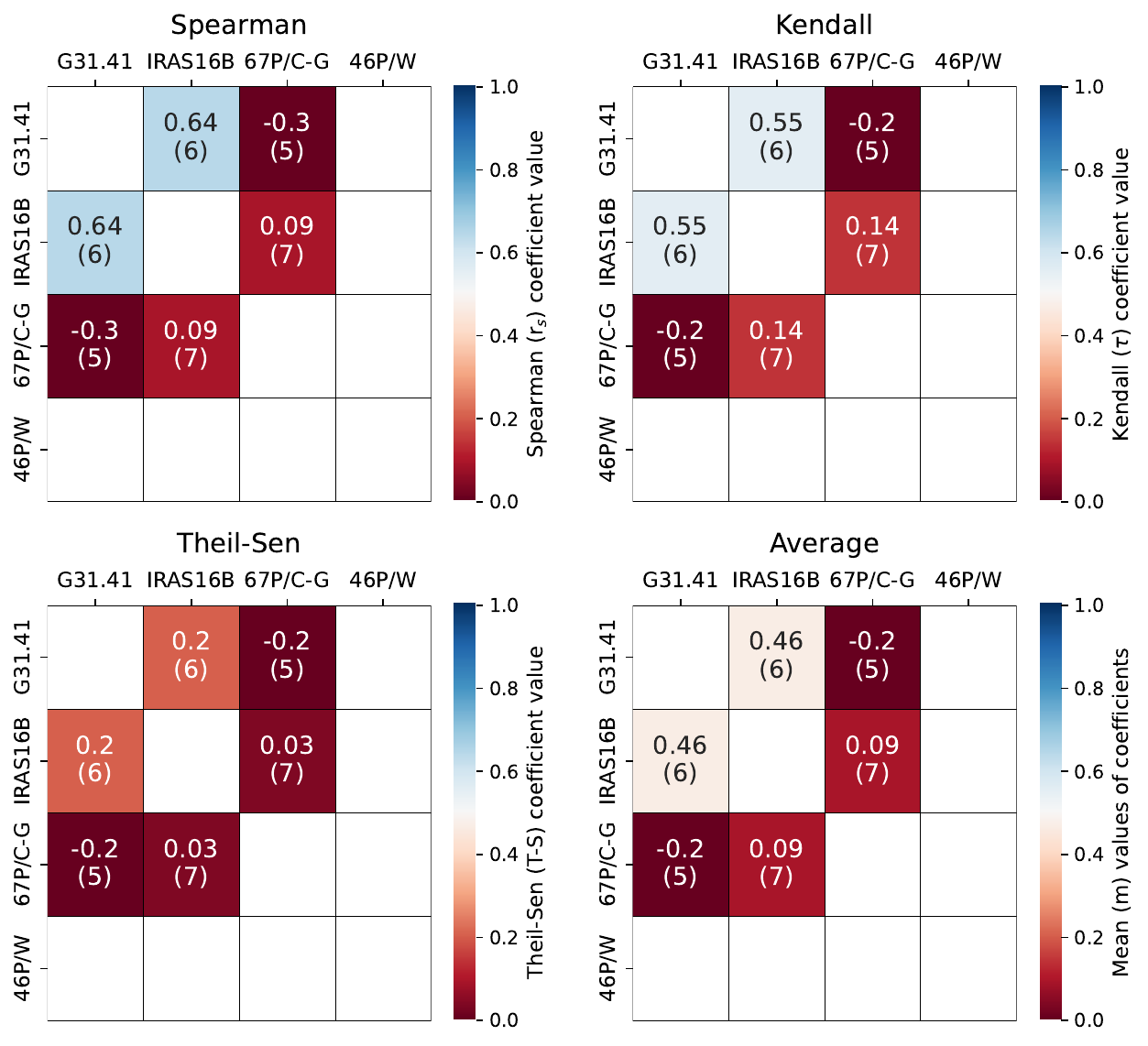}
    \caption{Correlation matrices for S-bearing molecules. The goodness of the correlation is color-coded, the bluer the color, the better the correlation. The value between brackets on each cell is the number of molecules considered.}
\label{fig:matriz_correlaciones_S}
\end{figure}

To analyse the results of the correlation tests we have computed four correlation matrices, one for each test, and one for the average correlation coefficient  ($m$) defined as:

\begin{equation}
    m = \frac{r_{\mathrm{s}} + \tau + \textrm{T-S}}{3} ,
\end{equation} 

where $r_{\mathrm{s}}$, $\tau$, and $\textrm{T-S}$ are Spearman, Kendall and Theil-Sen statistical tests, respectively.
Each matrix is divided in cells, with each cell corresponding to the correlation coefficient for a pair of sources.
The correlation matrices have been obtained for different groups of molecules: all molecules 
(Fig. \ref{fig:matriz_correlaciones_todas}), O-bearing (Fig. \ref{fig:matriz_correlaciones_O}), N-bearing (Fig. \ref{fig:matriz_correlaciones_N_con_O}), and S-bearing (Fig. \ref{fig:matriz_correlaciones_S}). 
As shown in these figures, the results obtained using the three different statistical tests and the average one ($m$) are overall similar, which stress the robustness of the correlation analysis. For this reason, we will restrict the discussion to the average correlation matrices.

\begin{figure*}
     \centering     
\includegraphics[width=\textwidth]{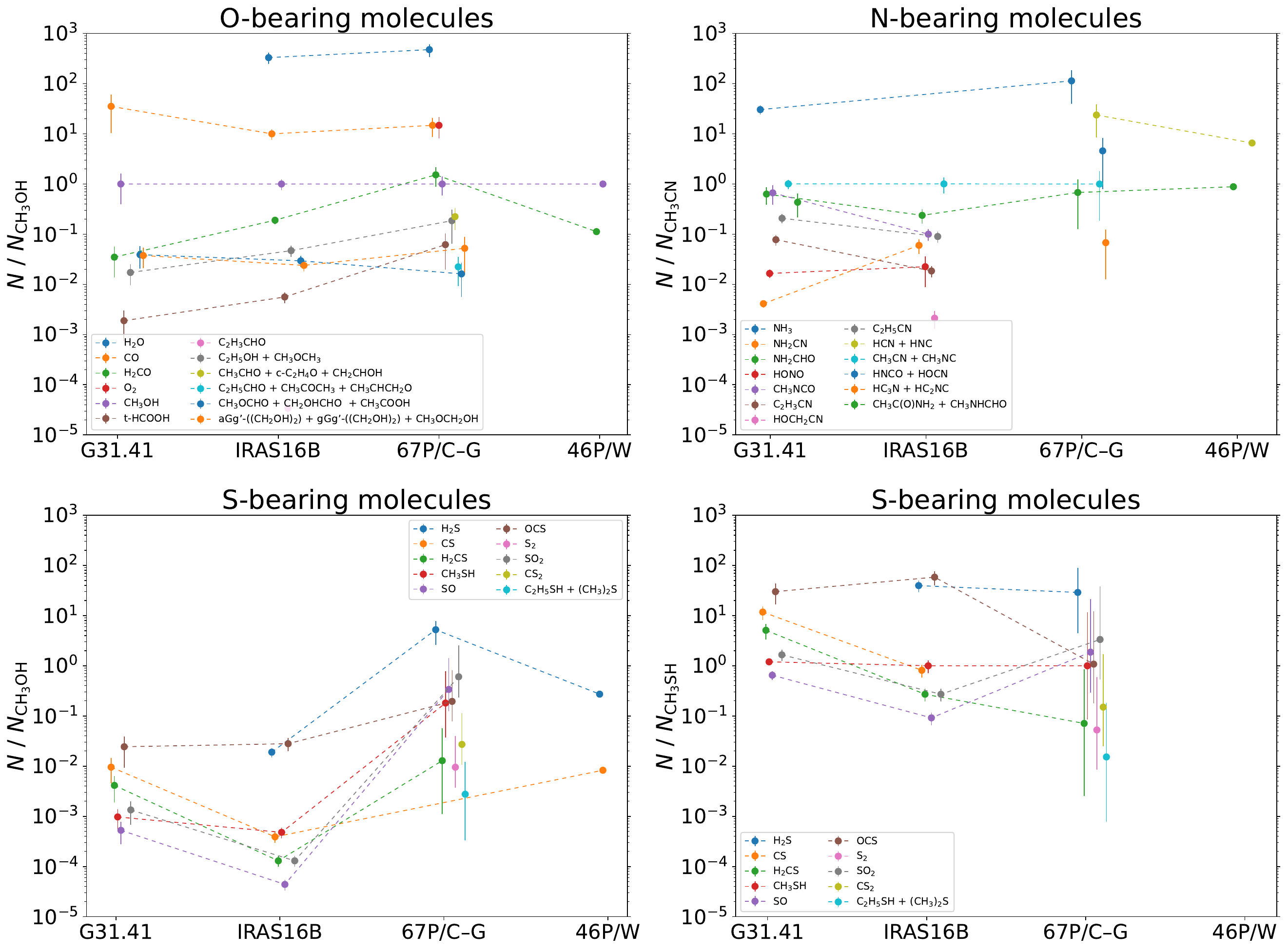}
\caption{Comparison of the molecular abundances of the different molecules detected towards the four astronomical sources analyzed in this work. 
Upper panels show the column densities of detected molecules of O- and N- bearing molecules with respect to \ch{CH3OH} and \ch{CH3CN} respectively. On the lower panels S-bearing are represented with respect to \ch{CH3OH} (left) and \ch{CH3SH} (right).}
\label{fig:evolution_mols_sources}
\end{figure*}

In order to help the interpretation of the correlation tests, we have arbitrarily defined different values of the average correlation coefficient: $m\geq$ 0.9 (excellent correlation), 0.9 $>m\geq$ 0.7 (very good correlation), 0.7 $>m\geq$ 0.5 (good correlation), 0.5 $>m\geq$ 0.3 (poor correlation), 0.3 $>m$ (no correlation). 

\subsubsection{Full chemical reservoir}

We first discuss the results found considering all the molecules analyzed. 

The average correlation matrix (Fig. \ref{fig:matriz_correlaciones_todas}) indicates that the pairs of sources with very good correlations are IRAS16B and 46P/W ($m$=0.78, $p$ values $<$ 0.01, based on 10 molecular species), and the two comets ($m$=0.7, $p$ = 0.1$-$0.23, indicating a low reliability because is based on only 5 species).
Good correlations are also found between G31.41 and IRAS16B ($m$=0.63, $p<$ 0.01 based on 35 molecular species), and G31.41 and 46P/W ($m$=0.54, $p$ = 0.07$-$0.12, based on 9 species).
The poorest correlations involve the 67P/C-G comet with the two star-forming regions: IRAS16B ($m$=0.45, $p<$ 0.1, based on 17 molecular species), and G31.41 ($m$=0.22, based on 15 molecular species, with high $p$ values ranging 0.21$-$0.7, which indicated a low reliability). 
As discussed in the next subsections, the main reason for these discrepancies are the different molecular abundance ratios for S-bearing species found in 67P/C-G compared to those in the star-forming regions. 

In the following subsections, we discuss the results obtained for the different families of molecules.

\subsubsection{O-bearing species}
\label{subsec:O-bearing species}

Considering only O-bearing species, the correlation matrix (Fig. \ref{fig:matriz_correlaciones_O}) indicates that this family of molecules has in general better correlations between sources than the full chemical reservoir (compare with Fig. \ref{fig:matriz_correlaciones_todas}). 
The best correlation is found between IRAS16B and comet 67P/C-G ($m$= 0.86, $p\leq$ 0.01, using 8 molecules); in agreement with the results previously found by \cite{Drozdovskaya2019}.
There is also a very good correlation between G31.41 and 46P/W ($m$= 0.84, $p<$ 0.1, based on 5 molecules). 

A very good correlation is also found between IRAS16B and 46P/W ($m$=0.72, using 5 molecules). The lower middle panel of Fig. \ref{fig:Correlaciones_fuentes_O} shows that the detected molecules are inside the grey shaded area around the 1:1 correlation line, although the $p$ values (0.06$-$0.23) indicate that the correlation should be taken with some caution.

The correlation between G31.41 and IRAS16B is good ($m$=0.65, $p<$ 0.001, based on 16 molecules).
Most of the molecular abundance ratios (with the only exceptions of \ch{CH3CHO} and \ch{CH3COOH}) fall within the area of $\pm$ 1 order of magnitude around the 1:1 line, confirming a very similar chemical composition of O-bearing molecules between these two star-forming sources. 
\ch{CH3CHO} is less abundant in G31.41 by a factor of 26.1, which might be partially due to opacity effects, since \citet{Mininni2023} derived that its emission is partially optically thick, with line opacities up to $\tau$=0.2.
\ch{CH3COOH} is more abundant in G31.41 than in IRAS16B by a factor of 40.5. However, this result is compatible with the ones obtained by \cite{Mininni2020}, in which \ch{CH3COOH} was more abundant in G31.41 compared with other 3 sources by a factor of 2 and also was more abundant on G31.41 than other 18 sources.

The poorest correlation found is between G31.41 and 67P/C-G ($m$=0.26, based on 7 groups with same molecular mass). The reason for this poor correlation (see upper middle panel of Fig. \ref{fig:Correlaciones_fuentes_O}) is the presence of two outliers: \ch{t-HCOOH} and \ch{H2CO}. 
In the case of \ch{t-HCOOH}, the transitions analyzed by \cite{Garcia-de-la-concepcion-2022} are slightly optically thick ($\tau$=0.05-0.2), so the true column density would be higher, and thus closer to the 1:1 region. 
These two outlier molecules might be either underabundant in G31.41 or overabundant in 67P/C-G. Comparing G31.41 with IRAS16B and the comet 46P/W, both molecules fall in the shaded area around the 1:1 line (Fig. \ref{fig:Correlaciones_fuentes_O}; being \ch{t-HCOOH} an upper limit in 46P/W).  
In addition, when comparing the two comets (67P/C-G and 46P/W), \ch{H2CO} and the upper limit of t-HCOOH are above the 1:1 line. Thus, these two species seem to be overabundant in comet 67P/C-G compared to the other sources. 

We did not perform correlation tests for the 67P/C-G and 46P/W pair because only 2 molecular abundance ratios are available (lower right panel of Fig. \ref{fig:Correlaciones_fuentes_O}).

\subsubsection{N-bearing species}
\label{subsec:N-bearing species}

The correlation matrices of N-bearing species are shown in Fig. \ref{fig:matriz_correlaciones_N_con_O}.
The comparison between the two star-forming regions (G31.41 and IRAS16B) shows a very good correlation ($m$=0.78, $p\leq$ 0.01, based on 11 molecules).

As shown in Fig. \ref{fig:Correlaciones_fuentes_N_con_O}, most of the molecular abundance ratios are within the area of $\pm$ 1 order of magnitude with respect to the 1:1 line, with only two outliers: \ch{HC3N} and \ch{NH2CN}. Both molecules fall only slightly outside of the $\pm$1 order of magnitude area. As already mentioned in Sect. \ref{mols:NH2CN}, this discrepancy in \ch{NH2CN} might be due to optical depth effects.

Only \ch{HCN}, \ch{HNC}, \ch{CH3CN}, and \ch{NH2CHO} are detected in both comets, and also \ch{NH3} and \ch{CH3NC} were reported in 67P/C-G. As a consequence, only three molecular abundance ratios are available for the comparisons of G31.41 vs. 67P/C-G,  G31.41 vs. 46P/W, and IRAS16B vs. 46P/W, and so the correlation tests performed should be taken with caution. The average correlation coefficients are in the three cases $m$=1 (Fig. \ref{fig:matriz_correlaciones_N_con_O}), but the low number of points (only three), make these correlations not fully reliable.

For the two comparisons left (IRAS16B vs. 67P/C-G and 67P/C-G vs. 46P/W), we did not preform statistical tests because only two molecular abundance ratios are available.

\subsubsection{S-, P- and Cl- bearing species}
\label{subsec:S-bearing species}

Figure \ref{fig:matriz_correlaciones_S} shows the correlation matrices for S-bearing molecules. The correlations are systematically poorer than those for O- and N-bearing molecules.

The comparison between the two star-forming regions, based on 6 molecules (upper left panel in Fig. \ref{fig:Correlaciones_fuentes_S_P_Cl}), gives a good correlation in the Spearman and Kendall correlation matrix ($r_{\mathrm{s}}$=0.64 and $\tau$=0.55 respectively). However, the Theil-Sen correlation matrix value gives no correlation (T-S=0.2). This is because Theil-Sen searches for a linear fit, while the other two methods check if the points are monotonically increasing. The low correlation derived by Theil-Sen decreases the average coefficient to $m$=0.46, which is slightly below to the threshold we have considered for a good correlation. In any case, the low number of molecules considered (6), and the high $p$-values ($p$=0.13-0.70), advice to take this result with some caution.

We cannot retrieve the correlation coefficients in the comparisons involving 46P/W due to the low number of S-bearing species detected (only \ch{H2S} and CS), and the non detection of the reference molecule, \ch{CH3SH}.

The comparison between the comet 67P/C-G and the two star-forming regions gives no correlation ($m$=-0.2 for G31.41, and $m$=0.09 for IRAS16B; Fig. \ref{fig:matriz_correlaciones_S}), confirming that the S chemical content in the comet significantly differs from that of the star-forming regions, unlike the O and N content, as already mentioned in Sect. \ref{sec:Comparison of molecular abundances}.

Regarding Cl- and P-bearing species, we did not perform correlation tests given the low number of species detected. Only CH$_3$Cl has been detected towards IRAS16B and 67P/C-G \citep{Fayolle2017}, and PO towards  67P/C-G (\citealt{Rivilla2020}), preventing us to compare among the different sources.

\section{Discussion: the chemical thread during star formation}
\label{sec:discuss}

We have compared in this work the chemical composition of four sources that cover initial and final steps of the formation of stars and planetary systems. On the one hand, the natal cores in a high-mass and a low-mass star-forming region (G31.41 and IRAS16B, respectively). On the other hand, two Solar System comets (67P/C-G and 46P/W), whose chemical composition is thought to be almost pristine, namely it informs us about the primitive molecular ingredients of our planetary system that could have been delivered to the young Earth \citep{Altwegg2016, Altwegg2019}.
In the following, we discuss the implications of our findings about the evolution of the chemical feedstock during the process of star and planet formation.

Complementing the correlations done in Sect. \ref{sec:comparison}, and to follow how the molecular abundance ratios vary from source to source, we have plotted them in Fig. \ref{fig:evolution_mols_sources}, separating O-bearing, N-bearing and S-bearing species. The two former are normalized by \ch{CH3OH} (upper right panel) and \ch{CH3CN (upper left panel), respectively. The latter is normalized} by \ch{CH3OH} (lower left panel) and \ch{CH3SH} (lower right panel).
%
As shown in  Fig. \ref{fig:evolution_mols_sources}, and already mentioned in Sect. \ref{sec:comparison}, we have found an overall very good correlation between the molecular abundance ratios of the two star-forming regions (G31.41 and IRAS16B), using a large sample of molecules (35 common detected species in both sources).
Most of the molecular ratio pairs in these two regions are consistent within an order of magnitude, with variation factors of 1.0$\pm$0.6 for O-bearing species, and of 2.3$\pm$1.6 for N-bearing molecules.
This similarity is remarkable because these two star-forming regions have very different physical properties. They differ in the masses of their gas envelopes (70 $M_{\odot}$ vs. 4 $M_{\odot}$; \citealt{Cesaroni2019} and \citealt{Jacobsen2018}; respectively), their protostellar masses  (15-26 M$_{\odot}$ vs. 0.1 M$_{\odot}$), and their luminosities (4.5$\times$10$^4$ $L_{\odot}$ vs. 3 $L_{\odot}$), from \cite{Beltran2021} and \cite{Jacobsen2018}, respectively. Their location within the Galaxy is also different, 
and the local environment clearly differs in terms of clustering: while G31.41 is a massive protostellar cluster (\citealt{Beltran2021}), IRAS16B is an isolated low-mass triple protostellar system (see e.g., \citealt{Maureira2020}).  This implies that the expected protostellar feedback (e.g. from protostellar molecular outflows, and UV radiation) is significantly different. As well, the protostellar heating timescales of the hot core G31.41 is expected to be much faster than that of the hot corino IRAS16B (\citealt{Viti2004, Awad2010}). Moreover, it should be noted that the spatial scales traced by the observations used in this work are very different: 4400 au for G31.41 and 60 au for IRAS16B (\citealt{Mininni2020} and \citealt{Jorgensen2018}, respectively). 
In principle, one should expect that all these above-mentioned differences in the physical conditions might have an imprint in the molecular ratios in the gas phase, due to e.g. different desorption histories (thermal and/or shock-induced), UV processing, or spatial scale effects.
However, we have found that the molecular ratios of O- and N-bearing species towards the hot core and hot corino of G31.41 and IRAS16B, respectively, are remarkably similar.
This result is in good agreement with previous works that have found similar chemical reservoirs comparing tens of star-forming regions, based on smaller samples of molecules (e.g. \citealt{Coletta2020,vanGelder2020,Colzi2021,Nazari2021,Nazari2022}; and references therein). 

%
%




When the comparison is extended to comets, we have found different results depending on the chemical family. O- and N-bearing species correlates well in general. As shown in the upper panels of Fig. \ref{fig:evolution_mols_sources}, most of the abundance ratios with respect to \ch{CH3OH} and \ch{CH3CN}, respectively, are consistent within one order of magnitude in the four sources.
This might imply that the molecular abundance ratios of O- and N-bearing species are set at early evolutionary stages of star formation, 
and that the subsequent evolution does not significantly alter them by a factor larger than one order of magnitude, as shown in Fig. \ref{fig:evolution_mols_sources}.
This is consistent with previous works that compared comets with several interstellar sources (\citealt{Bockelee-Morvan2000, Bianchi2019, Drozdovskaya2019, Coletta2020}), 
supporting the idea of chemical heritage of these chemical families during the different phases of star and planet formation. 

We note that this does not imply that all molecular abundance ratios remain completely fixed and unaltered during the process of star and planet formation. It is known that some molecular ratios can vary when comparing sources at different evolutionary stages. For instance, \citet{Sabatini2021}, using observations and predictions from chemical models of a large sample of massive star-forming clumps, found that while the median abundances of some species remain almost constant (\ch{H2CO} or \ch{CH3OH}), those of other species (\ch{CH3CN} or \ch{CH3CCH}) change by less than one order of magnitude. 
\citet{Urquhart2019} considered up to 120 line intensity ratios (which can be used as an approximate proxy of abundance ratios) towards massive clumps at different stages, and found that 13 of them show some trends with evolutionary stage (see their Fig. 20). The ratios change only by a factor of $\sim$2$-$4 at most, with only a couple of exceptions, for which the factor is $>$5.
These variations are in agreement with the results found in our analysis for the molecular ratios for N- and O-bearing species, when comparing the two star-forming regions (low and high-mass) and the two comets.
Although a few species appear as outliers, 
as discussed in Section \ref{sec:comparison}, most of the molecular ratios are in general well correlated in the four sources analyzed
, with variations less than  one order of magnitude (see Figs. \ref{fig:Correlaciones_fuentes_O}, \ref{fig:Correlaciones_fuentes_N_con_O} and \ref{fig:evolution_mols_sources}).

%

Regarding  S- and P-bearing species, clear differences were observed between the star-forming regions and the comets, as already mentioned in Sect. \ref{sec:comparison}. The two lower panels of Fig. \ref{fig:evolution_mols_sources} are very illustrative of the 
behaviour of the S-bearing family. 
The abundance ratios with respect to \ch{CH3OH} (lower left panel) have significantly higher values in 67P/C-G than in the two star-forming regions by 1-2 orders of magnitude (factors of 7-377). However, once normalized with \ch{CH3SH} (lower right panel), the molecular abundance ratios in 67P/C-G  are very similar to those of the two star-forming regions (\ch{CH3SH} was not detected towards 46P/W). 

The thiol/alcohol ratio \ch{CH3SH}/\ch{CH3OH} is 0.18 in 67P/C-G. This value is three orders of magnitude higher than those derived in G31.41 and IRAS16B: (8$\pm$4)$\times 10^{-4}$ and (4.8$\pm$0.8)$\times 10^{-4}$, respectively. This suggests an overabundance of S-bearing species, compared to O-bearing, in the comet 67P/C-G. 
This is in good agreement with the high S/O ratio of 0.015 found by \citet{Calmonte2016} in the coma of 67P/C-G, and in the comet C/1995 O1 Hale-Bopp of 0.02 (\citealt{Bockelee-Morvan2000}). Both values are very close to the cosmic abundance S/O ratio measured from solar photospheric abundances \citep{Anders1989, Lodders2010}. As discussed in \citet{Calmonte2016}, this strongly suggests that S is not depleted in these comets, in contrast to the results usually found in dense clouds and star-forming regions (\citealt{Gondhalekar1985, Jimenez-escobar2011, Laas2019, Fuente2023}), like G31.41 and IRAS16B.
This means that in star-forming regions most of the sulfur is trapped in ice grains, since S-bearing species are more refractory than O- and N-bearing species, and thus they are not easily thermally desorbed into the gas phase in the hot cores/corinos present in high- and low-mass star-forming regions, respectively. In contrast, the results found in comets indicate that these S-bearing species can escape the surface, and hence we can have access to the refractory S-bearing content.

A similar behaviour is also observed for P-bearing species, which are also thought to have a refractory character. In  this work, we have not performed the statistical tests for P-bearing molecules because only a single species, phosphorus monoxide (PO), has been detected towards only one of the sources considered, the comet 67P/C-G (\citealt{Rivilla2020}). 
The derived P/O ratio is (0.5$-$2.7)$\times$10$^{-4}$, close to the solar value of $\sim$6$\times$10$^{-4}$, which indicates that P, similarly to S, is only slightly or not depleted.
In contrast, the middle upper and left lower panels of Fig. \ref{fig:Correlaciones_fuentes_S_P_Cl} show that the PO upper limits derived towards G31.41 (this work) and IRAS16B (\citealt{Drozdovskaya2019}) are below the 1:1 line, indicating that this species is underabundant in the gas phase of star-forming regions compared to the 67P/C-G comet, similarly to the S-bearing species.

Indeed, P-bearing molecules are not detected in hot cores (\citealt{Rivilla2020}) or hot corinos (\citealt{Bergner2022}). The interferometric high-angular resolution of these works has shown that the emission of P-bearing molecules (PN and PO) does not arise from the hot gas surrounding the protostars, but from more distant gas affected by shocks, which are able to sputter the grains and release the P content of their surfaces. 
Very recently, \citet{Fontani2024} have confirmed that PN, and tentatively PO, are indeed detected in shocked gas associated with molecular outflows powered by the protostars embedded in the G31.41 cluster.

Finally, we note that the analysis performed here presents evident caveats that need to be addressed in the near future. We have considered the molecular ratios of only four sources (two star-forming regions and two comets), whose chemical composition is very well known. This has allowed to study tens of different molecular ratios, but with poor statistics in terms of number of sources. Therefore, we stress that it is mandatory to perform unbiased and sensitive spectral surveys towards large samples of 
low- and high-mass star-forming regions. This will allow an extensive analysis of their full chemical census, including tens of molecules, like the ones done for G31.41 and IRAS16B by the GUAPOS and PILS survey, respectively. 
This will complement previous efforts that have been focused on a handful of selected species towards large samples of hot corinos (e.g. \citealt{Bouvier2021} or \citealt{Yang2021}), and hot cores (e.g. \citealt{Chen2023}).


Moreover, there is still a key step in the process of star and planet formation whose chemical content is largely unknown: the planet-forming disk phase. Numerous efforts have been done in the last years (e.g. \citealt{Banzatti2020, Grant2021, Pegues2021, Booth2021, Brunken2022}), including the detection of complex molecules (\citealt{Brunken2022, Tabone2023}), which have suggested a direct inheritance between protoplanetary disks and previous phases e.g. \citep{Booth2021}. However, the number of species detected in these objects, which are the cradles of new planets, is still low compared with the ISM census (see \citealt{McGuire2022}). Hence, direct comparisons based on a large sample of molecules, like the ones done in this work, are not possible yet, and should be addressed in the forthcoming years. 
As well, current and future space {\it in-situ} and sample-return missions, targeting new comets and asteroids, like the on-going JAXA Hayabusa2 mission to the asteroid Ryugu (e.g. \citealt{Potiszil2023}), or the OSIRIS-REx mission to the asteroid Bennu (\citealt{Lauretta2017}), will significantly widen our current view of the chemical ingredient of Solar System objects, and thus they will allow us to attain a complete view of the chemical thread during the whole evolution of planetary systems.

\section{Summary and Conclusions}
\label{sec:conclusions}

In this work we provide a comprehensive view of the chemical feedstock of the massive star-forming protocluster G31.41+0.31. We have analyzed the emission of 34 molecules
using data from the GUAPOS project (G31.41+0.31 Unbiased ALMA sPectral Observational Survey).
We have derived the column densities and abundances of 18 molecular species, including 25 isotopologues and 3 tentative detections (HONO, \ch{C2H5CHO} and \ch{CH3OCH2OH}). For the 16 molecules not detected, we derived column density upper limits.
We have added these molecules to those already reported in previous GUAPOS works, to compile a total molecule sample of 57 species. Then, we have performed a comparative study of the G31.41 molecular abundance ratios with those of sources that represent different evolutionary stages of star formation, from the protostellar stage to Solar System objects: the prototypical Solar-like protostar (IRAS 16293-2422 B), and two comets (67P/Churyumov-Gerasimenko and 46P/Wirtanen).
To quantitatively study the correlation between the chemical content of the sources we performed three complementary statistical tests: 
Spearman, Kendall and Theil-Sen. 
Our results shows that the molecular abundances of the two star-forming regions, G31.41 (high-mass) and IRAS16B (low-mass), are in general well correlated, including O-, N- and S-bearing species. This similarity, based on a large sample of molecules, denotes that molecular abundance ratios do not vary by factors larger than one order of magnitude, despite of the very different physical properties of these two high- and low-mass star-forming regions, such as  
the mass or the luminosity, level of clustering, protostellar feedback, UV radiation, evolution timescales, or location within the Galaxy.


The comparison with the two comets (67P/C-G and 46P/W) shows that their O- and N-bearing molecular abundance ratios are also well correlated with those of the star-forming regions. This suggests that the abundances ratios of these chemical families are primarily set at initial phases of star formation, and subsequently inherited without variations larger than one order of magnitude during the process of star and planet formation.
However, the S-bearing molecules behave differently, being significantly more abundant in 67P/C-G and 46P/W than in the gas phase of star-forming regions. 
This result can be explained in terms of sulfur-depletion in the ISM, namely that sulfur is trapped on the icy surfaces of dust grains. On the contrary, S-bearing species are believed to have been freely desorbed into the coma of the comets, recovering the expected cosmic abundance. A similar behaviour is also suggested for P, based on the PO detection in the comet 67P/C-G, and the upper limits derived in the two star-forming regions.

\section*{Acknowledgements}

We thank the anonymous referee for a careful reading
of the article and her/his useful comments. The figures showing the molecular spectra of G31.41 and the table of molecular trasnsitions were made thanks to a code created by Andrés Megías (Github: https://github.com/andresmegias/madcuba-slim-scripts) and the Python library richvalues, respectively. We thank to Miguel Sanz-Novo for the help regarding the quantum numbers of each molecule. We also thank to M. Drozdovskaya for some discussion about the IRAS16B molecular abundances. 
This paper makes use of the following ALMA data: ADS/JAO.ALMA$\#$2017.1.00501.S. ALMA is a partnership of ESO (representing its member states), NSF (USA) and NINS (Japan), together with NRC (Canada), MOST and ASIAA (Taiwan), and KASI (Republic of Korea), in cooperation with the Republic of Chile. The Joint ALMA Observatory is operated by ESO, AUI/NRAO and NAOJ.
A. L-G. and V.M.R. have received support from the Comunidad de Madrid through the Atracci\'on de Talento Investigador Modalidad 1 (Doctores con experiencia) Grant (COOL: Cosmic Origins Of Life; 2019-T1/TIC-5379), the project RYC2020-029387-I funded by MICIU/AEI/10.13039/501100011033 and by “ESF, Investing in your future”, from the the Consejo Superior de Investigaciones Cient{\'i}ficas (CSIC) and the Centro de Astrobiolog{\'i}a (CAB) through the project 20225AT015 (Proyectos intramurales especiales del CSIC), and from the Spanish Ministry of Science through the project PID2022-136814NB-I00.
L.C. and I.J-S. acknowledge financial support through the Spanish grants PID2019-105552RB- C41 and PID2022-136814NB-I00 from the Spanish Ministry of Science and Innovation/State Agency of Research MICIU/AEI/10.13039/501100011033 and by “ERDF A way of making Europe”. C.M. acknowledges funding from the European Research Council (ERC) under the European Union's Horizon 2020 program through the ECOGAL Synergy grant (ID 855130). A.S-M. acknowledges support from the RyC2021-032892-I and PID2020-117710GB-I00 grants funded by MICIU/AEI/10.13039/501100011033 and by the European Union ‘Next GenerationEU’/PRTR, as well as the program Unidad de Excelencia María de Maeztu CEX2020-001058-M. S.V. ackowledges support from the European Research Council (ERC) grant MOPPEX ERC-833460.


\section*{Data Availability}

The data presented in this work are available upon reasonable request.
 



\bibliographystyle{mnras}
\bibliography{example} 

\begin{thebibliography}{}
\makeatletter
\relax
\def\mn@urlcharsother{\let\do\@makeother \do\$\do\&\do\#\do\^\do\_\do\%\do\~}
\def\mn@doi{\begingroup\mn@urlcharsother \@ifnextchar [ {\mn@doi@} {\mn@doi@[]}}
\def\mn@doi@[#1]#2{\def\@tempa{#1}\ifx\@tempa\@empty \href {http://dx.doi.org/#2} {doi:#2}\else \href {http://dx.doi.org/#2} {#1}\fi \endgroup}
\def\mn@eprint#1#2{\mn@eprint@#1:#2::\@nil}
\def\mn@eprint@arXiv#1{\href {http://arxiv.org/abs/#1} {{\tt arXiv:#1}}}
\def\mn@eprint@dblp#1{\href {http://dblp.uni-trier.de/rec/bibtex/#1.xml} {dblp:#1}}
\def\mn@eprint@#1:#2:#3:#4\@nil{\def\@tempa {#1}\def\@tempb {#2}\def\@tempc {#3}\ifx \@tempc \@empty \let \@tempc \@tempb \let \@tempb \@tempa \fi \ifx \@tempb \@empty \def\@tempb {arXiv}\fi \@ifundefined {mn@eprint@\@tempb}{\@tempb:\@tempc}{\expandafter \expandafter \csname mn@eprint@\@tempb\endcsname \expandafter{\@tempc}}}

\bibitem[\protect\citeauthoryear{{Adams}}{{Adams}}{2010}]{Adams2010}
{Adams} F.~C.,  2010, \mn@doi [\araa] {10.1146/annurev-astro-081309-130830}, \href {https://ui.adsabs.harvard.edu/abs/2010ARA&A..48...47A} {48, 47}

\bibitem[\protect\citeauthoryear{{Altwegg} et~al.,}{{Altwegg} et~al.}{2016}]{Altwegg2016}
{Altwegg} K.,  et~al., 2016, \mn@doi [Science Advances] {10.1126/sciadv.1600285}, \href {https://ui.adsabs.harvard.edu/abs/2016SciA....2E0285A} {2, e1600285}

\bibitem[\protect\citeauthoryear{{Altwegg} et~al.,}{{Altwegg} et~al.}{2017}]{Altwegg2017a}
{Altwegg} K.,  et~al., 2017, \mn@doi [Philosophical Transactions of the Royal Society of London Series A] {10.1098/rsta.2016.0253}, \href {https://ui.adsabs.harvard.edu/abs/2017RSPTA.37560253A} {375, 20160253}

\bibitem[\protect\citeauthoryear{{Altwegg}, {Balsiger}  \& {Fuselier}}{{Altwegg} et~al.}{2019}]{Altwegg2019}
{Altwegg} K.,  {Balsiger} H.,   {Fuselier} S.~A.,  2019, \mn@doi [\araa] {10.1146/annurev-astro-091918-104409}, \href {https://ui.adsabs.harvard.edu/abs/2019ARA&A..57..113A} {57, 113}

\bibitem[\protect\citeauthoryear{{Anders} \& {Grevesse}}{{Anders} \& {Grevesse}}{1989}]{Anders1989}
{Anders} E.,  {Grevesse} N.,  1989, \mn@doi [\gca] {10.1016/0016-7037(89)90286-X}, \href {https://ui.adsabs.harvard.edu/abs/1989GeCoA..53..197A} {53, 197}

\bibitem[\protect\citeauthoryear{{Awad}, {Viti}, {Collings}  \& {Williams}}{{Awad} et~al.}{2010}]{Awad2010}
{Awad} Z.,  {Viti} S.,  {Collings} M.~P.,   {Williams} D.~A.,  2010, \mn@doi [\mnras] {10.1111/j.1365-2966.2010.17077.x}, \href {https://ui.adsabs.harvard.edu/abs/2010MNRAS.407.2511A} {407, 2511}

\bibitem[\protect\citeauthoryear{{Banzatti} et~al.,}{{Banzatti} et~al.}{2020}]{Banzatti2020}
{Banzatti} A.,  et~al., 2020, \mn@doi [\apj] {10.3847/1538-4357/abbc1a}, \href {https://ui.adsabs.harvard.edu/abs/2020ApJ...903..124B} {903, 124}

\bibitem[\protect\citeauthoryear{{Beltr{\'a}n}, {Codella}, {Viti}, {Neri}  \& {Cesaroni}}{{Beltr{\'a}n} et~al.}{2009}]{Beltran2009}
{Beltr{\'a}n} M.~T.,  {Codella} C.,  {Viti} S.,  {Neri} R.,   {Cesaroni} R.,  2009, \mn@doi [\apjl] {10.1088/0004-637X/690/2/L93}, \href {https://ui.adsabs.harvard.edu/abs/2009ApJ...690L..93B} {690, L93}

\bibitem[\protect\citeauthoryear{{Beltr{\'a}n} et~al.,}{{Beltr{\'a}n} et~al.}{2018}]{Beltran2018}
{Beltr{\'a}n} M.~T.,  et~al., 2018, \mn@doi [\aap] {10.1051/0004-6361/201832811}, \href {https://ui.adsabs.harvard.edu/abs/2018A&A...615A.141B} {615, A141}

\bibitem[\protect\citeauthoryear{{Beltr{\'a}n} et~al.,}{{Beltr{\'a}n} et~al.}{2021}]{Beltran2021}
{Beltr{\'a}n} M.~T.,  et~al., 2021, \mn@doi [\aap] {10.1051/0004-6361/202040121}, \href {https://ui.adsabs.harvard.edu/abs/2021A&A...648A.100B} {648, A100}

\bibitem[\protect\citeauthoryear{{Bergner}, {Burkhardt}, {{\"O}berg}, {Rice}  \& {Bergin}}{{Bergner} et~al.}{2022}]{Bergner2022}
{Bergner} J.~B.,  {Burkhardt} A.~M.,  {{\"O}berg} K.~I.,  {Rice} T.~S.,   {Bergin} E.~A.,  2022, \mn@doi [\apj] {10.3847/1538-4357/ac47a2}, \href {https://ui.adsabs.harvard.edu/abs/2022ApJ...927....7B} {927, 7}

\bibitem[\protect\citeauthoryear{{Bianchi} et~al.,}{{Bianchi} et~al.}{2019}]{Bianchi2019}
{Bianchi} E.,  et~al., 2019, \mn@doi [\mnras] {10.1093/mnras/sty2915}, \href {https://ui.adsabs.harvard.edu/abs/2019MNRAS.483.1850B} {483, 1850}

\bibitem[\protect\citeauthoryear{{Biver} et~al.,}{{Biver} et~al.}{2021}]{Biver2021}
{Biver} N.,  et~al., 2021, \mn@doi [\aap] {10.1051/0004-6361/202040125}, \href {https://ui.adsabs.harvard.edu/abs/2021A&A...648A..49B} {648, A49}

\bibitem[\protect\citeauthoryear{{Bockel{\'e}e-Morvan} et~al.,}{{Bockel{\'e}e-Morvan} et~al.}{2000}]{Bockelee-Morvan2000}
{Bockel{\'e}e-Morvan} D.,  et~al., 2000, \aap, \href {https://ui.adsabs.harvard.edu/abs/2000A&A...353.1101B} {353, 1101}

\bibitem[\protect\citeauthoryear{{Bonato} et~al.,}{{Bonato} et~al.}{2018}]{Bonato2018}
{Bonato} M.,  et~al., 2018, \mn@doi [\mnras] {10.1093/mnras/sty1173}, \href {https://ui.adsabs.harvard.edu/abs/2018MNRAS.478.1512B} {478, 1512}

\bibitem[\protect\citeauthoryear{{Booth}, {Walsh}, {Terwisscha van Scheltinga}, {van Dishoeck}, {Ilee}, {Hogerheijde}, {Kama}  \& {Nomura}}{{Booth} et~al.}{2021}]{Booth2021}
{Booth} A.~S.,  {Walsh} C.,  {Terwisscha van Scheltinga} J.,  {van Dishoeck} E.~F.,  {Ilee} J.~D.,  {Hogerheijde} M.~R.,  {Kama} M.,   {Nomura} H.,  2021, \mn@doi [Nature Astronomy] {10.1038/s41550-021-01352-w}, \href {https://ui.adsabs.harvard.edu/abs/2021NatAs...5..684B} {5, 684}

\bibitem[\protect\citeauthoryear{{Bouvier}, {L{\'o}pez-Sepulcre}, {Ceccarelli}, {Sakai}, {Yamamoto}  \& {Yang}}{{Bouvier} et~al.}{2021}]{Bouvier2021}
{Bouvier} M.,  {L{\'o}pez-Sepulcre} A.,  {Ceccarelli} C.,  {Sakai} N.,  {Yamamoto} S.,   {Yang} Y.~L.,  2021, \mn@doi [\aap] {10.1051/0004-6361/202141157}, \href {https://ui.adsabs.harvard.edu/abs/2021A&A...653A.117B} {653, A117}

\bibitem[\protect\citeauthoryear{{Brinkman}, {den Hartogh}, {Doherty}, {Pignatari}  \& {Lugaro}}{{Brinkman} et~al.}{2021}]{Brinkman2021}
{Brinkman} H.~E.,  {den Hartogh} J.~W.,  {Doherty} C.~L.,  {Pignatari} M.,   {Lugaro} M.,  2021, \mn@doi [\apj] {10.3847/1538-4357/ac25ea}, \href {https://ui.adsabs.harvard.edu/abs/2021ApJ...923...47B} {923, 47}

\bibitem[\protect\citeauthoryear{{Brunken}, {Booth}, {Leemker}, {Nazari}, {van der Marel}  \& {van Dishoeck}}{{Brunken} et~al.}{2022}]{Brunken2022}
{Brunken} N. G.~C.,  {Booth} A.~S.,  {Leemker} M.,  {Nazari} P.,  {van der Marel} N.,   {van Dishoeck} E.~F.,  2022, \mn@doi [\aap] {10.1051/0004-6361/202142981}, \href {https://ui.adsabs.harvard.edu/abs/2022A&A...659A..29B} {659, A29}

\bibitem[\protect\citeauthoryear{{Calcutt} et~al.,}{{Calcutt} et~al.}{2018}]{Calcutt2018}
{Calcutt} H.,  et~al., 2018, \mn@doi [\aap] {10.1051/0004-6361/201732289}, \href {https://ui.adsabs.harvard.edu/abs/2018A&A...616A..90C} {616, A90}

\bibitem[\protect\citeauthoryear{{Calcutt} et~al.,}{{Calcutt} et~al.}{2019}]{Calcutt2019}
{Calcutt} H.,  et~al., 2019, \mn@doi [\aap] {10.1051/0004-6361/201936323}, \href {https://ui.adsabs.harvard.edu/abs/2019A&A...631A.137C} {631, A137}

\bibitem[\protect\citeauthoryear{{Calmonte} et~al.,}{{Calmonte} et~al.}{2016}]{Calmonte2016}
{Calmonte} U.,  et~al., 2016, \mn@doi [\mnras] {10.1093/mnras/stw2601}, \href {https://ui.adsabs.harvard.edu/abs/2016MNRAS.462S.253C} {462, S253}

\bibitem[\protect\citeauthoryear{{Carpenter}}{{Carpenter}}{2000}]{Carpenter2000}
{Carpenter} J.~M.,  2000, \mn@doi [\aj] {10.1086/316845}, \href {https://ui.adsabs.harvard.edu/abs/2000AJ....120.3139C} {120, 3139}

\bibitem[\protect\citeauthoryear{{Ceccarelli} et~al.,}{{Ceccarelli} et~al.}{2023}]{Ceccarelli2023}
{Ceccarelli} C.,  et~al., 2023, in {Inutsuka} S.,  {Aikawa} Y.,  {Muto} T.,  {Tomida} K.,   {Tamura} M.,  eds,  Astronomical Society of the Pacific Conference Series Vol. 534, Astronomical Society of the Pacific Conference Series. p.~379

\bibitem[\protect\citeauthoryear{{Cesaroni}}{{Cesaroni}}{2019}]{Cesaroni2019}
{Cesaroni} R.,  2019, \mn@doi [\aap] {10.1051/0004-6361/201936334}, \href {https://ui.adsabs.harvard.edu/abs/2019A&A...631A..65C} {631, A65}

\bibitem[\protect\citeauthoryear{{Cesaroni}, {Churchwell}, {Hofner}, {Walmsley}  \& {Kurtz}}{{Cesaroni} et~al.}{1994}]{Cesaroni1994}
{Cesaroni} R.,  {Churchwell} E.,  {Hofner} P.,  {Walmsley} C.~M.,   {Kurtz} S.,  1994, \aap, \href {https://ui.adsabs.harvard.edu/abs/1994A&A...288..903C} {288, 903}

\bibitem[\protect\citeauthoryear{{Chen} et~al.,}{{Chen} et~al.}{2023}]{Chen2023}
{Chen} Y.,  et~al., 2023, \mn@doi [\aap] {10.1051/0004-6361/202346491}, \href {https://ui.adsabs.harvard.edu/abs/2023A&A...678A.137C} {678, A137}

\bibitem[\protect\citeauthoryear{{Cleeves}, {Bergin}, {Alexander}, {Du}, {Graninger}, {{\"O}berg}  \& {Harries}}{{Cleeves} et~al.}{2014}]{Cleeves2014}
{Cleeves} L.~I.,  {Bergin} E.~A.,  {Alexander} C. M.~O.~D.,  {Du} F.,  {Graninger} D.,  {{\"O}berg} K.~I.,   {Harries} T.~J.,  2014, \mn@doi [Science] {10.1126/science.1258055}, \href {https://ui.adsabs.harvard.edu/abs/2014Sci...345.1590C} {345, 1590}

\bibitem[\protect\citeauthoryear{{Coletta}, {Fontani}, {Rivilla}, {Mininni}, {Colzi}, {S{\'a}nchez-Monge}  \& {Beltr{\'a}n}}{{Coletta} et~al.}{2020}]{Coletta2020}
{Coletta} A.,  {Fontani} F.,  {Rivilla} V.~M.,  {Mininni} C.,  {Colzi} L.,  {S{\'a}nchez-Monge} {\'A}.,   {Beltr{\'a}n} M.~T.,  2020, \mn@doi [\aap] {10.1051/0004-6361/202038212}, \href {https://ui.adsabs.harvard.edu/abs/2020A&A...641A..54C} {641, A54}

\bibitem[\protect\citeauthoryear{{Colzi}, {Fontani}, {Rivilla}, {S{\'a}nchez-Monge}, {Testi}, {Beltr{\'a}n}  \& {Caselli}}{{Colzi} et~al.}{2018}]{Colzi2018}
{Colzi} L.,  {Fontani} F.,  {Rivilla} V.~M.,  {S{\'a}nchez-Monge} A.,  {Testi} L.,  {Beltr{\'a}n} M.~T.,   {Caselli} P.,  2018, \mn@doi [\mnras] {10.1093/mnras/sty1027}, \href {https://ui.adsabs.harvard.edu/abs/2018MNRAS.478.3693C} {478, 3693}

\bibitem[\protect\citeauthoryear{{Colzi} et~al.,}{{Colzi} et~al.}{2021}]{Colzi2021}
{Colzi} L.,  et~al., 2021, \mn@doi [\aap] {10.1051/0004-6361/202141573}, \href {https://ui.adsabs.harvard.edu/abs/2021A&A...653A.129C} {653, A129}

\bibitem[\protect\citeauthoryear{{Coutens} et~al.,}{{Coutens} et~al.}{2018}]{Coutens2018}
{Coutens} A.,  et~al., 2018, \mn@doi [\aap] {10.1051/0004-6361/201732346}, \href {https://ui.adsabs.harvard.edu/abs/2018A&A...612A.107C} {612, A107}

\bibitem[\protect\citeauthoryear{{Coutens} et~al.,}{{Coutens} et~al.}{2019}]{Coutens2019}
{Coutens} A.,  et~al., 2019, \mn@doi [\aap] {10.1051/0004-6361/201935040}, \href {https://ui.adsabs.harvard.edu/abs/2019A&A...623L..13C} {623, L13}

\bibitem[\protect\citeauthoryear{{Dhooghe} et~al.,}{{Dhooghe} et~al.}{2017}]{Dhooghe2017}
{Dhooghe} F.,  et~al., 2017, \mn@doi [\mnras] {10.1093/mnras/stx1911}, \href {https://ui.adsabs.harvard.edu/abs/2017MNRAS.472.1336D} {472, 1336}

\bibitem[\protect\citeauthoryear{{Dickman}}{{Dickman}}{1978}]{Dickman1978}
{Dickman} R.~L.,  1978, \mn@doi [\apjs] {10.1086/190535}, \href {https://ui.adsabs.harvard.edu/abs/1978ApJS...37..407D} {37, 407}

\bibitem[\protect\citeauthoryear{{Drozdovskaya}, {van Dishoeck}, {Rubin}, {J{\o}rgensen}  \& {Altwegg}}{{Drozdovskaya} et~al.}{2019}]{Drozdovskaya2019}
{Drozdovskaya} M.~N.,  {van Dishoeck} E.~F.,  {Rubin} M.,  {J{\o}rgensen} J.~K.,   {Altwegg} K.,  2019, \mn@doi [\mnras] {10.1093/mnras/stz2430}, \href {https://ui.adsabs.harvard.edu/abs/2019MNRAS.490...50D} {490, 50}

\bibitem[\protect\citeauthoryear{{Duan}, {Li}, {Pagani}, {Goldsmith}, {Ching}, {Wang}  \& {Xie}}{{Duan} et~al.}{2023}]{Duan2023}
{Duan} Y.,  {Li} D.,  {Pagani} L.,  {Goldsmith} P.~F.,  {Ching} T.-C.,  {Wang} C.,   {Xie} J.,  2023, \mn@doi [Research in Astronomy and Astrophysics] {10.1088/1674-4527/acd7bd}, \href {https://ui.adsabs.harvard.edu/abs/2023RAA....23i5006D} {23, 095006}

\bibitem[\protect\citeauthoryear{{Dukes} \& {Krumholz}}{{Dukes} \& {Krumholz}}{2012}]{Dukes2012}
{Dukes} D.,  {Krumholz} M.~R.,  2012, \mn@doi [\apj] {10.1088/0004-637X/754/1/56}, \href {https://ui.adsabs.harvard.edu/abs/2012ApJ...754...56D} {754, 56}

\bibitem[\protect\citeauthoryear{{Endres}, {Schlemmer}, {Schilke}, {Stutzki}  \& {M{\"u}ller}}{{Endres} et~al.}{2016}]{Endres2016}
{Endres} C.~P.,  {Schlemmer} S.,  {Schilke} P.,  {Stutzki} J.,   {M{\"u}ller} H. S.~P.,  2016, \mn@doi [Journal of Molecular Spectroscopy] {10.1016/j.jms.2016.03.005}, \href {https://ui.adsabs.harvard.edu/abs/2016JMoSp.327...95E} {327, 95}

\bibitem[\protect\citeauthoryear{{Fayolle} et~al.,}{{Fayolle} et~al.}{2017}]{Fayolle2017}
{Fayolle} E.~C.,  et~al., 2017, \mn@doi [Nature Astronomy] {10.1038/s41550-017-0237-7}, \href {https://ui.adsabs.harvard.edu/abs/2017NatAs...1..703F} {1, 703}

\bibitem[\protect\citeauthoryear{{Fontani} et~al.,}{{Fontani} et~al.}{2024}]{Fontani2024}
{Fontani} F.,  et~al., 2024, \mn@doi [\aap] {10.1051/0004-6361/202348219}, \href {https://ui.adsabs.harvard.edu/abs/2024A&A...682A..74F} {682, A74}

\bibitem[\protect\citeauthoryear{{Fuente} et~al.,}{{Fuente} et~al.}{2023}]{Fuente2023}
{Fuente} A.,  et~al., 2023, \mn@doi [\aap] {10.1051/0004-6361/202244843}, \href {https://ui.adsabs.harvard.edu/abs/2023A&A...670A.114F} {670, A114}

\bibitem[\protect\citeauthoryear{{Garc{\'\i}a de la Concepci{\'o}n} et~al.,}{{Garc{\'\i}a de la Concepci{\'o}n} et~al.}{2022}]{Garcia-de-la-concepcion-2022}
{Garc{\'\i}a de la Concepci{\'o}n} J.,  et~al., 2022, \mn@doi [\aap] {10.1051/0004-6361/202142287}, \href {https://ui.adsabs.harvard.edu/abs/2022A&A...658A.150G} {658, A150}

\bibitem[\protect\citeauthoryear{{Gondhalekar}}{{Gondhalekar}}{1985}]{Gondhalekar1985}
{Gondhalekar} P.~M.,  1985, \mn@doi [\mnras] {10.1093/mnras/217.3.585}, \href {https://ui.adsabs.harvard.edu/abs/1985MNRAS.217..585G} {217, 585}

\bibitem[\protect\citeauthoryear{{Grant} et~al.,}{{Grant} et~al.}{2021}]{Grant2021}
{Grant} S.~L.,  et~al., 2021, \mn@doi [\apj] {10.3847/1538-4357/abf432}, \href {https://ui.adsabs.harvard.edu/abs/2021ApJ...913..123G} {913, 123}

\bibitem[\protect\citeauthoryear{{Hacar}, {Bosman}  \& {van Dishoeck}}{{Hacar} et~al.}{2020}]{Hacar2020}
{Hacar} A.,  {Bosman} A.~D.,   {van Dishoeck} E.~F.,  2020, \mn@doi [\aap] {10.1051/0004-6361/201936516}, \href {https://ui.adsabs.harvard.edu/abs/2020A&A...635A...4H} {635, A4}

\bibitem[\protect\citeauthoryear{{Hadraoui} et~al.,}{{Hadraoui} et~al.}{2019}]{Hadraoui2019}
{Hadraoui} K.,  et~al., 2019, \mn@doi [\aap] {10.1051/0004-6361/201935018}, \href {https://ui.adsabs.harvard.edu/abs/2019A&A...630A..32H} {630, A32}

\bibitem[\protect\citeauthoryear{{Herbst} \& {van Dishoeck}}{{Herbst} \& {van Dishoeck}}{2009}]{Herbst_van_dishoeck_2009}
{Herbst} E.,  {van Dishoeck} E.~F.,  2009, \mn@doi [\araa] {10.1146/annurev-astro-082708-101654}, \href {https://ui.adsabs.harvard.edu/abs/2009ARA&A..47..427H} {47, 427}

\bibitem[\protect\citeauthoryear{{Immer}, {Li}, {Quiroga-Nu{\~n}ez}, {Reid}, {Zhang}, {Moscadelli}  \& {Rygl}}{{Immer} et~al.}{2019}]{Immer2019}
{Immer} K.,  {Li} J.,  {Quiroga-Nu{\~n}ez} L.~H.,  {Reid} M.~J.,  {Zhang} B.,  {Moscadelli} L.,   {Rygl} K.~L.~J.,  2019, \mn@doi [\aap] {10.1051/0004-6361/201834208}, \href {https://ui.adsabs.harvard.edu/abs/2019A&A...632A.123I} {632, A123}

\bibitem[\protect\citeauthoryear{{Jacobsen} et~al.,}{{Jacobsen} et~al.}{2018}]{Jacobsen2018}
{Jacobsen} S.~K.,  et~al., 2018, \mn@doi [\aap] {10.1051/0004-6361/201731668}, \href {https://ui.adsabs.harvard.edu/abs/2018A&A...612A..72J} {612, A72}

\bibitem[\protect\citeauthoryear{{Jim{\'e}nez-Escobar} \& {Mu{\~n}oz Caro}}{{Jim{\'e}nez-Escobar} \& {Mu{\~n}oz Caro}}{2011}]{Jimenez-escobar2011}
{Jim{\'e}nez-Escobar} A.,  {Mu{\~n}oz Caro} G.~M.,  2011, \mn@doi [\aap] {10.1051/0004-6361/201014821}, \href {https://ui.adsabs.harvard.edu/abs/2011A&A...536A..91J} {536, A91}

\bibitem[\protect\citeauthoryear{{J{\o}rgensen} et~al.,}{{J{\o}rgensen} et~al.}{2016}]{Jorgensen2016}
{J{\o}rgensen} J.~K.,  et~al., 2016, \mn@doi [\aap] {10.1051/0004-6361/201628648}, \href {https://ui.adsabs.harvard.edu/abs/2016A&A...595A.117J} {595, A117}

\bibitem[\protect\citeauthoryear{{J{\o}rgensen} et~al.,}{{J{\o}rgensen} et~al.}{2018}]{Jorgensen2018}
{J{\o}rgensen} J.~K.,  et~al., 2018, \mn@doi [\aap] {10.1051/0004-6361/201731667}, \href {https://ui.adsabs.harvard.edu/abs/2018A&A...620A.170J} {620, A170}

\bibitem[\protect\citeauthoryear{Korschinek, Faestermann, Poutivtsev, Arazi, Knie, Rugel  \& Wallner}{Korschinek et~al.}{2020}]{Korschinek2020}
Korschinek G.,  Faestermann T.,  Poutivtsev M.,  Arazi A.,  Knie K.,  Rugel G.,   Wallner A.,  2020, \mn@doi [Phys. Rev. Lett.] {10.1103/PhysRevLett.125.031101}, 125, 031101

\bibitem[\protect\citeauthoryear{{Laas} \& {Caselli}}{{Laas} \& {Caselli}}{2019}]{Laas2019}
{Laas} J.~C.,  {Caselli} P.,  2019, \mn@doi [\aap] {10.1051/0004-6361/201834446}, \href {https://ui.adsabs.harvard.edu/abs/2019A&A...624A.108L} {624, A108}

\bibitem[\protect\citeauthoryear{{Lada} \& {Lada}}{{Lada} \& {Lada}}{2003}]{Lada2003}
{Lada} C.~J.,  {Lada} E.~A.,  2003, \mn@doi [\araa] {10.1146/annurev.astro.41.011802.094844}, \href {https://ui.adsabs.harvard.edu/abs/2003ARA&A..41...57L} {41, 57}

\bibitem[\protect\citeauthoryear{{Lauretta} et~al.,}{{Lauretta} et~al.}{2017}]{Lauretta2017}
{Lauretta} D.~S.,  et~al., 2017, \mn@doi [\ssr] {10.1007/s11214-017-0405-1}, \href {https://ui.adsabs.harvard.edu/abs/2017SSRv..212..925L} {212, 925}

\bibitem[\protect\citeauthoryear{{Le Roy} et~al.,}{{Le Roy} et~al.}{2015}]{LeRoy2015}
{Le Roy} L.,  et~al., 2015, \mn@doi [\aap] {10.1051/0004-6361/201526450}, \href {https://ui.adsabs.harvard.edu/abs/2015A&A...583A...1L} {583, A1}

\bibitem[\protect\citeauthoryear{{Ligterink} et~al.,}{{Ligterink} et~al.}{2017}]{Ligterink2017}
{Ligterink} N.~F.~W.,  et~al., 2017, \mn@doi [\mnras] {10.1093/mnras/stx890}, \href {https://ui.adsabs.harvard.edu/abs/2017MNRAS.469.2219L} {469, 2219}

\bibitem[\protect\citeauthoryear{{Ligterink}, {Terwisscha van Scheltinga}, {Taquet}, {J{\o}rgensen}, {Cazaux}, {van Dishoeck}  \& {Linnartz}}{{Ligterink} et~al.}{2018}]{Ligterink2018}
{Ligterink} N.~F.~W.,  {Terwisscha van Scheltinga} J.,  {Taquet} V.,  {J{\o}rgensen} J.~K.,  {Cazaux} S.,  {van Dishoeck} E.~F.,   {Linnartz} H.,  2018, \mn@doi [\mnras] {10.1093/mnras/sty2066}, \href {https://ui.adsabs.harvard.edu/abs/2018MNRAS.480.3628L} {480, 3628}

\bibitem[\protect\citeauthoryear{{Lis} \& {Goldsmith}}{{Lis} \& {Goldsmith}}{1988}]{Lis1988}
{Lis} D.~C.,  {Goldsmith} P.~F.,  1988, in {Dickman} R.~L.,  {Snell} R.~L.,   {Young} J.~S.,  eds, , Vol.~315, Molecular Clouds, Milky-Way and External Galaxies.
p.~191, \mn@doi{10.1007/3-540-50438-9_266}

\bibitem[\protect\citeauthoryear{{Lodders}}{{Lodders}}{2010}]{Lodders2010}
{Lodders} K.,  2010, in Principles and Perspectives in Cosmochemistry. p.~379 (\mn@eprint {arXiv} {1010.2746}), \mn@doi{10.1007/978-3-642-10352-0_8}

\bibitem[\protect\citeauthoryear{{Loomis}, {Cleeves}, {{\"O}berg}, {Aikawa}, {Bergner}, {Furuya}, {Guzman}  \& {Walsh}}{{Loomis} et~al.}{2018}]{Loomis2018}
{Loomis} R.~A.,  {Cleeves} L.~I.,  {{\"O}berg} K.~I.,  {Aikawa} Y.,  {Bergner} J.,  {Furuya} K.,  {Guzman} V.~V.,   {Walsh} C.,  2018, \mn@doi [\apj] {10.3847/1538-4357/aac169}, \href {https://ui.adsabs.harvard.edu/abs/2018ApJ...859..131L} {859, 131}

\bibitem[\protect\citeauthoryear{{Lykke} et~al.,}{{Lykke} et~al.}{2017}]{Lykke2017}
{Lykke} J.~M.,  et~al., 2017, \mn@doi [\aap] {10.1051/0004-6361/201629180}, \href {https://ui.adsabs.harvard.edu/abs/2017A&A...597A..53L} {597, A53}

\bibitem[\protect\citeauthoryear{{Manigand} et~al.,}{{Manigand} et~al.}{2020}]{Manigand2020}
{Manigand} S.,  et~al., 2020, \mn@doi [\aap] {10.1051/0004-6361/201936299}, \href {https://ui.adsabs.harvard.edu/abs/2020A&A...635A..48M} {635, A48}

\bibitem[\protect\citeauthoryear{{Manigand} et~al.,}{{Manigand} et~al.}{2021}]{Manigand2021}
{Manigand} S.,  et~al., 2021, \mn@doi [\aap] {10.1051/0004-6361/202038113}, \href {https://ui.adsabs.harvard.edu/abs/2021A&A...645A..53M} {645, A53}

\bibitem[\protect\citeauthoryear{{Mart{\'\i}n-Dom{\'e}nech}, {Rivilla}, {Jim{\'e}nez-Serra}, {Qu{\'e}nard}, {Testi}  \& {Mart{\'\i}n-Pintado}}{{Mart{\'\i}n-Dom{\'e}nech} et~al.}{2017}]{Martin-Domenech2017}
{Mart{\'\i}n-Dom{\'e}nech} R.,  {Rivilla} V.~M.,  {Jim{\'e}nez-Serra} I.,  {Qu{\'e}nard} D.,  {Testi} L.,   {Mart{\'\i}n-Pintado} J.,  2017, \mn@doi [\mnras] {10.1093/mnras/stx915}, \href {https://ui.adsabs.harvard.edu/abs/2017MNRAS.469.2230M} {469, 2230}

\bibitem[\protect\citeauthoryear{{Mart{\'\i}n}, {Mart{\'\i}n-Pintado}, {Blanco-S{\'a}nchez}, {Rivilla}, {Rodr{\'\i}guez-Franco}  \& {Rico-Villas}}{{Mart{\'\i}n} et~al.}{2019}]{Martin2019}
{Mart{\'\i}n} S.,  {Mart{\'\i}n-Pintado} J.,  {Blanco-S{\'a}nchez} C.,  {Rivilla} V.~M.,  {Rodr{\'\i}guez-Franco} A.,   {Rico-Villas} F.,  2019, \mn@doi [\aap] {10.1051/0004-6361/201936144}, \href {https://ui.adsabs.harvard.edu/abs/2019A&A...631A.159M} {631, A159}

\bibitem[\protect\citeauthoryear{{Mauersberger}, {Henkel}, {Langer}  \& {Chin}}{{Mauersberger} et~al.}{1996}]{Mauersberger1996}
{Mauersberger} R.,  {Henkel} C.,  {Langer} N.,   {Chin} Y.~N.,  1996, \aap, \href {https://ui.adsabs.harvard.edu/abs/1996A&A...313L...1M} {313, L1}

\bibitem[\protect\citeauthoryear{{Maureira}, {Pineda}, {Segura-Cox}, {Caselli}, {Testi}, {Lodato}, {Loinard}  \& {Hern{\'a}ndez-G{\'o}mez}}{{Maureira} et~al.}{2020}]{Maureira2020}
{Maureira} M.~J.,  {Pineda} J.~E.,  {Segura-Cox} D.~M.,  {Caselli} P.,  {Testi} L.,  {Lodato} G.,  {Loinard} L.,   {Hern{\'a}ndez-G{\'o}mez} A.,  2020, \mn@doi [\apj] {10.3847/1538-4357/ab960b}, \href {https://ui.adsabs.harvard.edu/abs/2020ApJ...897...59M} {897, 59}

\bibitem[\protect\citeauthoryear{{McGuire}}{{McGuire}}{2022}]{McGuire2022}
{McGuire} B.~A.,  2022, \mn@doi [\apjs] {10.3847/1538-4365/ac2a48}, \href {https://ui.adsabs.harvard.edu/abs/2022ApJS..259...30M} {259, 30}

\bibitem[\protect\citeauthoryear{{McGuire} et~al.,}{{McGuire} et~al.}{2017}]{McGuire2017}
{McGuire} B.~A.,  et~al., 2017, \mn@doi [\apjl] {10.3847/2041-8213/aaa0c3}, \href {https://ui.adsabs.harvard.edu/abs/2017ApJ...851L..46M} {851, L46}

\bibitem[\protect\citeauthoryear{{McMullin}, {Waters}, {Schiebel}, {Young}  \& {Golap}}{{McMullin} et~al.}{2007}]{McMullin2007}
{McMullin} J.~P.,  {Waters} B.,  {Schiebel} D.,  {Young} W.,   {Golap} K.,  2007, in {Shaw} R.~A.,  {Hill} F.,   {Bell} D.~J.,  eds,  Astronomical Society of the Pacific Conference Series Vol. 376, Astronomical Data Analysis Software and Systems XVI. p.~127

\bibitem[\protect\citeauthoryear{{Milam}, {Savage}, {Brewster}, {Ziurys}  \& {Wyckoff}}{{Milam} et~al.}{2005}]{Milam2005}
{Milam} S.~N.,  {Savage} C.,  {Brewster} M.~A.,  {Ziurys} L.~M.,   {Wyckoff} S.,  2005, \mn@doi [\apj] {10.1086/497123}, \href {https://ui.adsabs.harvard.edu/abs/2005ApJ...634.1126M} {634, 1126}

\bibitem[\protect\citeauthoryear{{Mininni} et~al.,}{{Mininni} et~al.}{2020}]{Mininni2020}
{Mininni} C.,  et~al., 2020, \mn@doi [\aap] {10.1051/0004-6361/202038966}, \href {https://ui.adsabs.harvard.edu/abs/2020A&A...644A..84M} {644, A84}

\bibitem[\protect\citeauthoryear{{Mininni} et~al.,}{{Mininni} et~al.}{2023}]{Mininni2023}
{Mininni} C.,  et~al., 2023, \mn@doi [\aap] {10.1051/0004-6361/202245277}, \href {https://ui.adsabs.harvard.edu/abs/2023A&A...677A..15M} {677, A15}

\bibitem[\protect\citeauthoryear{{M{\"u}ller}, {Thorwirth}, {Roth}  \& {Winnewisser}}{{M{\"u}ller} et~al.}{2001}]{Muller2001}
{M{\"u}ller} H.~S.~P.,  {Thorwirth} S.,  {Roth} D.~A.,   {Winnewisser} G.,  2001, \mn@doi [\aap] {10.1051/0004-6361:20010367}, \href {https://ui.adsabs.harvard.edu/abs/2001A&A...370L..49M} {370, L49}

\bibitem[\protect\citeauthoryear{{M{\"u}ller}, {Schl{\"o}der}, {Stutzki}  \& {Winnewisser}}{{M{\"u}ller} et~al.}{2005}]{Muller2005}
{M{\"u}ller} H. S.~P.,  {Schl{\"o}der} F.,  {Stutzki} J.,   {Winnewisser} G.,  2005, \mn@doi [Journal of Molecular Structure] {10.1016/j.molstruc.2005.01.027}, \href {https://ui.adsabs.harvard.edu/abs/2005JMoSt.742..215M} {742, 215}

\bibitem[\protect\citeauthoryear{{Nazari} et~al.,}{{Nazari} et~al.}{2021}]{Nazari2021}
{Nazari} P.,  et~al., 2021, \mn@doi [\aap] {10.1051/0004-6361/202039996}, \href {https://ui.adsabs.harvard.edu/abs/2021A&A...650A.150N} {650, A150}

\bibitem[\protect\citeauthoryear{{Nazari} et~al.,}{{Nazari} et~al.}{2022}]{Nazari2022}
{Nazari} P.,  et~al., 2022, \mn@doi [\aap] {10.1051/0004-6361/202243788}, \href {https://ui.adsabs.harvard.edu/abs/2022A&A...668A.109N} {668, A109}

\bibitem[\protect\citeauthoryear{{{\"O}berg}, {Boogert}, {Pontoppidan}, {van den Broek}, {van Dishoeck}, {Bottinelli}, {Blake}  \& {Evans}}{{{\"O}berg} et~al.}{2011}]{Oberg2011}
{{\"O}berg} K.~I.,  {Boogert} A.~C.~A.,  {Pontoppidan} K.~M.,  {van den Broek} S.,  {van Dishoeck} E.~F.,  {Bottinelli} S.,  {Blake} G.~A.,   {Evans} Neal~J. I.,  2011, \mn@doi [\apj] {10.1088/0004-637X/740/2/109}, \href {https://ui.adsabs.harvard.edu/abs/2011ApJ...740..109O} {740, 109}

\bibitem[\protect\citeauthoryear{{{\"O}berg}, {Guzm{\'a}n}, {Furuya}, {Qi}, {Aikawa}, {Andrews}, {Loomis}  \& {Wilner}}{{{\"O}berg} et~al.}{2015}]{Oberg2015}
{{\"O}berg} K.~I.,  {Guzm{\'a}n} V.~V.,  {Furuya} K.,  {Qi} C.,  {Aikawa} Y.,  {Andrews} S.~M.,  {Loomis} R.,   {Wilner} D.~J.,  2015, \mn@doi [\nat] {10.1038/nature14276}, \href {https://ui.adsabs.harvard.edu/abs/2015Natur.520..198O} {520, 198}

\bibitem[\protect\citeauthoryear{{Osorio}, {Anglada}, {Lizano}  \& {D'Alessio}}{{Osorio} et~al.}{2009}]{Osorio2009}
{Osorio} M.,  {Anglada} G.,  {Lizano} S.,   {D'Alessio} P.,  2009, \mn@doi [\apj] {10.1088/0004-637X/694/1/29}, \href {https://ui.adsabs.harvard.edu/abs/2009ApJ...694...29O} {694, 29}

\bibitem[\protect\citeauthoryear{{Pegues} et~al.,}{{Pegues} et~al.}{2021}]{Pegues2021}
{Pegues} J.,  et~al., 2021, \mn@doi [\apj] {10.3847/1538-4357/abe870}, \href {https://ui.adsabs.harvard.edu/abs/2021ApJ...911..150P} {911, 150}

\bibitem[\protect\citeauthoryear{{Pfalzner} \& {Vincke}}{{Pfalzner} \& {Vincke}}{2020}]{PfalznerVincke2020}
{Pfalzner} S.,  {Vincke} K.,  2020, \mn@doi [\apj] {10.3847/1538-4357/ab9533}, \href {https://ui.adsabs.harvard.edu/abs/2020ApJ...897...60P} {897, 60}

\bibitem[\protect\citeauthoryear{{Pickett}, {Poynter}, {Cohen}, {Delitsky}, {Pearson}  \& {M{\"u}ller}}{{Pickett} et~al.}{1998}]{Pinckett1998}
{Pickett} H.~M.,  {Poynter} R.~L.,  {Cohen} E.~A.,  {Delitsky} M.~L.,  {Pearson} J.~C.,   {M{\"u}ller} H.~S.~P.,  1998, \mn@doi [\jqsrt] {10.1016/S0022-4073(98)00091-0}, \href {https://ui.adsabs.harvard.edu/abs/1998JQSRT..60..883P} {60, 883}

\bibitem[\protect\citeauthoryear{{Porras}, {Christopher}, {Allen}, {Di Francesco}, {Megeath}  \& {Myers}}{{Porras} et~al.}{2003}]{Porras2003}
{Porras} A.,  {Christopher} M.,  {Allen} L.,  {Di Francesco} J.,  {Megeath} S.~T.,   {Myers} P.~C.,  2003, \mn@doi [\aj] {10.1086/377623}, \href {https://ui.adsabs.harvard.edu/abs/2003AJ....126.1916P} {126, 1916}

\bibitem[\protect\citeauthoryear{{Potiszil} et~al.,}{{Potiszil} et~al.}{2023}]{Potiszil2023}
{Potiszil} C.,  et~al., 2023, \mn@doi [Nature Communications] {10.1038/s41467-023-37107-6}, \href {https://ui.adsabs.harvard.edu/abs/2023NatCo..14.1482P} {14, 1482}

\bibitem[\protect\citeauthoryear{{Reach} et~al.,}{{Reach} et~al.}{2009}]{Reach2009}
{Reach} W.~T.,  et~al., 2009, \mn@doi [\apj] {10.1088/0004-637X/690/1/683}, \href {https://ui.adsabs.harvard.edu/abs/2009ApJ...690..683R} {690, 683}

\bibitem[\protect\citeauthoryear{{Rivilla}, {Beltr{\'a}n}, {Cesaroni}, {Fontani}, {Codella}  \& {Zhang}}{{Rivilla} et~al.}{2017}]{Rivilla2017}
{Rivilla} V.~M.,  {Beltr{\'a}n} M.~T.,  {Cesaroni} R.,  {Fontani} F.,  {Codella} C.,   {Zhang} Q.,  2017, \mn@doi [\aap] {10.1051/0004-6361/201628373}, \href {https://ui.adsabs.harvard.edu/abs/2017A&A...598A..59R} {598, A59}

\bibitem[\protect\citeauthoryear{{Rivilla} et~al.,}{{Rivilla} et~al.}{2020}]{Rivilla2020}
{Rivilla} V.~M.,  et~al., 2020, \mn@doi [\mnras] {10.1093/mnras/stz3336}, \href {https://ui.adsabs.harvard.edu/abs/2020MNRAS.492.1180R} {492, 1180}

\bibitem[\protect\citeauthoryear{{Rubin} et~al.,}{{Rubin} et~al.}{2019}]{Rubin2019}
{Rubin} M.,  et~al., 2019, \mn@doi [\mnras] {10.1093/mnras/stz2086}, \href {https://ui.adsabs.harvard.edu/abs/2019MNRAS.489..594R} {489, 594}

\bibitem[\protect\citeauthoryear{{Sabatini} et~al.,}{{Sabatini} et~al.}{2021}]{Sabatini2021}
{Sabatini} G.,  et~al., 2021, \mn@doi [\aap] {10.1051/0004-6361/202140469}, \href {https://ui.adsabs.harvard.edu/abs/2021A&A...652A..71S} {652, A71}

\bibitem[\protect\citeauthoryear{{S{\'a}nchez-Monge}, {Schilke}, {Ginsburg}, {Cesaroni}  \& {Schmiedeke}}{{S{\'a}nchez-Monge} et~al.}{2018}]{Sanchez-Monge2018}
{S{\'a}nchez-Monge} {\'A}.,  {Schilke} P.,  {Ginsburg} A.,  {Cesaroni} R.,   {Schmiedeke} A.,  2018, \mn@doi [\aap] {10.1051/0004-6361/201730425}, \href {https://ui.adsabs.harvard.edu/abs/2018A&A...609A.101S} {609, A101}

\bibitem[\protect\citeauthoryear{{Schuhmann} et~al.,}{{Schuhmann} et~al.}{2019}]{Schuhmann2019}
{Schuhmann} M.,  et~al., 2019, \mn@doi [ACS Earth and Space Chemistry] {10.1021/acsearthspacechem.9b00094}, \href {https://ui.adsabs.harvard.edu/abs/2019ESC.....3.1854S} {3, 1854}

\bibitem[\protect\citeauthoryear{{Suzuki} et~al.,}{{Suzuki} et~al.}{2023}]{Suzuki2023}
{Suzuki} T.,  et~al., 2023, \mn@doi [\apj] {10.3847/1538-4357/acdb6d}, \href {https://ui.adsabs.harvard.edu/abs/2023ApJ...954..189S} {954, 189}

\bibitem[\protect\citeauthoryear{{Tabone} et~al.,}{{Tabone} et~al.}{2023}]{Tabone2023}
{Tabone} B.,  et~al., 2023, \mn@doi [Nature Astronomy] {10.1038/s41550-023-01965-3}, \href {https://ui.adsabs.harvard.edu/abs/2023NatAs...7..805T} {7, 805}

\bibitem[\protect\citeauthoryear{{Tercero}, {Cuadrado}, {L{\'o}pez}, {Brouillet}, {Despois}  \& {Cernicharo}}{{Tercero} et~al.}{2018}]{Tercero2018}
{Tercero} B.,  {Cuadrado} S.,  {L{\'o}pez} A.,  {Brouillet} N.,  {Despois} D.,   {Cernicharo} J.,  2018, \mn@doi [\aap] {10.1051/0004-6361/201834417}, \href {https://ui.adsabs.harvard.edu/abs/2018A&A...620L...6T} {620, L6}

\bibitem[\protect\citeauthoryear{{Tobin} et~al.,}{{Tobin} et~al.}{2023}]{Tobin2023}
{Tobin} J.~J.,  et~al., 2023, \mn@doi [\nat] {10.1038/s41586-022-05676-z}, \href {https://ui.adsabs.harvard.edu/abs/2023Natur.615..227T} {615, 227}

\bibitem[\protect\citeauthoryear{{Urquhart} et~al.,}{{Urquhart} et~al.}{2019}]{Urquhart2019}
{Urquhart} J.~S.,  et~al., 2019, \mn@doi [\mnras] {10.1093/mnras/stz154}, \href {https://ui.adsabs.harvard.edu/abs/2019MNRAS.484.4444U} {484, 4444}

\bibitem[\protect\citeauthoryear{{Visser}, {van Dishoeck}  \& {Black}}{{Visser} et~al.}{2009}]{Visser2009}
{Visser} R.,  {van Dishoeck} E.~F.,   {Black} J.~H.,  2009, \mn@doi [\aap] {10.1051/0004-6361/200912129}, \href {https://ui.adsabs.harvard.edu/abs/2009A&A...503..323V} {503, 323}

\bibitem[\protect\citeauthoryear{{Viti}, {Collings}, {Dever}, {McCoustra}  \& {Williams}}{{Viti} et~al.}{2004}]{Viti2004}
{Viti} S.,  {Collings} M.~P.,  {Dever} J.~W.,  {McCoustra} M. R.~S.,   {Williams} D.~A.,  2004, \mn@doi [\mnras] {10.1111/j.1365-2966.2004.08273.x}, \href {https://ui.adsabs.harvard.edu/abs/2004MNRAS.354.1141V} {354, 1141}

\bibitem[\protect\citeauthoryear{{Wilson}}{{Wilson}}{1999}]{Wilson1999}
{Wilson} T.~L.,  1999, \mn@doi [Reports on Progress in Physics] {10.1088/0034-4885/62/2/002}, \href {https://ui.adsabs.harvard.edu/abs/1999RPPh...62..143W} {62, 143}

\bibitem[\protect\citeauthoryear{{Yan} et~al.,}{{Yan} et~al.}{2019}]{Yan2019}
{Yan} Y.~T.,  et~al., 2019, \mn@doi [\apj] {10.3847/1538-4357/ab17d6}, \href {https://ui.adsabs.harvard.edu/abs/2019ApJ...877..154Y} {877, 154}

\bibitem[\protect\citeauthoryear{{Yang} et~al.,}{{Yang} et~al.}{2021}]{Yang2021}
{Yang} Y.-L.,  et~al., 2021, \mn@doi [\apj] {10.3847/1538-4357/abdfd6}, \href {https://ui.adsabs.harvard.edu/abs/2021ApJ...910...20Y} {910, 20}

\bibitem[\protect\citeauthoryear{{Yu} et~al.,}{{Yu} et~al.}{2020}]{Yu2020}
{Yu} H.~Z.,  et~al., 2020, \mn@doi [\apj] {10.3847/1538-4357/aba8f1}, \href {https://ui.adsabs.harvard.edu/abs/2020ApJ...899..145Y} {899, 145}

\bibitem[\protect\citeauthoryear{{Zeng}, {Qu{\'e}nard}, {Jim{\'e}nez-Serra}, {Mart{\'\i}n-Pintado}, {Rivilla}, {Testi}  \& {Mart{\'\i}n-Dom{\'e}nech}}{{Zeng} et~al.}{2019}]{Zeng2019}
{Zeng} S.,  {Qu{\'e}nard} D.,  {Jim{\'e}nez-Serra} I.,  {Mart{\'\i}n-Pintado} J.,  {Rivilla} V.~M.,  {Testi} L.,   {Mart{\'\i}n-Dom{\'e}nech} R.,  2019, \mn@doi [\mnras] {10.1093/mnrasl/slz002}, \href {https://ui.adsabs.harvard.edu/abs/2019MNRAS.484L..43Z} {484, L43}

\bibitem[\protect\citeauthoryear{{van Gelder} et~al.,}{{van Gelder} et~al.}{2020}]{vanGelder2020}
{van Gelder} M.~L.,  et~al., 2020, \mn@doi [\aap] {10.1051/0004-6361/202037758}, \href {https://ui.adsabs.harvard.edu/abs/2020A&A...639A..87V} {639, A87}

\bibitem[\protect\citeauthoryear{{van der Marel}, {Booth}, {Leemker}, {van Dishoeck}  \& {Ohashi}}{{van der Marel} et~al.}{2021}]{vanderMarel2021}
{van der Marel} N.,  {Booth} A.~S.,  {Leemker} M.,  {van Dishoeck} E.~F.,   {Ohashi} S.,  2021, \mn@doi [\aap] {10.1051/0004-6361/202141051}, \href {https://ui.adsabs.harvard.edu/abs/2021A&A...651L...5V} {651, L5}

\makeatother
\end{thebibliography}




\appendix

\onecolumn

\section{SLIM fits of other isotopologues}

\setlength{\tabcolsep}{5pt}

\begin{longtable}{cccccccccc}
\caption{\label{tab:Fits_isotopologues_molecules_on_G31} Results of the SLIM fits of the isotopologues of the molecules analysed towards G31.41 that were not used to derive the abundances of the main isotopologues, ordered by increasing molecular mass. The resulting physical parameters, along with their associated uncertainties, are presented. The values of the parameters fixed to perform some fits are shown without uncertainties. In the case of non detections, we indicated the upper limit in the column density with <. When $T_{\mathrm{ex}}$ is fixed, we calculate in what factor the value of $N$ changes when using the different temperatures $T_{\mathrm{min}}$ and $T_{\mathrm{max}}$: 25-75 K and 100-200 K for fixed temperatures of 50 K and 150 K, respectively.
We also present the maximum line opacity of the transitions of each molecule ($\tau_{max}$). The transitions used to obtain these values are in Table \ref{tab:Transitions_Molecules_on_G31}. 
}
\\

    \hline
     Formula &  $\textit{T}_{\textrm{ex}}$ & \textit{N} & \textit{N(T}$_{\textrm{min}}$)/\textit{N-} & $v-v_{\rm 0}$ & FWHM & $\tau_{max}^a$ & Fig. \\ 
     & (K) & ($\times$10$^{16}$ cm$^{-2}$) & \textit{N(T}$_{\textrm{max}}$)/\textit{N} & (km s$^{-1})$ & (km s$^{-1})$ & & \\ \hline \endfirsthead

     \caption{Continued.}\\
     \hline
     Formula & $\textit{T}_{\textrm{ex}}$ & \textit{N} & \textit{N(T}$_{\textrm{min}}$)/\textit{N-} & $v-v_{\rm 0}$ & FWHM & $\tau_{max}^a$ & Fig. \\ 
     & (K) & ($\times$10$^{16}$ cm$^{-2}$) & \textit{N(T}$_{\textrm{max}}$)/\textit{N} & (km s$^{-1})$ & (km s$^{-1})$ & & \\ \hline \endhead   
     
     \hline
     \endfoot

     \hline \noalign{\smallskip}
    \multicolumn{10}{p{11.5cm}}{\textbf{Notes.} $^{a}$ Maximum $\tau$ of all the transitions detected in each species.}
     \endlastfoot

    \multicolumn{9}{c}{Detected molecules}  \tabularnewline \hline
    C$^{17}$O &  50 & <5 & 0.7-1.6 &  0 & 7 & 0.02 &  \ref{fig:CO}\\
    C$^{18}$O  & 50 & 14 $\pm$ 5 & 0.7-1.4 & 0 & 7 & 0.15 $\pm$ 0.10 & \ref{fig:CO}\\ \hline
    H$_{2}$CO & 50 & 62 $\pm$ 22 & 3.8-0.6 & 0.8 $\pm$ 0.4 & 7.4 $\pm$ 1.0 & 4 $\pm$ 14 &  \ref{fig:H2CO}\\
    H$_{2}^{13}$CO & 50 & 3.0 $\pm$ 0.4 & 2.2-1.0 & 1.4 $\pm$ 0.4 & 6.3 $\pm$ 0.9 & 0.24 $\pm$ 0.05 &  \ref{fig:H2CO}\\
    H$_{2}$C$^{17}$O & 50 & 4.9 $\pm$ 1.3 & 2.8-1.0 & 0 & 7 & 0.4 $\pm$ 0.3 & \ref{fig:H2CO}\\ \hline
    $^{13}$CS & 50 & 1.21 $\pm$ 0.14 & 1.0-1.0 & 1.2 $\pm$ 0.3 & 7.7 $\pm$ 0.5 & 2.5 $\pm$ 1.0 &  \ref{fig:CS}\\
    $^{13}$C$^{34}$S & 50 & 0.23 $\pm$ 0.02 & 0.7-1.3 & -2.0 $\pm$ 0.5 & 11.3 $\pm$ 1.2 & 0.32 $\pm$ 0.06 & \ref{fig:CS}\\
\hline
    
\end{longtable}

\section{Sample of molecules used for comparisons between sources}

The complete list of molecules used in this work (detections, tentative detections and upper limits) is given in Table \ref{tab:Molecules_used_each_source} for the four sources used in this work (G31.41, IRAS16B, 67P/C-G and 46P/W). The species are ordered by increasing molecular mass.

\begin{longtable}{c|c|c|c|c|c|c}
\caption{\label{tab:Molecules_used_each_source} Molecules used in this work. Molecules with a [\checkmark] implies that they are detected; [$\sim$ \checkmark] indicates a tentative detection; and [ < ] denotes non detections. The molecules indicated with * in the 67P/C-G column are members of isomeric groups}. \\
\hline
     Formula & Name & Mol. mass & G31.41 & IRAS16B & 67P/C-G & 46P/W \\ \hline 
     \endfirsthead
    \caption{Continued.}\\
\hline
     Formula & Name & Mol. mass & G31.41 & IRAS16B & 67P/C-G & 46P/W \\  
   \endhead
    \endfoot
    \hline \noalign{\smallskip}
    \multicolumn{7}{p{13cm}}{\textbf{References.} a) \cite{Cesaroni1994}; b) \cite{Mininni2020}; c) \cite{Colzi2021}; d) \cite{Mininni2023}; e) This work; f) \cite{Drozdovskaya2019} and references therein; g) \cite{Garcia-de-la-concepcion-2022}; h) \cite{Calcutt2018}; i) \cite{Calcutt2019}; j) \cite{Coutens2018}; k) \cite{Coutens2019}; l) \cite{Manigand2021}; m) \cite{Martin-Domenech2017}; n) \cite{Zeng2019}; o) \cite{Ligterink2018}; p) \cite{Biver2021}.}
     \endlastfoot
    \hline
        \ch{NH3} & Ammonia & 17 & \checkmark$^a$ & <$^f$ & \checkmark$^f$ & - \\ \hline
        \ch{H2O} & Water & 18 & - & \checkmark$^f$ & \checkmark$^f$ & - \\ \hline
        HCN & Hydrogen cyanide & 27 & \checkmark$^e$ & \checkmark$^f$ & \checkmark*$^f$ & \checkmark$^p$ \\ \hline
        HNC & Hydrogen isocyanide & 27 & <$^e$ & <$^f$ & \checkmark*$^f$ & \checkmark$^p$ \\ \hline
        CO & Carbon monoxide & 28 & \checkmark$^e$ & \checkmark$^f$ & \checkmark$^f$ & <$^p$ \\ \hline
        H$_{2}$CO & Formaldehyde & 30 & \checkmark$^e$ & \checkmark$^f$ & \checkmark$^f$ & \checkmark$^p$ \\ \hline
        \ch{O2} & Molecular oxygen & 32 & - & <$^f$ & \checkmark$^f$ & - \\ \hline
        \ch{CH3OH} & Methanol & 32 & \checkmark$^d$ & \checkmark$^f$ & \checkmark$^f$ & \checkmark$^p$ \\
\hline
    
\end{longtable}

\section{Spectroscopic catalog entries}

Table \ref{tab:Entries_CDMS_JPL_molecules_G31} lists the spectroscopic catalog entries of all the molecules analysed in this work towards G31.41. The species are ordered by increasing molecular mass.

\setlength{\tabcolsep}{2pt}

\begin{longtable}{ccccccc}
\caption{Spectroscopic entries used to perform the analysis, including the molecular catalog, tag of the species, version number, and date of the version.}
\label{tab:Entries_CDMS_JPL_molecules_G31}\\
\hline
     Formula & Name & Mol. mass & Catalogue & Tag & Version & Date version\\ 
     \endfirsthead
    \caption{Continued.}\\
\hline
     Formula & Name & Mol. mass & Catalog & TAG & Version & Date version\\
     \hline 
   \endhead
    \endfoot
    \hline
        \multicolumn{7}{c}{Detected molecules}  \tabularnewline \hline
        HCN & Hydrogen cyanide & 27 & CDMS & 27501 & 4 & May 2007\\
        HC$^{15}$N & ~ & ~ & CDMS & 28506 & 2 & Dec. 2017\\ \hline 
        C$^{18}$O & ~ & ~ & CDMS & 30502 & 1 & June 2001\\ 
        $^{13}$C$^{18}$O & ~ & ~ & CDMS & 31502 & 1 & July 2000\\ \hline
        H$_{2}$CO & Formaldehyde & 30 & CDMS & 30501 & 3 & Feb. 2017\\
        H$_{2}^{13}$CO & ~ & ~ & CDMS & 31503 & 1 & May 2000\\ 
        H$_{2}$C$^{17}$O & ~ & ~ & CDMS & 31514 & 1 & Feb. 2017\\
\end{longtable}

\section{Molecular transitions used in the analysis}

The transitions used for the MADCUBA fits of the molecular emission of the analysed species in this work towards G31.41+0.31 (see Table \ref{tab:Poperties_molecules_on_G31} and Table \ref{tab:Fits_isotopologues_molecules_on_G31}) are listed in Table \ref{tab:Transitions_Molecules_on_G31}.

\begin{longtable}{cccccccc}

\caption{List of the transitions of the molecules analysed in this work that were used to perform the MADCUBA fits to obtain their physical parameters (see Table \ref{tab:Poperties_molecules_on_G31} and Table \ref{tab:Fits_isotopologues_molecules_on_G31}). Frequency is the rest frequency of the transitions in GHz; Quantum numbers indicates the quantum number used to label the rotational energy levels; log $I$ is the logarithm of the line intensity on nm$^{2}$ MHz; $E\mathrm{_{up}}$ is the energy of the state of higher energy in K; Area is the integrated intensity of the line in K km s$^{-1}$ and $\tau$ is the optical depth of the transition. In case of upper limits, the transitions listed are only those used to retrieve the upper limit (see Fig \ref{fig:upper-limits}). The quantum numbers used are: $J$ which indicates the total angular momentum (fine structure effects imply the use of both $N$ and $J$), $K_\mathrm{a}$ and $K_\mathrm{c}$ are the projections of $J$ onto the $a$ and $c$ inertial axis (if the molecule is symmetric top only $K$ is needed), $v$ specifies the vibrational states, $v_\mathrm{t}$ is the torsional state and $F$ designates the spin quanta. The $A$ and $E$ labels refers to the A- and E-symmetry states, respectively, if present. We label as $0^+$ and $0^-$ the symmetric and antisymmetric tunneling states, respectively, if present. The molecules are in the ground state by default in this table unless specified.}

\label{tab:Transitions_Molecules_on_G31}\\
\hline
     Molecule & Frequency & Quantum & Transition & log $I$ & $E\mathrm{_{up}}$ & Area & $\tau$ \\ 
     & (GHz) & numbers &  &(nm$^{2}$ MHz) & (K) & (K km s$^{-1}$) & \\ \hline 
     \endfirsthead
    \caption{Continued.}\\
\hline
     Molecule & Frequency & Quantum & Transition & log $I$ & $E\mathrm{_{up}}$ & Area & $\tau$ \\ 
     & (GHz) & numbers & & (nm$^{2}$ MHz) & (K) & (K km s$^{-1}$) & \\ \hline 
   \endhead
    \endfoot

\multicolumn{8}{c}{Detected molecules}  \tabularnewline \hline
HC$^{15}$N & $86.05497$ & $J$ & $1_{} \rightarrow 0_{}$ & $-2.5525$ & $4.1$ & $478 \pm 6$ & $1.7 \pm 0.2$ \tabularnewline 
\hline C$^{18}$O & $109.78217$ & $J$ & $1_{} \rightarrow 0_{}$ & $-5.0708$ & $5.3$ & $49 \pm 9$ & $0.15 \pm 0.10$ \tabularnewline 
\hline $^{13}$C$^{18}$O & $104.71140$ & $J_\mathrm{F+1/2}$ & $1_{1} \rightarrow 0_{1}$ & $-5.6079$ & $5.0$ & $29 \pm 12$ & $0.07 \pm 0.11$ \tabularnewline 
$^{13}$C$^{18}$O & $104.71140$ & & $1_{2} \rightarrow 0_{1}$ & $-5.3069$ & $5.0$ & $56 \pm 12$ & $0.13 \pm 0.11$ \tabularnewline 
\hline H$_2$CO & $89.56506$ & $J_\mathrm{{K_a, K_c}}$ & $13_{2,11} \rightarrow 13_{2,12}$ & $-4.7850$ & $370$ & $3 \pm 18$ & $(8.0 \pm 162.7)\cdot10^{-3}$ \tabularnewline 
H$_2$CO & $101.33299$ & & $6_{1,5} \rightarrow 6_{1,6}$ & $-4.0441$ & $88$ & $596 \pm 20$ & $4 \pm 14$ \tabularnewline  
\hline H$_2^{13}$CO & $96.37575$ & $J_\mathrm{{K_a, K_c}}$ & $6_{1,5} \rightarrow 6_{1,6}$ & $-4.0960$ & $86$ & $71 \pm 4$ & $0.24 \pm 0.05$ \tabularnewline
\hline  
 
\end{longtable}

\section{Statistical tests}
\label{sec:statistical_tests}

We have used three complementary correlation tests, Spearman ($r_{\mathrm{s}}$), Kendall ($\tau$) and Theil-Sen (T-S), to perform a quantitative analysis of the comparison of the chemical content in the different interstellar sources. We explain here the basics of these correlation tests, and how to interpret their results.

\subsection{Spearman}
Spearman correlation test ($r_{\mathrm{s}}$) is a non-parametric test or a range correlation test. This test orders the $x_i$ and $y_i$ data from lower to higher separately. Once the data of $x_i$ and $y_i$ is reordered, each data are replaced by its range, i.e. an integer number starting by 1 for $x_i$ and $y_i$ following the order. If two or more data have the same value ($x_i$ or $y_i$), the number assigned is the average number of the ranges. The formula which calculates the Spearman correlation coefficient depends on the range of both data, in this case the range of $x_i$ and $y_i$ is $R_i$ and $S_i$ respectively. Therefore, this is the formula:
\begin{equation}
    r_{\mathrm{s}} = \frac{\sum^n_{i=1}(R_i-\overline{R})(S_i-\overline{S})}{\sqrt{\sum^n_{i=1}(R_i-\overline{R})^2} \: \sqrt{\sum^n_{i=1}(S_i-\overline{S})^2}}
\end{equation}
The range of this correlation test is from -1 to 1. -1 is perfect anti-correlation, 0 no correlation at all and 1 perfect correlation.

\subsection{Kendall}
Kendall correlation test ($\tau$) is based on Kendall correlation coefficient  which is a non-parametrical correlation test.
In this case, we calculate by pairs of $n$ data ($x_i$, $y_i$). We have the $n$ data ($x$, $y$) with all the possible pair of points ($x_i$, $y_i$) - ($x_i$, $y_i$) having $n(n-1)/2$ different pairs, which are classified in two different categories. Any pair of data ($x_i$, $y_i$) and ($x_j$, $y_j$) where $i \neq j$ are concordant if $x_i > x_j$ and $y_i > y_j$ or $x_i < x_j$ and $y_i < y_j$. Otherwise, the pair of data are discordant. 

To calculate the value of this coefficient the equations is the following:
\begin{equation}
    \tau = \frac{2(n_\mathrm{C}-n_\mathrm{D})}{n(n-1)}    
\end{equation}
Where $n_\mathrm{C}$ is the concordant pair, $n_\mathrm{D}$ is the discordant one and $n$ is the number of the data ($x$, $y$). This coefficient has the same range and interpretation as Spearman correlation coefficient.

\subsection{Theil-Sen}
The Theil-Sen estimator (T-S) is an alternative for the conventional linear regression (i.e. Pearson correlation coefficient). This method provides us with a non-parametric linear regression which is very robust. It is very effective against outliers. The idea is to calculate the slope ($b_{ij}$) for each possible pair of points, using the following formula: 
\begin{equation}
    b_{ij} = \frac{y_j-y_i}{x_j-x_i}
\end{equation}
The final estimation for the slope ($b_{\mathrm{T-S}}$) of the linear regression is the median of all the slopes. The intercept point is calculated by doing the median of all the intercept points of the data.

Once we have the slope, we use formula which associates the slope of a linear regression with a correlation coefficient (this is used to obtain Pearson correlation coefficient), Theil-Sen in this case. The formula is the following:
\begin{equation}
    \mathrm{T\!-\!S} = \sqrt{b_{\mathrm{T-S}}(y,x) \times b_{\mathrm{T-S}}(x,y)}
\end{equation}

By doing so, we obtain a correlation coefficient from a linear regression, but more robust and less dependent on outliers. This correlation coefficient has the same range and interpretation as Spearman and Kendall. The difference is that this coefficient is evaluating if the points are close to a straight line, and the other coefficients evaluates is the points are monotonically increasing or decreasing.

To calculate the $p$-value of Theil-Sen, we use the $p$-value of Pearson correlation coefficient, due to this estimator is a modified linear regression (Pearson), so for $p$-value we used the t distribution:
\begin{equation}
\label{dens_prob}
    t = \frac{r\sqrt{n-2}}{\sqrt{1-r^2}}
\end{equation}
where $r$ is the coefficient value and $n$ is the number of observations. 
To calculate the $p$-value the formula is 2 x $P(T>t)$ where $T$ follows a t distribution with $n-2$ degrees of freedom. $P$ is the probability of a certain value of being higher than $t$, $P$ follows the density distribution of a $t$ distribution. We integrated the probability density from $t$ to infinite to obtain the $P$, the expression used is the following:
\begin{equation}
\label{int_prob}
    P_{x>t} = \int_t^\infty \frac{\Gamma (\frac{n+1}{2})}{\sqrt{n\pi}\Gamma  (\frac{n}{2})} \left(1 + \frac{x^2}{n}\right) ^{-\frac{n+1}{2}}dx
\end{equation}

Where $n$ is the degrees of freedom, $t$ is the t distribution described in equation \ref{dens_prob}, and $x$ is $T$. 

To obtain the $p$-value we solve the equation \ref{int_prob} and we multiply the result by 2 because the $p$-value is from two tails of the Gaussian.

\section{REMAINING SPECTRA OF DETECTED MOLECULES}
We show here the spectra of the detected molecules towards G31.41 analyzed in this work that have not been presented in the main text. The results of the parameters obtained from the fits are summarized in Table \ref{tab:Poperties_molecules_on_G31} and Table \ref{tab:Fits_isotopologues_molecules_on_G31}, and the information about the transitions used are listed in Table \ref{tab:Transitions_Molecules_on_G31}.

\begin{figure*}
\begin{subfigure}{0.24\textwidth}
\centering
\includegraphics[scale=0.5]{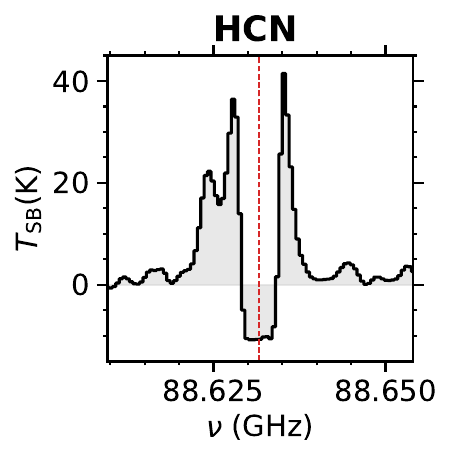}
\end{subfigure}
\begin{subfigure}{0.24\textwidth}
\centering
\includegraphics[scale=0.5]{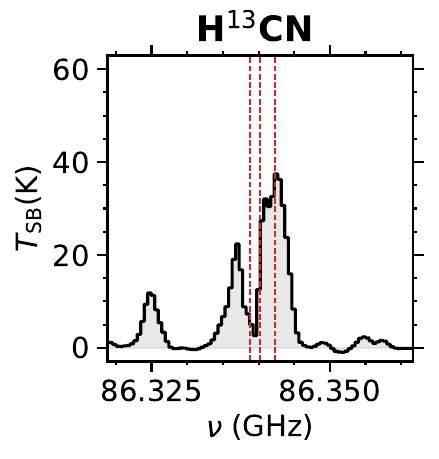}
\end{subfigure}
\begin{subfigure}{0.24\textwidth}
\centering
\includegraphics[scale=0.5]{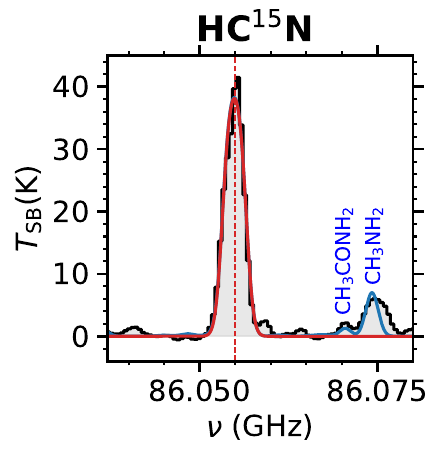}
\end{subfigure}

\begin{subfigure}{0.24\textwidth}
\centering
\includegraphics[scale=0.5]{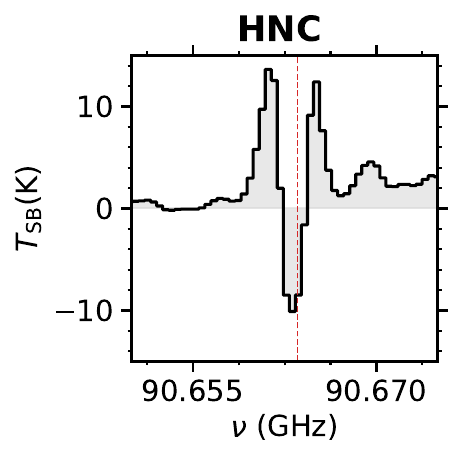}
\end{subfigure}
\begin{subfigure}{0.24\textwidth}
\centering
\includegraphics[scale=0.5]{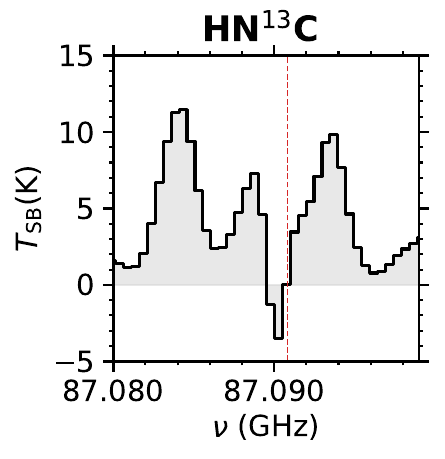}
\end{subfigure}
\begin{subfigure}{0.24\textwidth}
\centering
\includegraphics[scale=0.5]{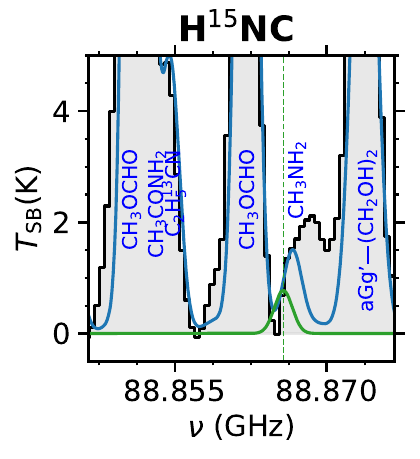}
\end{subfigure}
\caption{HCN and HNC isotopologues described in Sect. \ref{mols:HCN} and \ref{mols:HNC}. The black histogram and its grey shadow are the observational spectrum. The red curve is the fit of the individual transition, the green curve is the upper limit obtained using that transition and the blue curve is the cumulative fit considering all detected species. The red/green dashed lines indicate the frequency of the molecular transitions.}
\label{fig:HCN_HNC}
\end{figure*}

\begin{figure*}
     \centering
      \begin{subfigure}{0.24\textwidth}
          \centering
     \includegraphics[scale=0.5]{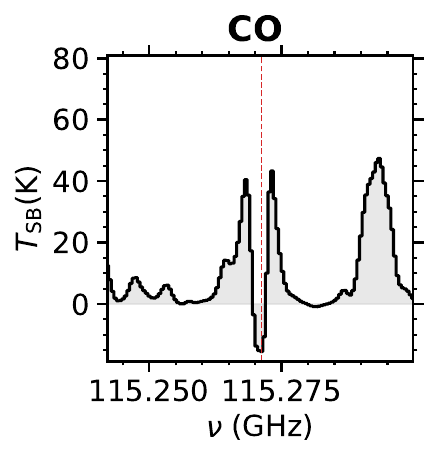}
     \end{subfigure}
   \begin{subfigure}{0.24\textwidth}
    \centering
    \includegraphics[scale=0.5]{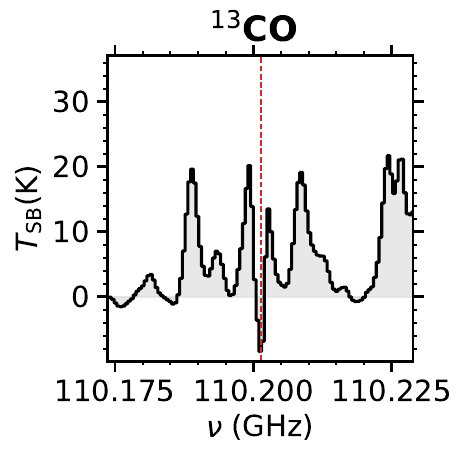}
     \end{subfigure}
     \begin{subfigure}{0.24\textwidth}
            \centering
         \includegraphics[scale=0.5]{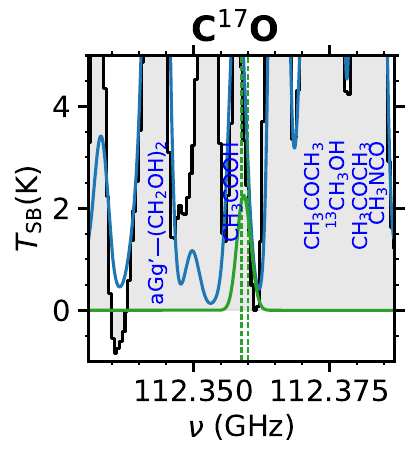}
     \end{subfigure}
     \begin{subfigure}{0.24\textwidth}
         \centering
         \includegraphics[scale=0.5]{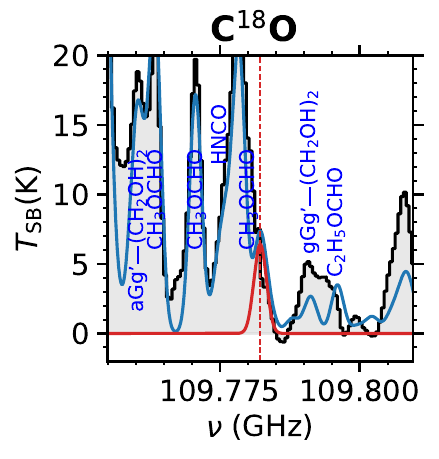}
     \end{subfigure}
     \begin{subfigure}{0.24\textwidth}
         \centering
         \includegraphics[scale=0.5]{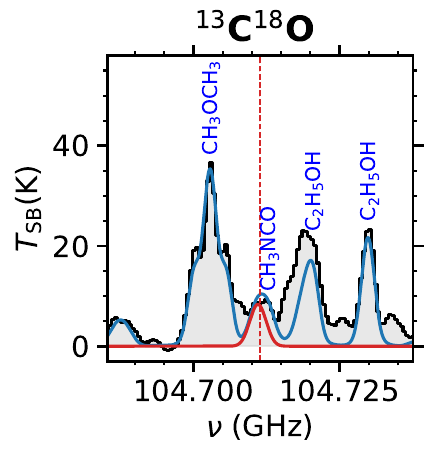}
     \end{subfigure}
        \caption{CO isotopolgues described in Sect. \ref{mols:CO}. The black histogram and its grey shadow are the observational spectrum. The red curve is the fit of the individual transition, the green curve is the upper limit obtained using that transition and the blue curve is the cumulative fit considering all detected species. The red/green dashed lines indicate the frequency of the molecular transitions.}
        \label{fig:CO}
\end{figure*}

\begin{figure*}
     \centering
     \begin{subfigure}{0.48\textwidth}
              \centering
        \includegraphics[scale=0.5]{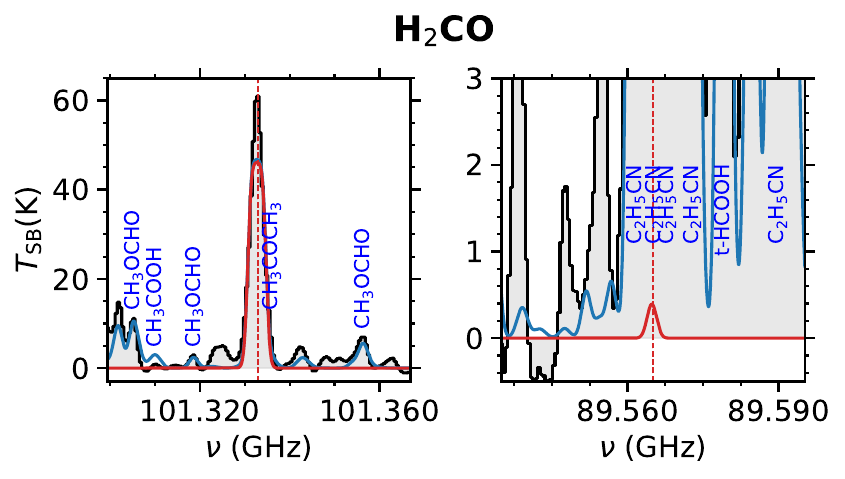}
     \end{subfigure}
          \hspace{-1.5cm}
     \begin{subfigure}{0.48\textwidth}
         \centering
        \includegraphics[scale=0.5]{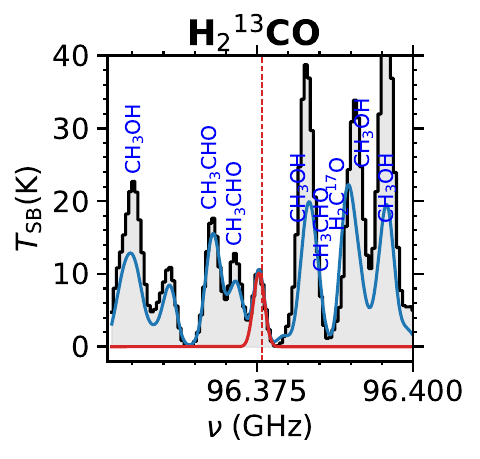}
     \end{subfigure}
     
     \begin{subfigure}{0.48\textwidth}
         \centering
    \includegraphics[scale=0.5]{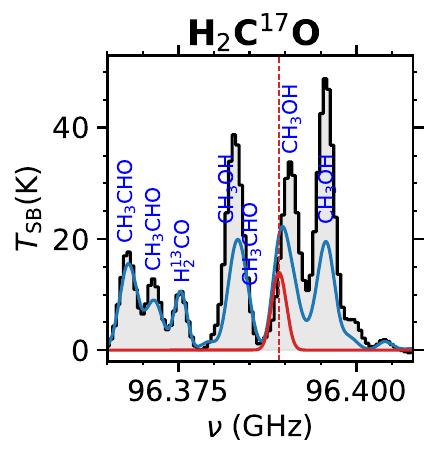}
     \end{subfigure}
          \hspace{-2.5cm}
      \begin{subfigure}{0.48\textwidth}
     \centering
    \includegraphics[scale=0.5]{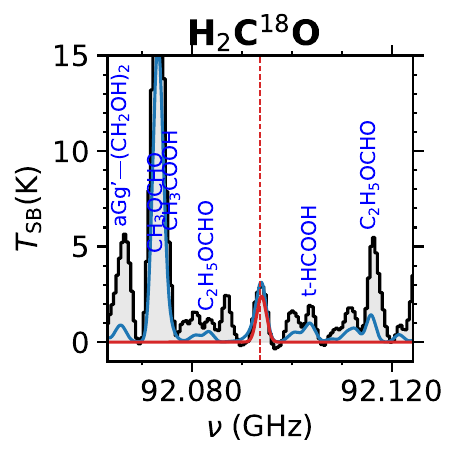}
     \end{subfigure}
        \caption{\ch{H2CO} isotopolgues described in Sect. \ref{mols:H2CO}. The black histogram and its grey shadow are the observational spectrum. The red curve is the fit of the individual transition and the blue curve is the cumulative fit considering all detected species. The red dashed lines indicate the frequency of the molecular transitions. The plots are sorted by decreasing intensity of the corresponding species (red line).}
        \label{fig:H2CO}
\end{figure*}

\begin{figure*}
\includegraphics[scale=0.5]{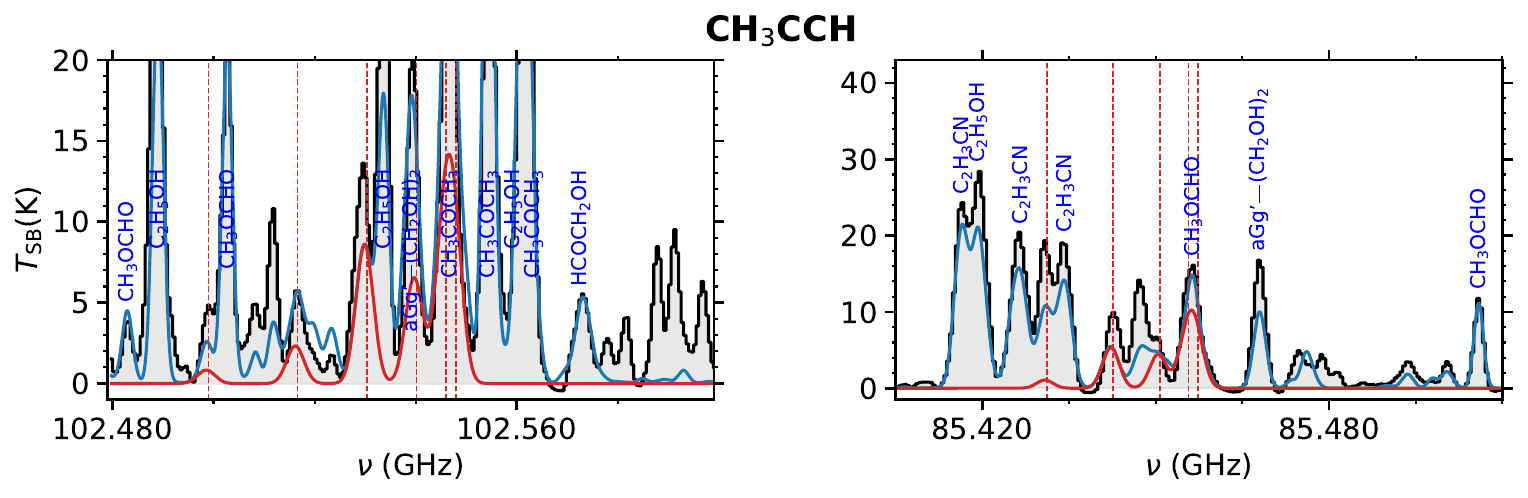}
\caption{\ch{CH3CCH} described in Sect. \ref{mols:CH3CCH}. The black histogram and its grey shadow are the observational spectrum. The red curve is the fit of the individual transition and the blue curve is the cumulative fit considering all detected species. The red dashed lines indicate the frequency of the molecular transitions. The plots are sorted by decreasing intensity of the corresponding species (red line).}
\label{fig:CH3CCH}
\end{figure*}

\begin{figure*}
\includegraphics[scale=0.5]{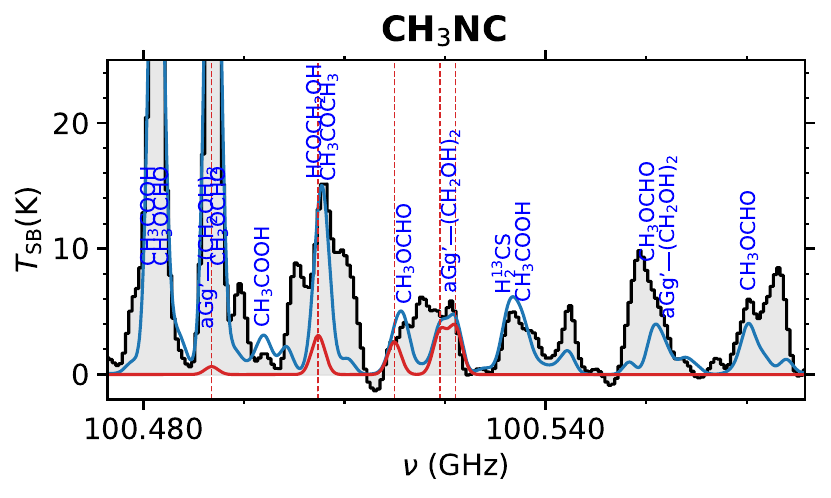}
\caption{\ch{CH3NC} described in Sect. \ref{mols:CH3NC}. The black histogram and its grey shadow are the observational spectrum. The red curve is the fit of the individual transition and the blue curve is the cumulative fit considering all detected species. The red dashed lines indicate the frequency of the molecular transitions.}
\label{fig:CH3NC}
\end{figure*}

\begin{figure*}
     \centering
     \begin{subfigure}{\textwidth}
              \centering
         \includegraphics[scale=0.5]{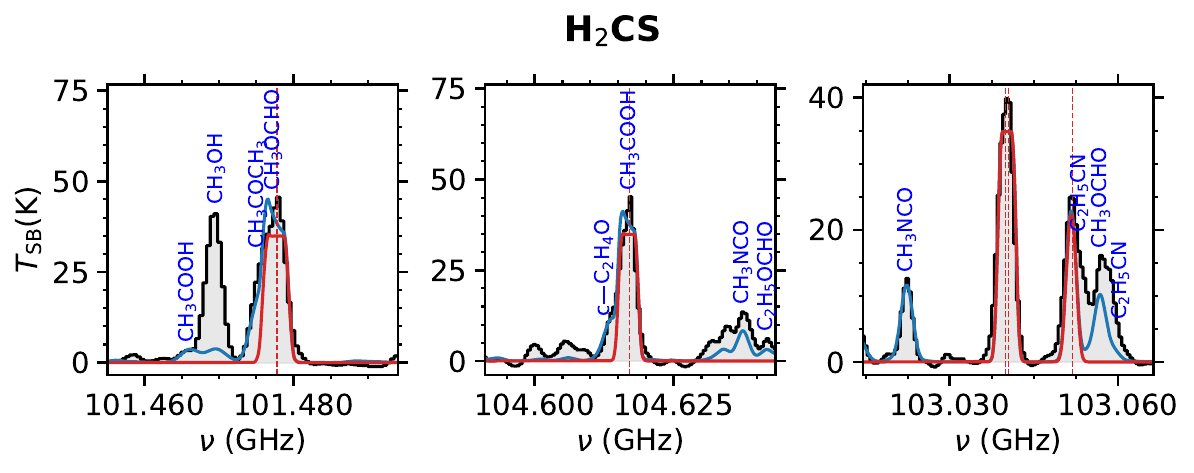}
     \end{subfigure}

     \begin{subfigure}{\textwidth}
         \centering
        \includegraphics[scale=0.5]{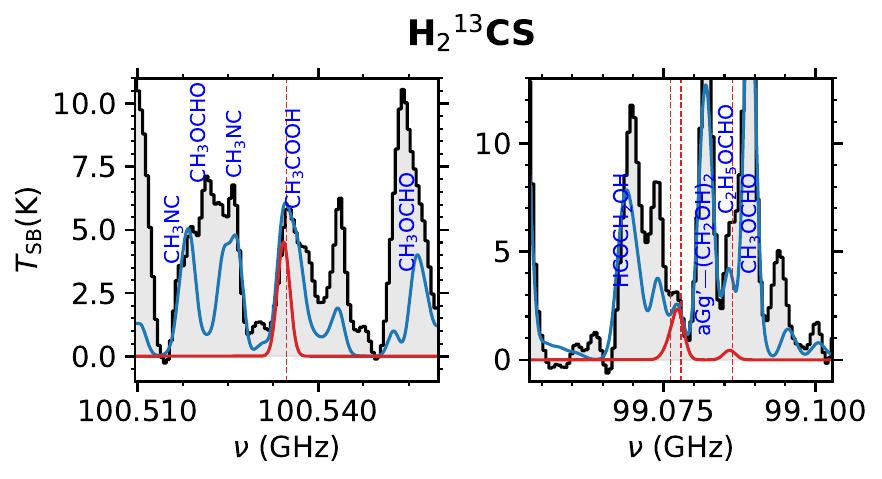}
     \end{subfigure}

     \begin{subfigure}{\textwidth}
         \centering
\includegraphics[scale=0.5]{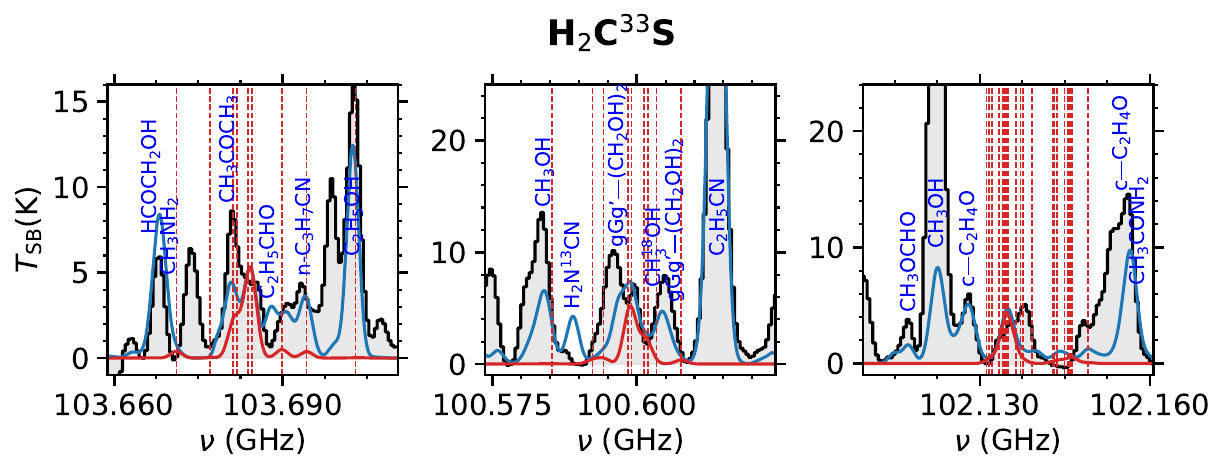}
     \end{subfigure}

          \begin{subfigure}{\textwidth}
         \centering
\includegraphics[scale=0.5]{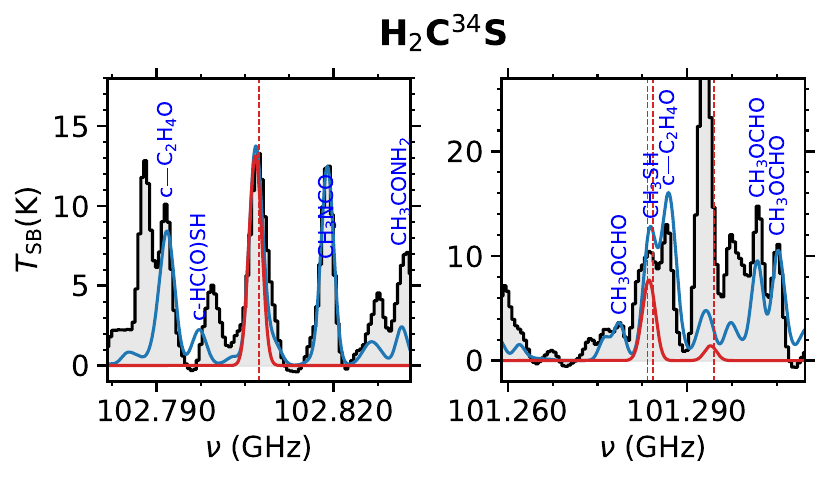}
     \end{subfigure}
        \caption{\ch{H2}CS isotopologues described in Sect. \ref{mols:H2CS}. The black histogram and its grey shadow are the observational spectrum. The red curve is the fit of the individual transition and the blue curve is the cumulative fit considering all detected species. The red dashed lines indicate the frequency of the molecular transitions. The plots are sorted by decreasing intensity of the corresponding species (red line).}
        \label{fig:H2CS}
\end{figure*}

 \begin{figure*}
\includegraphics[scale=0.5]{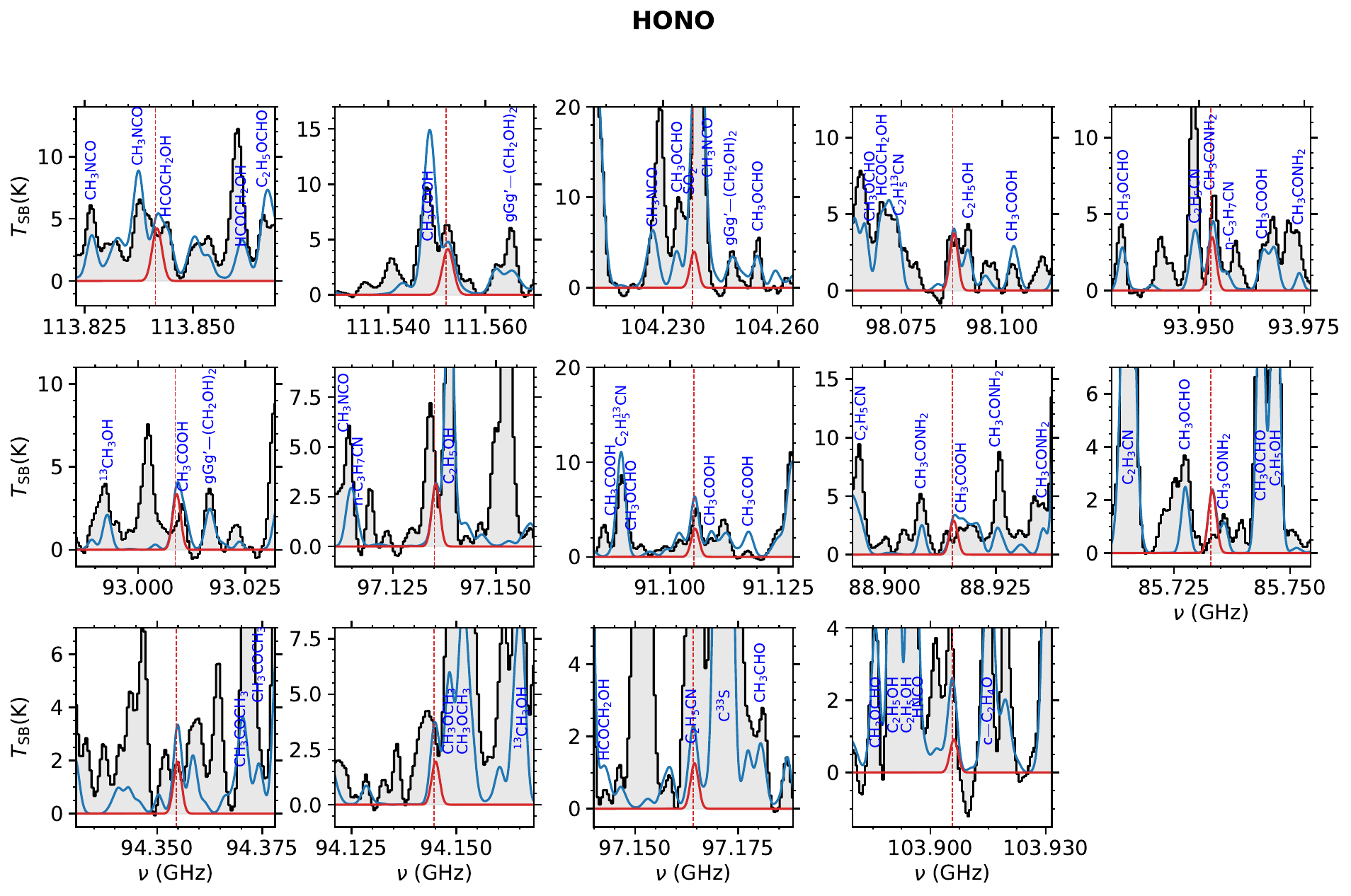}
\caption{\ch{HONO} described in Sect. \ref{mols:HONO}. The black histogram and its grey shadow are the observational spectrum. The red curve is the fit of the individual transition and the blue curve is the cumulative fit considering all detected species. The red dashed lines indicate the frequency of the molecular transitions. The plots are sorted by decreasing intensity of the corresponding species (red line).}
        \label{fig:HONO}
\end{figure*}

\begin{figure*}
     \centering
     \begin{subfigure}{\textwidth}
              \centering
         \includegraphics[scale=0.5]{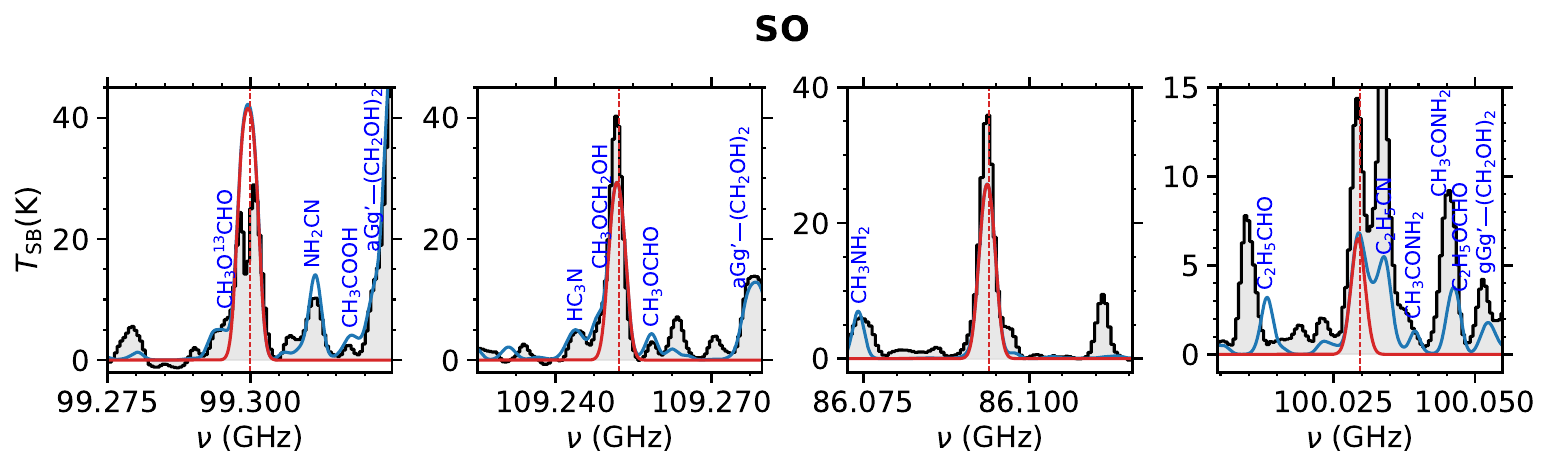}
     \end{subfigure}

     \begin{subfigure}{\textwidth}
         \centering
        \includegraphics[scale=0.5]{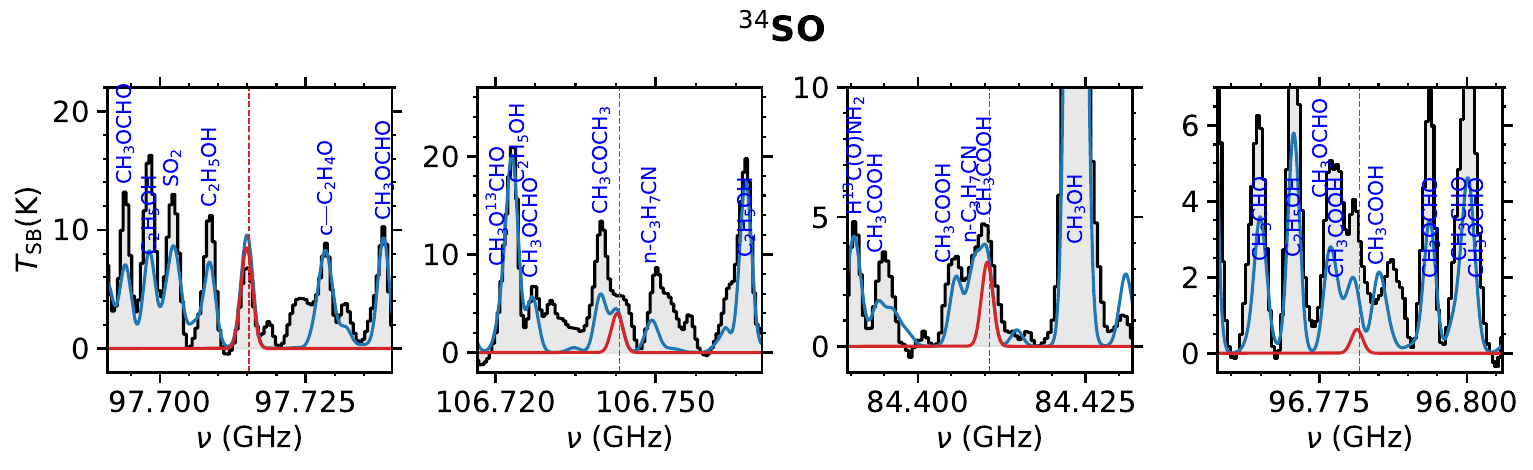}
     \end{subfigure}
             \caption{SO isotopologue described in Sect. \ref{mols:SO}. The black histogram and its grey shadow are the observational spectrum. The red curve is the fit of the individual transition and the blue curve is the cumulative fit considering all detected species. The red dashed lines indicate the frequency of the molecular transitions. The plots are sorted by decreasing intensity of the corresponding species (red line).}
                     \label{fig:SO}
\end{figure*}

\begin{figure*}
     \centering
     \begin{subfigure}{\textwidth}
              \centering
         \includegraphics[scale=0.5]{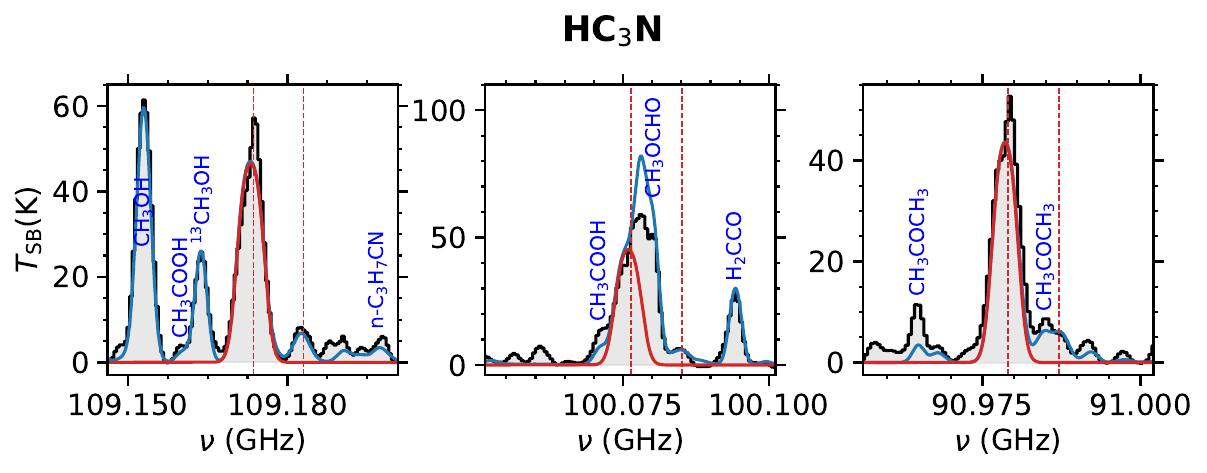}
     \end{subfigure}
     \centering
     \begin{subfigure}{\textwidth}
              \centering
         \includegraphics[scale=0.5]{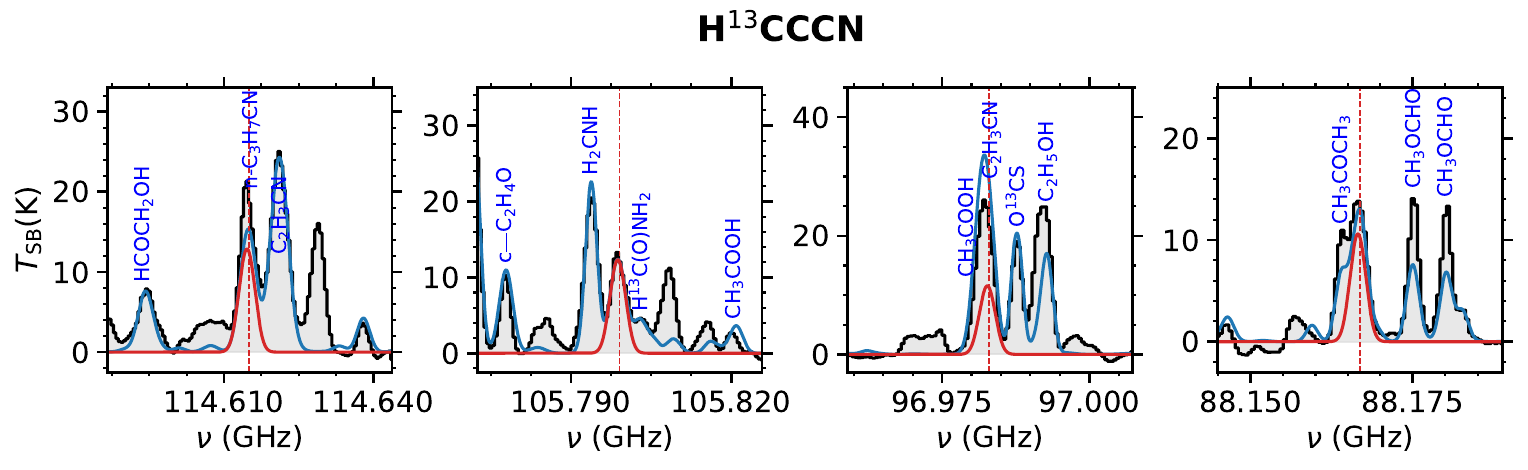}
     \end{subfigure}

     \begin{subfigure}{\textwidth}
         \centering
        \includegraphics[scale=0.5]{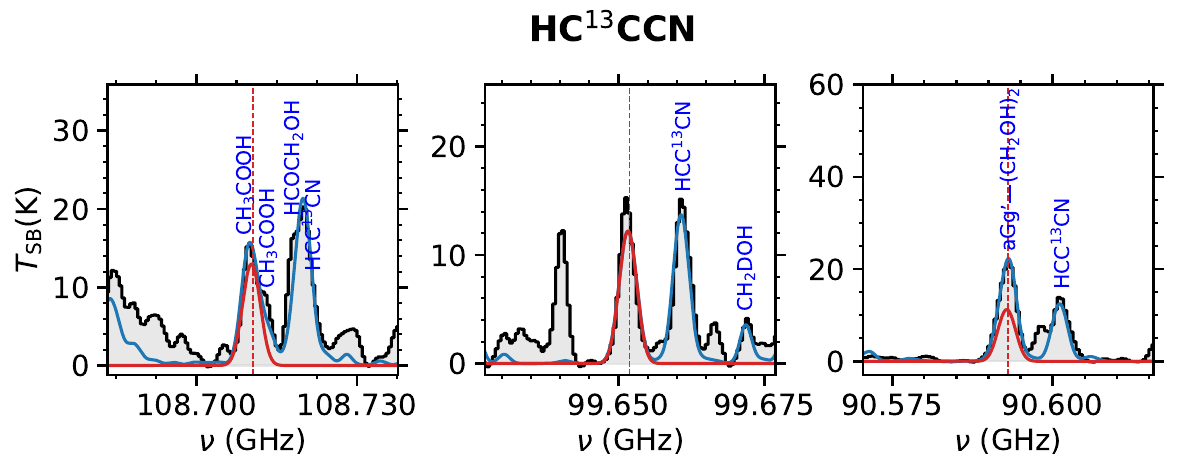}
     \end{subfigure}

     \begin{subfigure}{\textwidth}
         \centering
\includegraphics[scale=0.5]{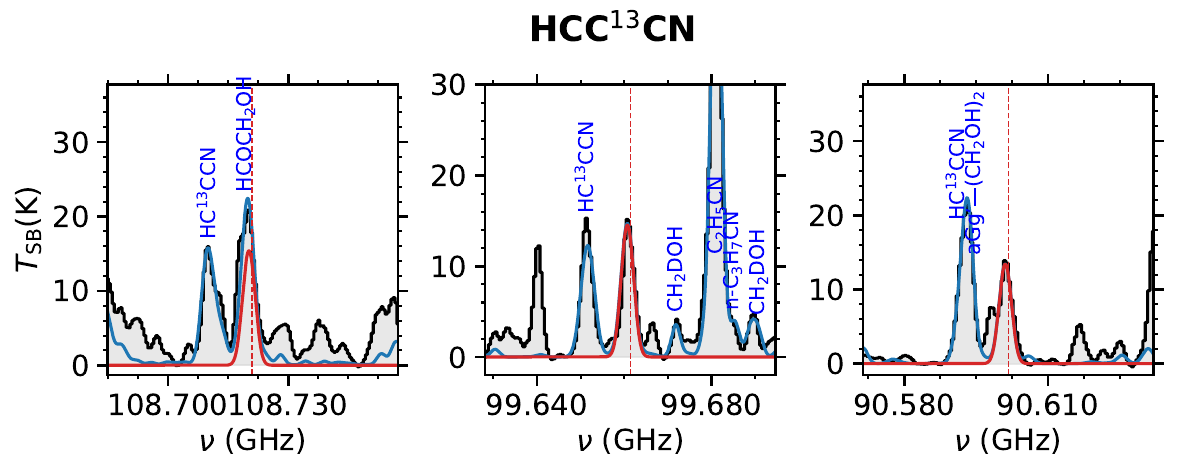}
     \end{subfigure}
        \caption{\ch{HC3N} isotopologues described in Sect. \ref{mols:HC3N}. The black histogram and its grey shadow are the observational spectrum. The red curve is the fit of the individual transition and the blue curve is the cumulative fit considering all detected species. The red dashed lines indicate the frequency of the molecular transitions. The plots are sorted by decreasing intensity of the corresponding species (red line). }
        \label{fig:HC3N}
\end{figure*}

\begin{figure*}
         \centering
         \includegraphics[scale=0.5]{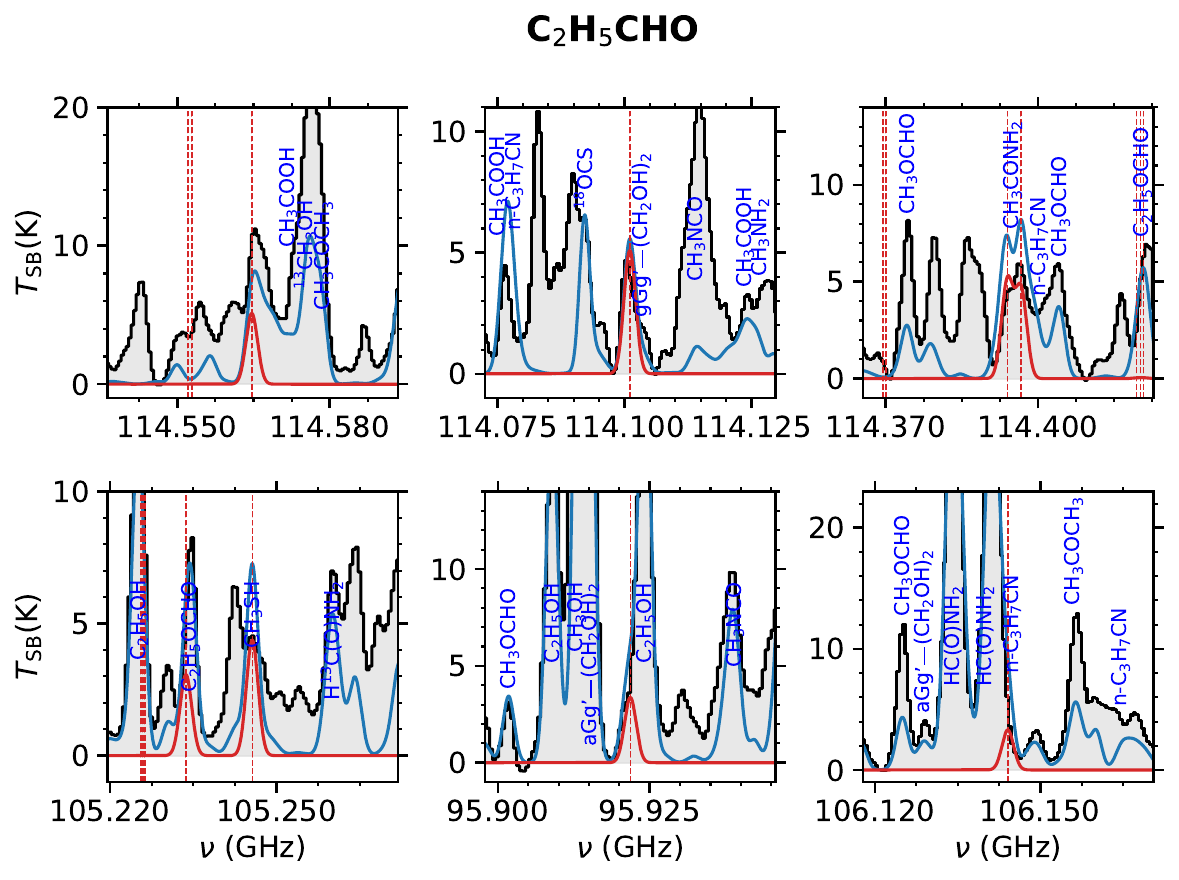}
        \caption{\ch{C2H5CHO} described in Sect. \ref{mols:C2H5CHO}. The black histogram and its grey shadow are the observational spectrum. The red curve is the fit of the individual transition and the blue curve is the cumulative fit considering all detected species. The red dashed lines indicate the frequency of the molecular transitions. The plots are sorted by decreasing intensity of the corresponding species (red line).}
         \label{fig:C2H5CHO}
\end{figure*}

\begin{figure*}
     \centering
     \begin{subfigure}{\textwidth}
              \centering
        \includegraphics[scale=0.5]{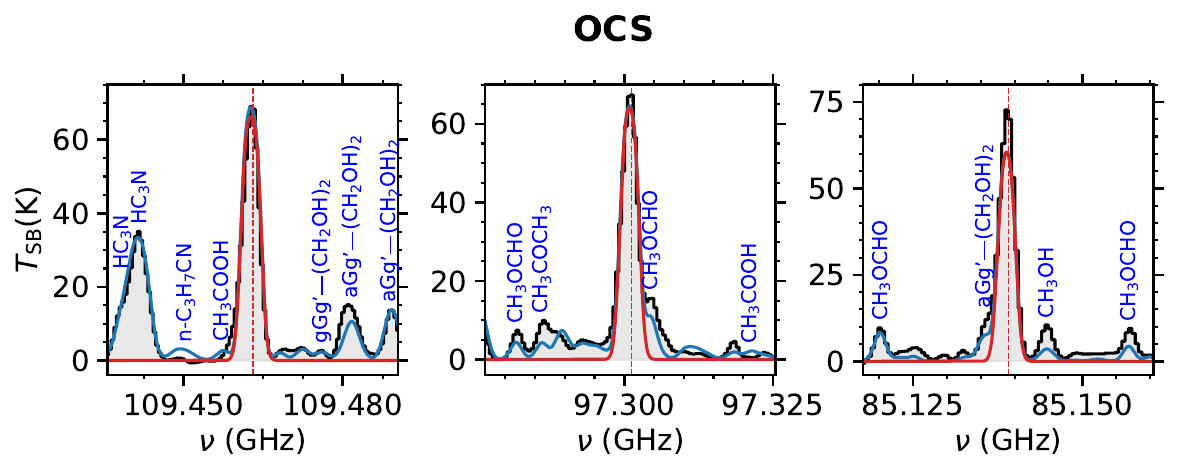}
     \end{subfigure}

     \begin{subfigure}{\textwidth}
         \centering
\includegraphics[scale=0.5]{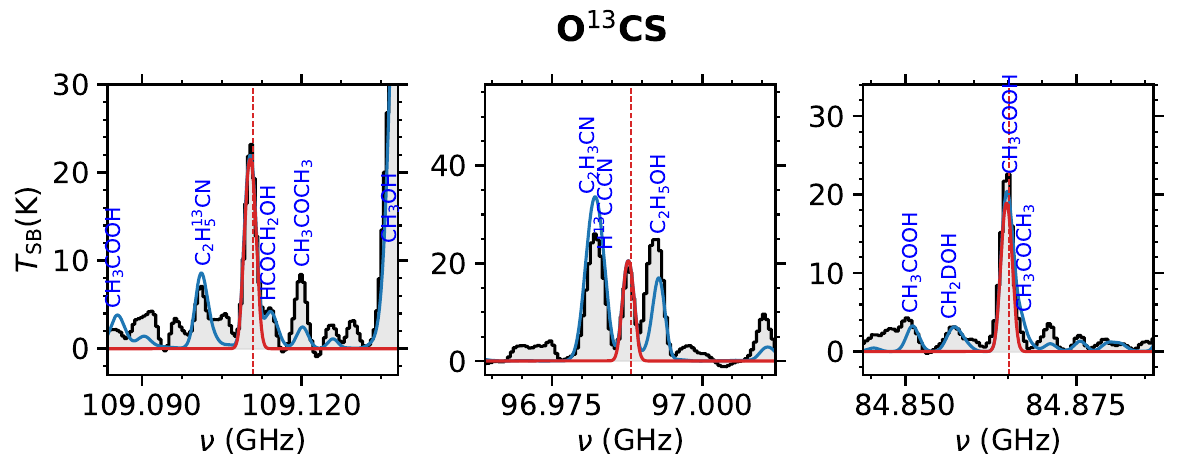}
     \end{subfigure}

     \begin{subfigure}{\textwidth}
         \centering
\includegraphics[scale=0.5]{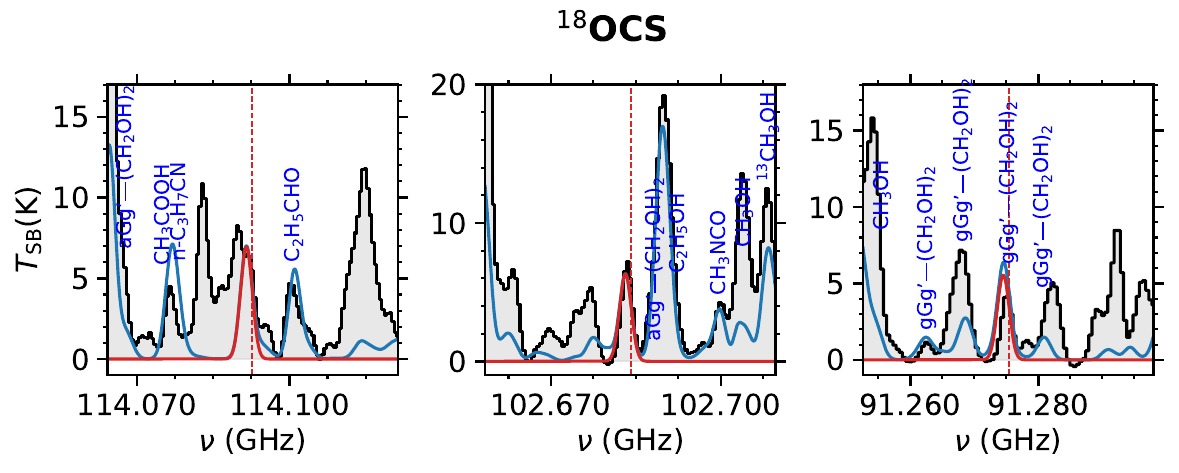}
     \end{subfigure}

          \begin{subfigure}{0.49\textwidth}
         \centering
\includegraphics[scale=0.5]{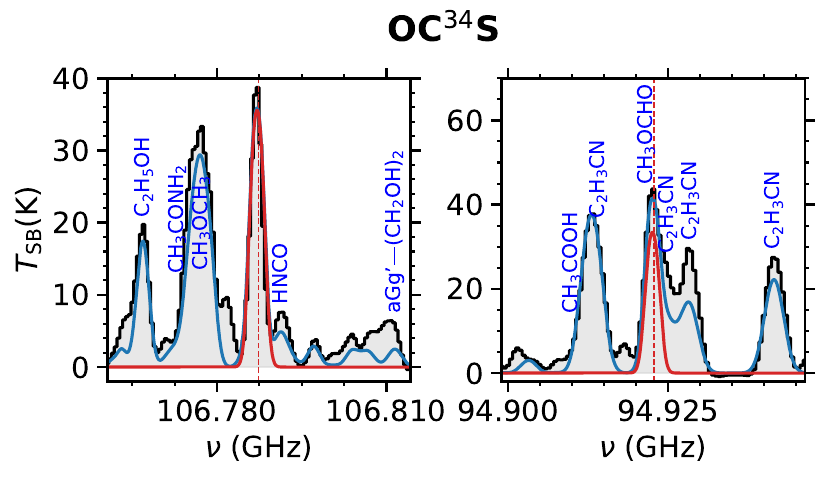}
     \end{subfigure}
          \hspace{-1cm}
          \begin{subfigure}{0.49\textwidth}
         \centering
\includegraphics[scale=0.5]{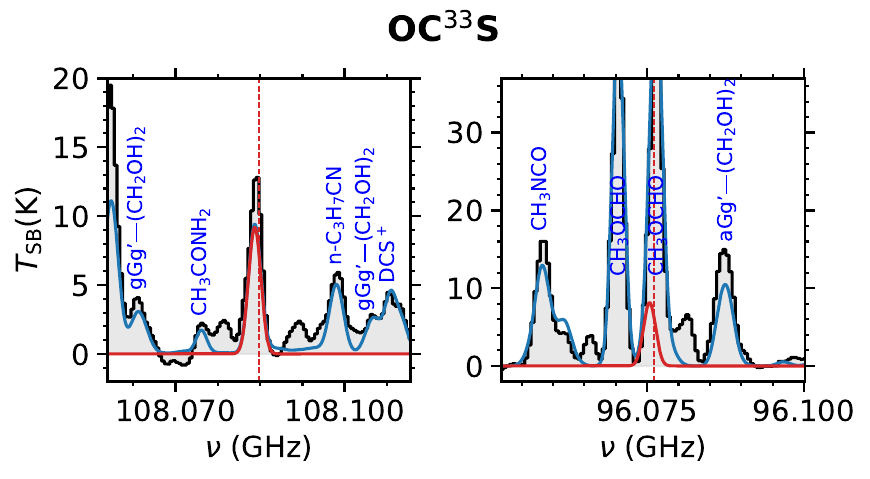}
     \end{subfigure}
        \caption{OCS isotopologues described in Sect. \ref{mols:OCS}. The black histogram and its grey shadow are the observational spectrum. The red curve is the fit of the individual transition and the blue curve is the cumulative fit considering all detected species. The red dashed lines indicate the frequency of the molecular transitions. The plots are sorted by decreasing intensity of the corresponding species (red line).}
        \label{fig:OCS}
\end{figure*}

\begin{figure*}
\includegraphics[scale=0.5]{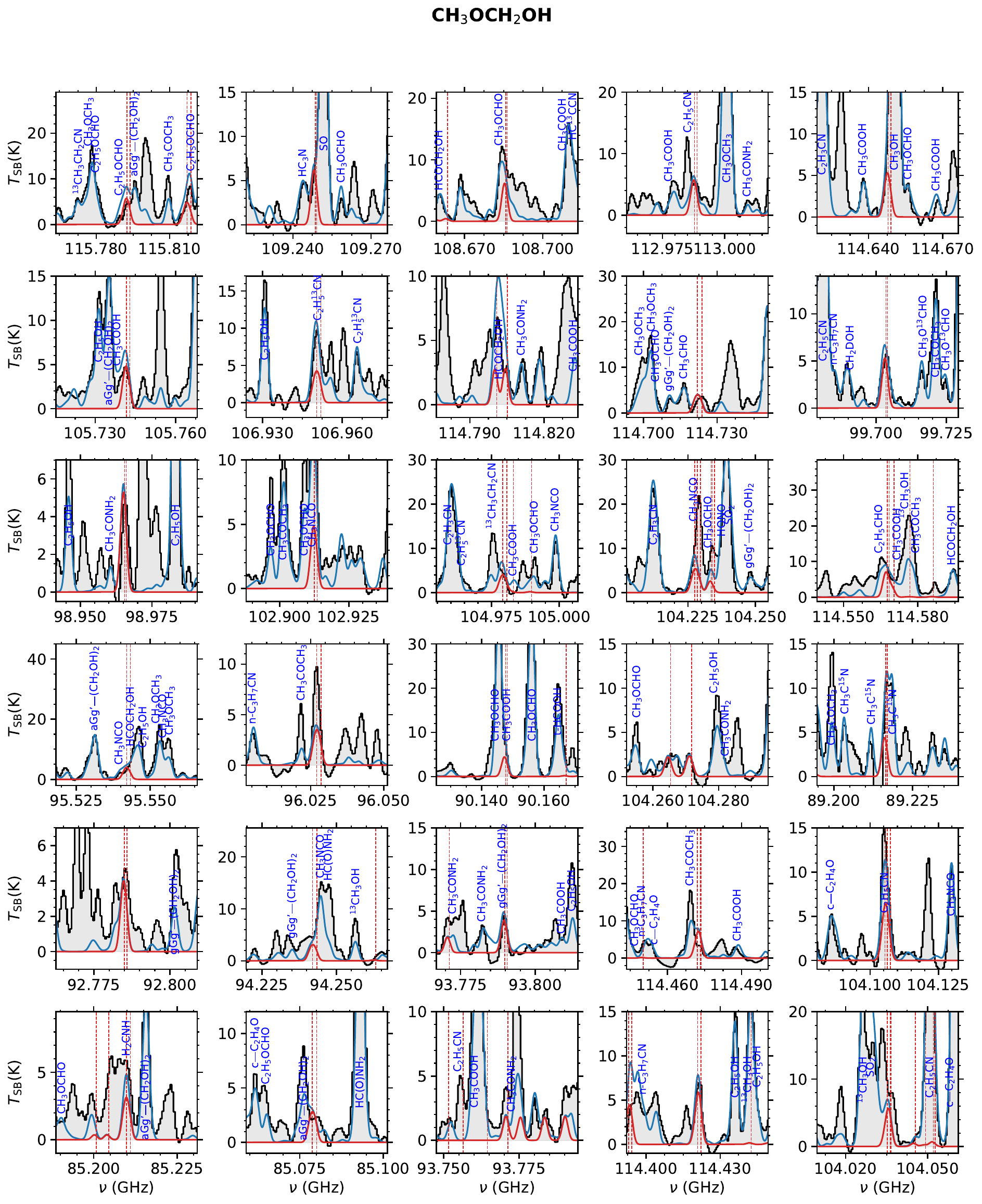}
\caption{\ch{CH3OCH2OH} described in Sect. \ref{mols:CH3OCH2OH}. The black histogram and its grey shadow are the observational spectrum. The red curve is the fit of the individual transition and the blue curve is the cumulative fit considering all detected species. The red dashed lines indicate the frequency of the molecular transitions. The plots are sorted by decreasing intensity of the corresponding species (red line).}
\label{fig:CH3OCH2OH}
\end{figure*}

 \begin{figure*}
     \centering
     \begin{subfigure}{\textwidth}
              \centering
\includegraphics[scale=0.5]{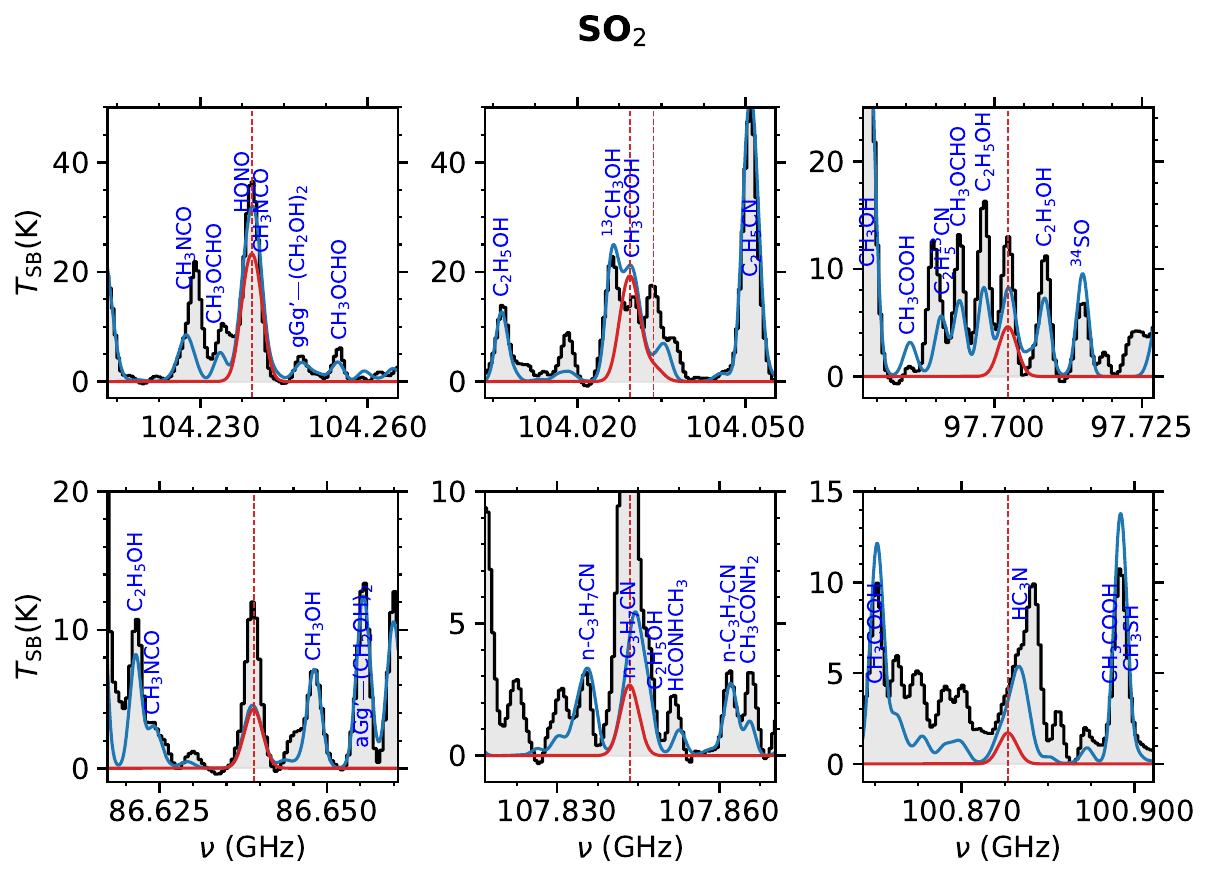}
     \end{subfigure}
     \begin{subfigure}{\textwidth}
         \centering
\includegraphics[scale=0.5]{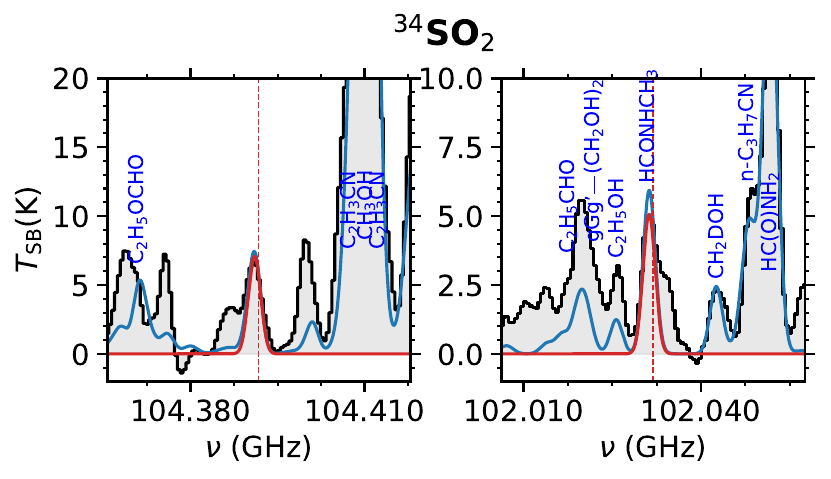}
     \end{subfigure}
     \begin{subfigure}{\textwidth}
         \centering
\includegraphics[scale=0.5]{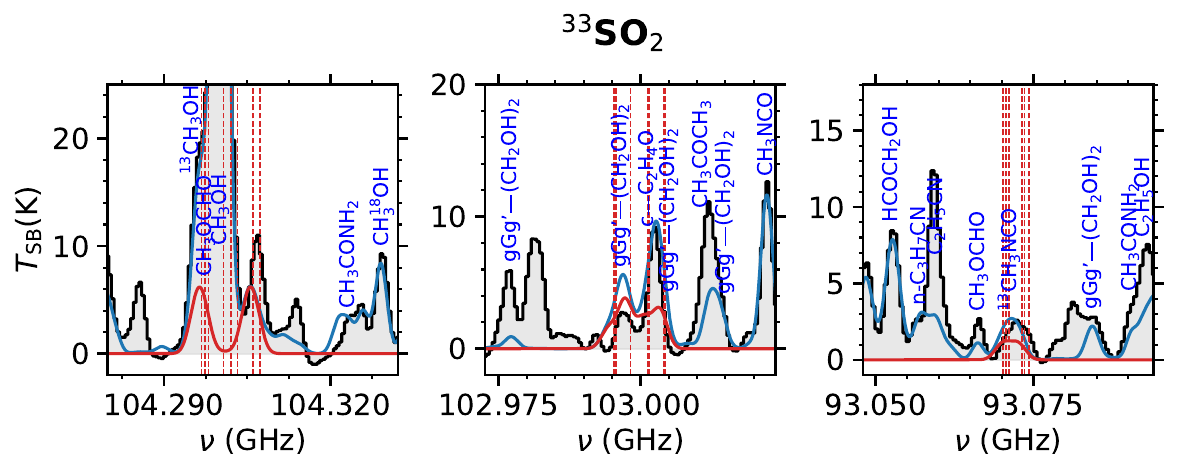}
     \end{subfigure}
        \caption{\ch{SO2} isotopologues described in Sect. \ref{mols:SO2}. The black histogram and its grey shadow are the observational spectrum. The red curve is the fit of the individual transition and the blue curve is the cumulative fit considering all detected species. The red dashed lines indicate the frequency of the molecular transitions. The plots are sorted by decreasing intensity of the corresponding species (red line).}
        \label{fig:SO2}
\end{figure*}

\section{Molecules not detected towards G31.41+0.31}
\label{sec:Molecules_not_detected_G31}

\begin{figure*}
     \centering
     \begin{subfigure}{0.24\textwidth}
              \centering
         \includegraphics[scale=0.5]{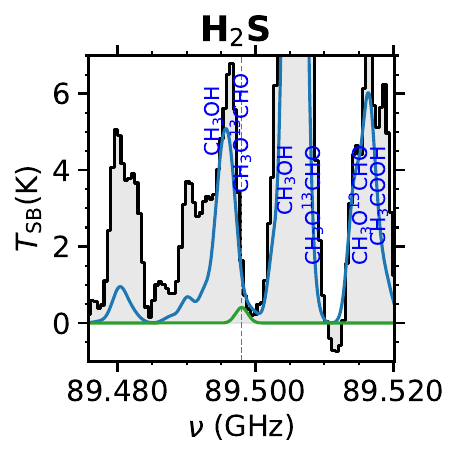}
     \end{subfigure}
     \begin{subfigure}{0.24\textwidth}
         \centering
        \includegraphics[scale=0.5]{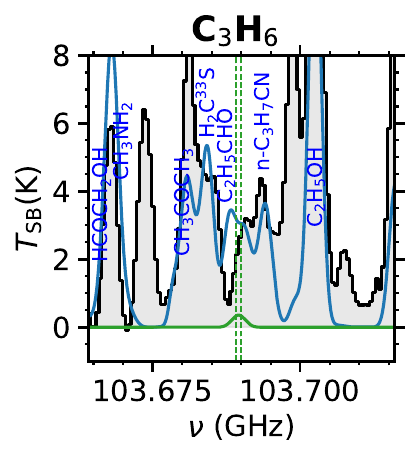}
     \end{subfigure}
     \begin{subfigure}{0.24\textwidth}
         \centering
         \includegraphics[scale=0.5]{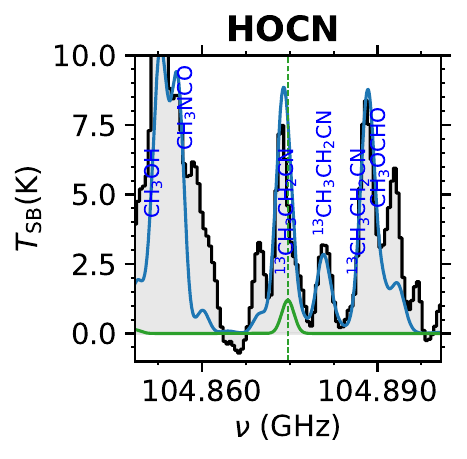}
     \end{subfigure}
               \begin{subfigure}{0.24\textwidth}
         \centering
         \includegraphics[scale=0.5]{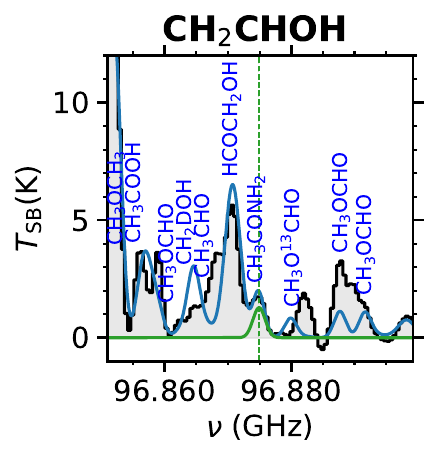}
     \end{subfigure}
          \begin{subfigure}{0.24\textwidth}
         \centering
         \includegraphics[scale=0.5]{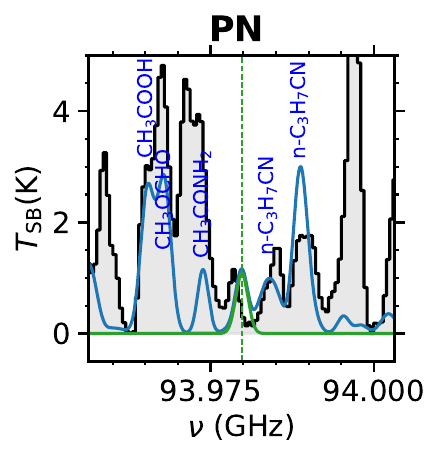}
     \end{subfigure}
     \begin{subfigure}{0.24\textwidth}
         \centering
         \includegraphics[scale=0.5]{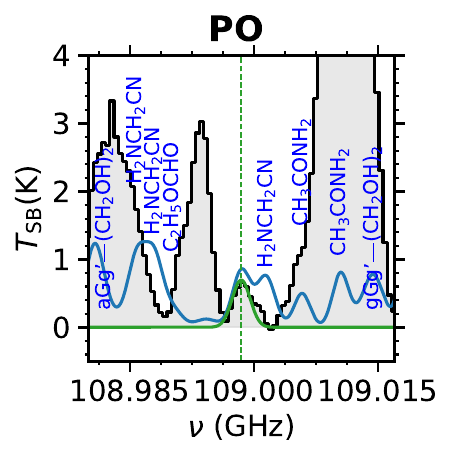}
     \end{subfigure}
          \begin{subfigure}{0.24\textwidth}
         \centering
         \includegraphics[scale=0.5]{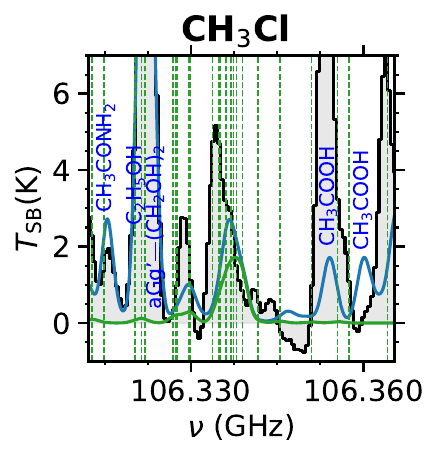}
     \end{subfigure}
               \begin{subfigure}{0.24\textwidth}
         \centering
         \includegraphics[scale=0.5]{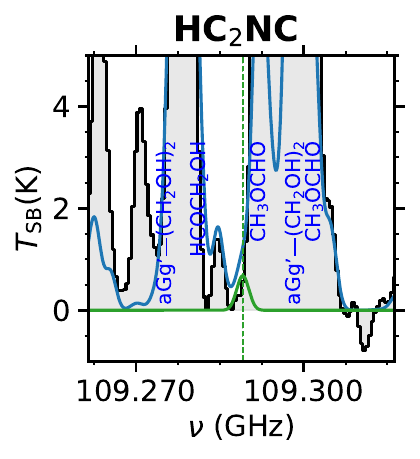}
     \end{subfigure}
          \begin{subfigure}{0.24\textwidth}
         \centering
         \includegraphics[scale=0.5]{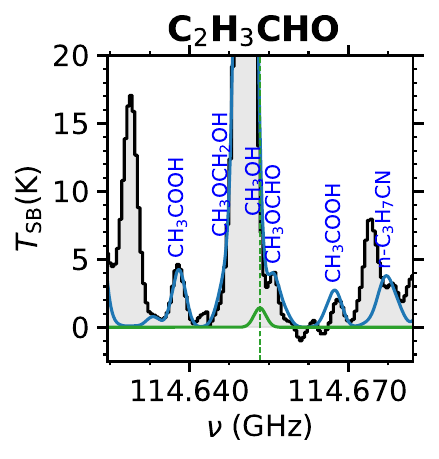}
     \end{subfigure}
     \begin{subfigure}{0.24\textwidth}
         \centering
         \includegraphics[scale=0.5]{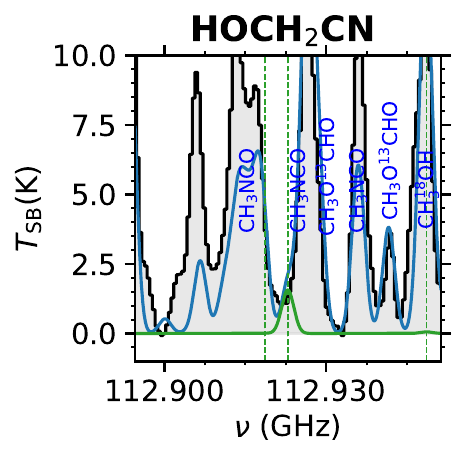}
              \end{subfigure}
               \begin{subfigure}{0.24\textwidth}
         \centering
         \includegraphics[scale=0.5]{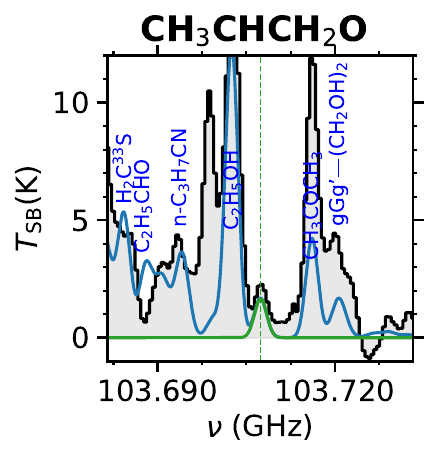}
     \end{subfigure}
          \begin{subfigure}{0.24\textwidth}
         \centering
         \includegraphics[scale=0.5]{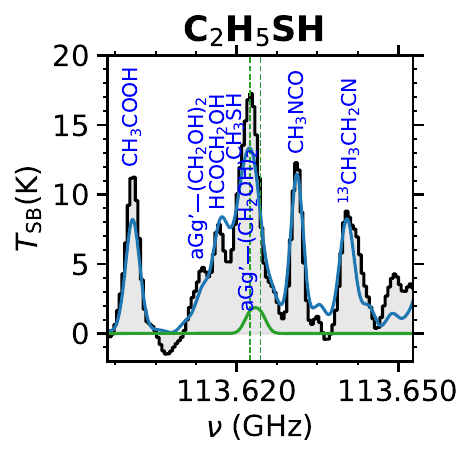}
     \end{subfigure}
       \begin{subfigure}{0.24\textwidth}
         \centering
         \includegraphics[scale=0.5]{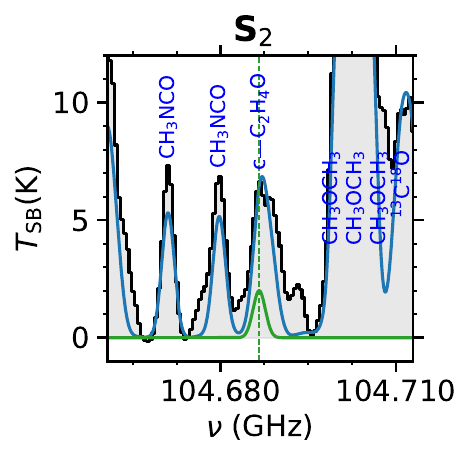}
     \end{subfigure}
          \begin{subfigure}{0.24\textwidth}
         \centering
         \includegraphics[scale=0.5]{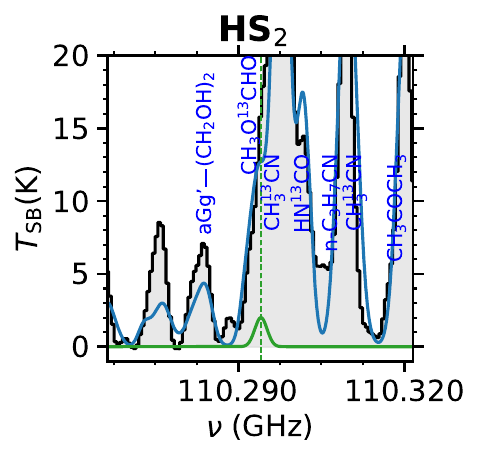}
     \end{subfigure}
               \begin{subfigure}{0.24\textwidth}
         \centering
         \includegraphics[scale=0.5]{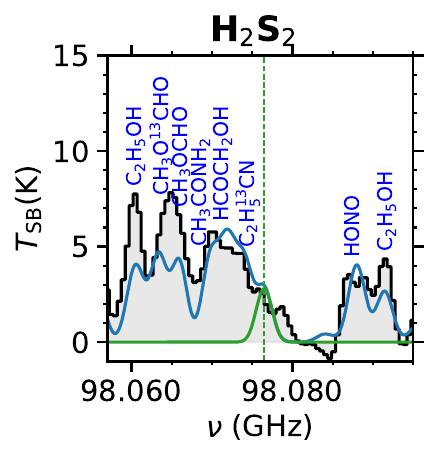}
     \end{subfigure}
          \begin{subfigure}{0.24\textwidth}
         \centering
         \includegraphics[scale=0.5]{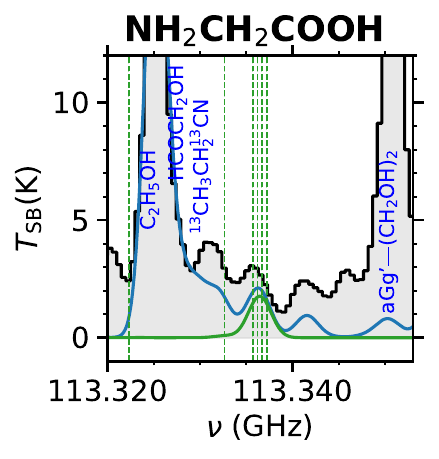}
     \end{subfigure}
        \caption{Molecules not detected in G31.41. The black histogram and its grey shadow are the observational spectrum. The LTE synthetic spectra using the derived upper limits of $N$ are indicated with green curves. To compute the upper limits of their molecular abundances we have used the brightest and less blended transitions, which are shown here. The green dashed lines indicate the frequency of the molecular transitions. The transitions of this plots are in Table \ref{tab:Transitions_Molecules_on_G31} on the not detected molecules section.}
        \label{fig:upper-limits}
\end{figure*}

\begin{figure}
     \centering  
\includegraphics[width=0.45\textwidth]{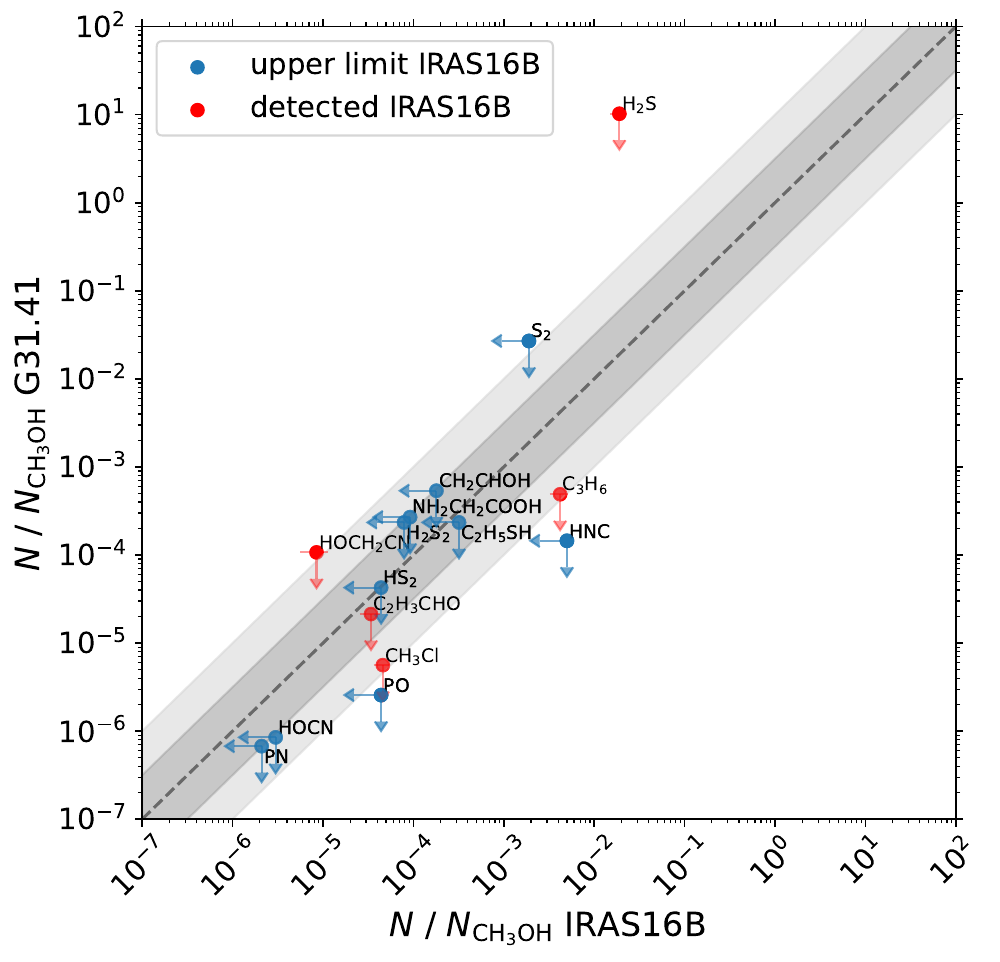}
\caption{Molecular abundances of G31.41 upper limits with respect to \ch{CH3OH} compared with their respective values of IRAS16B. The errorbars marked with arrows are the upper limits. Grey dotted line is the 1:1 relation on each plot. Dark and light grey are half and one order of magnitude difference with respect to the grey dotted line.}
\label{fig:uplims_g31_iras}
\end{figure}

The spectra of the transitions used to obtain the column density upper limits towards G31.41 presented in Table \ref{tab:Poperties_molecules_on_G31} are shown in Fig. \ref{fig:upper-limits}, and the information about the transitions used are included in Table \ref{tab:Transitions_Molecules_on_G31}. The upper limits are calculated by using the brightest transitions according to the LTE model that are not heavily blended, and by performing a visual inspection to make the upper limits compatible with the observed spectra (explained in more detail in Sect \ref{sec:not_detceted_mols}).
To assess how constraining (or not) are the derived upper limits, in Fig. \ref{fig:uplims_g31_iras} we show how they compare, normalized to \ch{CH3OH}, with the fractional abundances or upper limits derived towards IRAS16B.
Most of the molecules not detected towards G31.41 were also reported as upper limits towards IRAS16B, and they fall within or very close to the region of $\pm$1 order of magnitude around the 1:1 line (light grey shaded area of Fig. \ref{fig:uplims_g31_iras}). Only five species of the sample are not detected towards G31.41 but detected in IRAS16B: \ch{H2S}, \ch{HOCH2CN}, \ch{C2H3CHO}, \ch{CH3Cl}, and \ch{C3H6}. The upper limit of \ch{H2S} we derived in G31.41 is not constraining because there are not adequate transitions of this species in the ALMA Band 3. As shown in Table \ref{tab:Transitions_Molecules_on_G31}, the transition used to derive the upper limit, although it is the best in the spectra analyzed, it is relatively weak and has a high energy level; and as a consequence the derived upper limit is high. The upper limit derived for \ch{HOCH2CN} is not particularly constraining either,  around one order of magnitude higher than the value derived from its detection in IRAS16B. In contrast, for \ch{C2H3CHO}, \ch{CH3Cl} and \ch{C3H6}, the upper limits are below the values derived in IRAS16B, indicating that these species are less abundant in G31.41.

\bsp	
\label{lastpage}
\end{document}